\documentclass[useAMS,usenatbib]{mnras}

\usepackage{amsmath}
\usepackage{amssymb}
\usepackage{txfonts}
\usepackage{graphicx}
\usepackage{soul}
\usepackage{subcaption}
\usepackage{caption}
\usepackage{tablefootnote}
\usepackage{physics}

\newcommand{\ud}{\ensuremath{\mathrm{d}}}

\title[CCSN models  with Rotation and Magnetic Fields]{Gravitational Wave Signals from Two-Dimensional Core-Collapse Supernova Models with Rotation and Magnetic Fields}

\author[Jardine, Powell \& M\"uller]{
Rylan~Jardine$^1$,
Jade Powell$^2$, and
Bernhard~M\"uller$^1$\thanks{bernhard.mueller@monash.edu}
\\
$^{1}$School of Physics and Astronomy,
Monash University, Clayton, VIC 3800 Australia\\
$^{2}$OzGrav, Centre for Astrophysics and Supercomputing, Swinburne University of Technology, Hawthorn, VIC 3122, Australia
}

\begin{document}

\maketitle

\begin{abstract}
We investigate the impact of rotation and magnetic fields on the dynamics and gravitational wave emission in 2D core-collapse supernova simulations with neutrino transport. We simulate 17 different models of
$15\,M_\odot$ and $39\,M_\odot$ progenitor stars with various initial rotation profiles and initial magnetic fields strengths up to $10^{12}\, \mathrm{G}$, assuming a dipolar field geometry in the progenitor. Strong magnetic fields generally prove conducive to shock revival, though this trend is not without exceptions. The impact of rotation on the post-bounce dynamics is more variegated, in line with previous studies. A significant impact on the time-frequency structure of the gravitational wave signal is found only for rapid rotation or strong initial fields. For rapid rotation, the angular momentum gradient at the proto-neutron star surface can appreciably affect the frequency of the dominant mode, so that known analytic relations for the high-frequency emission band no longer hold. In case of two magnetorotational explosion models, the deviation from these analytic relations is even more pronounced. One of the magnetorotational explosions has been evolved to more than half a second after the onset of the explosion and shows a subsidence of high-frequency emission at late times. Its most conspicuous gravitational wave signature is a high-amplitude tail signal.
We also estimate the maximum detection distances for our waveforms. The magnetorotational models do not stick out for higher detectability during the post-bounce and explosion phase.
\end{abstract}


\section{Introduction}
After the groundbreaking detection of gravitational waves (GWs) from merging compact binaries \citep{abbott2016firstgw,abbott2017nsmerger}, the new field of GW astronomy still faces further challenges. Among the yet undetected sources of GWs, explosions of massive stars as core-collapse supernovae (CCSNe) are a coveted target. The detection of a CCSN in our own galaxy or its satellites could shed significant insights on the inner workings of the explosions,  the rotational state of the progenitor stars, the structure of the newly born compact remnant, and the nuclear equation of state
\citep[for a review, see][]{kotake_17,kalogera_19,abdikamalov_gravitational_2020}. 

In order to optimise the chances of a detection and maximise the scientific insights from a prospective nearby event, it is critical to thoroughly investigate the structure and physical dependencies of the CCSN GW signal from the collapse to the post-explosion phase based on a broad exploration of factors that influence the dynamics in the supernova core, such as progenitor mass, rotation, magnetic fields, and the nuclear equation of state. The first phase of GW emission in CCSNe, the signal from the collapse and bounce of rotating iron cores is already well understood \citep{2008PhRvD..78f4056D, 2014PhRvD..90d4001A, 2017PhRvD..95f3019R} to the point that it could be used to quantitatively constrain the rotation of the progenitor core in the case of a Galactic event. The impact of magnetic fields on the bounce signal has also been investigated \citep{obergaulinger_06,Scheidegger2008}.
Although less apparent at first glance, the GW signal from the post-bounce phase has also been shown to contain clear fingerprints of the structure of the newly born proto-neutron star (PNS) and the dynamics in the supernova core \citep[e.g.,][]{muller_2013_gw,cerda_13,sotani_16,Morogw2018, 2018ApJ...865...81O,torres_18, andresen_19, 2019MNRAS.487.1178P, radice_characterizing_2019,2020arXiv201002453P, mezzacappa_2020}.  The time-frequency structure of the post-bounce signal shows distinct emission bands that reveal the frequencies of quadrupolar PNS oscillation modes -- the most prominent one being an $l=2$ g- mode, later often replaced by the fundamental f- mode frequency after several hundred milliseconds \citep{Morogw2018,torres_18,sotani_20} -- and in some models 
\citep{kuroda_16,abjgw2017,powell_20}
the standing accretion shock instability \citep[SASI;][]{0004-637X-584-2-971, 2006ApJ...642..401B, 2007ApJ...654.1006F}, and possibly
a triaxial corotation instability in the case of extremely
rapid rotation
\citep{2014PhRvD..89d4011K}.  In recent years, it has become possible to rigorously identify the nature of underlying oscillation modes in the spectrograms by means of a linear eigenmode analysis. While the frequency of the characteristic emission bands reveal the \emph{structure} of the PNS and its environment, the amplitudes of the various signal components reflect dynamics
in the supernova core, i.e., the violence of oscillatory instabilities and of the turbulent flow that excites PNS oscillations \citep{radice_characterizing_2019,2019MNRAS.487.1178P}.

For CCSNe of non-rotating progenitors, the f/g-mode emission band and the SASI emission band can be described by simple, ``universal'' scaling laws in terms of the PNS mass, radius, and surface temperature, and the shock radius \citep{torres-forne_universal_2019}. In the case of a GW detection with a sufficiently high signal-to-noise ratio, these relations could potentially be exploited to quantitatively constrain PNS and shock parameters. For non-rotating models, the overall signal strength (as quantified by the peak amplitude and
overall energy emitted in GWs) also shows correlations with the progenitor core mass and a trend towards stronger GW emission in the case of successful explosions \citep{muller_2013_gw,radice_characterizing_2019,2019MNRAS.487.1178P}. This may provide further qualitative clues about the progenitor and the explosion dynamics in the case of a Galactic supernovae. These patterns emerge fairly consistently among the manifold 2D 
\citep{muller_2004_artrot,marek_gw_2009,yakunin_gravitational_2010,muller_2013_gw,yakunin_16,Morogw2018,pajkos_support}
and 3D \citep{kuroda_16,abjgw2017,2019MNRAS.487.1178P, radice_characterizing_2019,mezzacappa_2020,powell_20,powell_21} studies of the GW signal based on modern neutrino radiation hydrodynamics simulations of non-rotating progenitors despite differences in detail in mode frequencies and GW amplitudes.

However, this picture of GW emission from the post-bounce phase may be substantially affected by rotation and magnetic fields. Already on their own rapid rotation \citep{2014PhRvD..89d4011K,2018MNRAS.475L..91T,summa_rotation-supported_2018} and strong magnetic fields  \citep{obergaulinger_14,mueller_20b,matsumoto_20} can each have a significant impact on CCSN dynamics. Acting in tandem, rapid rotation and strong magnetic fields can give rise to powerful magnetorotational explosions, which have already been explored very actively by means of 2D and 3D simulations \citep{burrows_07,winteler_12,moesta_14b,obergaulinger_17,moesta_18,obergaulinger_20,kuroda_magnetorotational_2020,obergaulinger_21}. The major impact of rapid rotation and/or magnetic fields is bound to leave an imprint on the GW signal as well. Several studies have already shown that (in addition to producing the characteristic bounce signal) rotation can  substantially alter the mode frequencies, affect the amplitudes of different components of the GW signal, and give rise to new, powerful signal features \citep{andresen_19,powell_20,pajkos_support,2014PhRvD..89d4011K,shibagaki_20}. The combined impact of rotation and magnetic fields on the GW signal still merits further investigation, however. Studies of GW emission from rotating magnetised stellar cores have so far been confined to the
collapse, bounce, and early post-bounce phase
\citep{obergaulinger_06,2010A&A...514A..51S,takiwaki_11}. 
Waveform predictions from anelastic long-time magnetohydrodynamic (MHD) simulations of PNS convection
in rapidly rotating neutron stars have recently also become available \citep{raynaud_21}.
The impact of magnetic fields in non-rotating models \citep{obergaulinger_14} and the combined impact
of rotation and magnetic fields \citep{obergaulinger_20} on supernova dynamics have been explored more systematically across a wider range of the parameter space already by means of long-time simulations including MHD and neutrino transport, but for the GW signal, such an exploration of parameter space is still lacking.

In this paper, we therefore conduct a suite of
17 axisymmetric (2D) MHD simulations
with the \textsc{CoCoNuT-FMT} code
to produce gravitational waveforms over a range of initial magnetic fields and rotation rates for two different CCSN progenitors, a $15\,M_\odot$ red supergiant \citep{heger_2005_rotprofile} and a $39\,M_\odot$ helium star \citep{2018ApJm39}. We analyse the combined effect of rotation and magnetic fields on the overall  strength of the GW emission and the characteristic features in the GW spectrogram, and estimate detection distances for our models. We also investigate the impact of rotation and magnetic fields on the pre-explosion and explosion dynamics to complement and corroborate the aforementioned parameter studies of \citet{obergaulinger_14,matsumoto_20}.

Our paper is structured as follows.
In Section~\ref{sec:numerics} we describe the numerical methods and microphysics used in our simulations. The progenitor models and our choice
of initial conditions for the pre-collapse rotation profiles and magnetic fields are discussed in
Section~\ref{sec:progmodels}. In Section~\ref{sec:results}, we first review the dynamical evolution of our models, and then analyse their GW emission. We summarise our findings, discuss their implications, and outline questions for future work in Section~\ref{sec:conclusions}.

\section{Numerical Methods}
\label{sec:numerics}
We analyse the GW emission in several 2D simulations performed with the Newtonian version of the neutrino MHD code \textsc{CoCoNuT-FMT}. The equations of Newtonian MHD are solved in spherical polar coordinates using a finite-volume scheme employing higher-order reconstruction and the HLLC solver of \cite{gurski_hllc-type_2004,miyoshi_multi-state_2005}, in conjunction with hyperbolic divergence cleaning \citep{dedner_hyperbolic_2002}.
The  MHD equations 
for the density $\rho$, magnetic field $\mathbf{B}$, total energy density $e$, velocity $\mathbf{v}$ and Lagrangian multiplier $\psi$ are expressed in Equations~\eqref{eq:cont}--\eqref{eq:divclean}
in Gaussian units including divergence
cleaning terms as
\begin{align}
\partial_{t} \rho+\nabla \cdot(\rho \mathbf{v}) &=0,   \label{eq:cont}\\
\label{eq:mhd2}
\partial_{t}(\rho \mathbf{v})+\nabla \cdot\left[\rho \mathbf{v} \mathbf{v}^{\mathrm{T}}+\left(p+\frac{\mathbf{B}^{2}}{8\pi}\right) \mathcal{I}-\frac{\mathbf{B} \mathbf{B}^{\mathrm{T}}}{4\pi}\right] &=\rho \mathbf{g}+\mathbf{Q}_\mathrm{m}
-\frac{(\nabla \cdot \mathbf{B}) \mathbf{B}}{4\pi}, \\
\partial_{t} \mathbf{B}+\nabla \cdot\left(\mathbf{v} \mathbf{B}^{\mathrm{T}}-\mathbf{B} \mathbf{v}^{\mathrm{T}}+\psi \mathcal{I}\right) &=0,  \\
\label{eq:mhd4}
\partial_{t} e+\nabla \cdot\left[\left(e+p+\frac{\mathbf{B}^{2}}{8\pi} \right) \mathbf{v}-\mathbf{B}(\mathbf{v} \cdot \mathbf{B})\right] &=
\rho \mathbf{v}\cdot \mathbf{g}+Q_\mathrm{e}\\
&+\mathbf{Q}_\mathrm{m}\cdot\mathbf{v}
-\frac{\mathbf{B} \cdot \nabla (c_\mathrm{h} \psi)}{4\pi},  
\nonumber
\\
\partial_{t} \psi+c_\mathrm{h} \nabla \cdot \mathbf{B} &=-\frac{\psi}{\tau} . 
\label{eq:divclean}
\end{align}
Here $c_\mathrm{h}$ denotes the hyperbolic cleaning speed, $p$ the gas pressure, $\tau$ the damping time for
the Lagrangian multiplier, and $Q_\mathrm{e}$ and $\mathbf{Q}_\mathrm{m}$ are the neutrino energy
and momentum source terms.
The cleaning speed $c_\mathrm{h}$ is identified with the fast magnetosonic velocity, and the damping
time is set to eight times the magnetosonic crossing time of a cell.
Note that we use a symmetrised form
of the cleaning terms, which gives
$\psi$ the same dimension as the magnetic
field and has superior stability properties. 
One of our models (m15afB12, Section~\ref{sec:progmodels}) was recalculated with
the cleaning scheme of \citet{mueller_20b} to improve
stability during the explosion phase and eliminate
occasional rapid oscillations in the time step limit.
We utilise the effective relativistic potential of case `Arot' in \cite{effectivepotentialsmuller2008}.

The fast multi-group transport (FMT) method 
of \citet{fmtmethod2015} is used for the neutrino transport,
to obtain the neutrino source terms $\mathbf{Q}_\mathrm{m}$,
$Q_\mathrm{e}$, and
$Q_{Y_\mathrm{e}}$ in the momentum and energy
equations (\ref{eq:mhd2},\ref{eq:mhd4}), and the electron number
source term in the equation for the electron fraction $ Y_\mathrm{e}$,
\begin{equation}
 \partial_{t} (\rho  Y_\mathrm{e}) +\nabla \cdot(\rho \mathbf{v} Y_\mathrm{e}) =Q_{Y_\mathrm{e}}.
\end{equation}
We use a spatial resolution of $550\times 128 $ 
in radius $r$ (with non-equidistant spacing) and angle $\theta$, and energy groups for our neutrino transport. At high densities, we employ the equation of state of \citet{lattimer_swesty_1991}   with a bulk incompressibility of $K = 220\, \mathrm{MeV}$. The low-density equation of state accounts for an ideal gas of nuclei alongside photons, electrons and positrons combined with a flashing treatment for nuclear reactions \citep{rampp_radiation_2002}. 

The innermost $10\, \mathrm{km}$ of the grid are treated using a spherical 1D grid. In the 1D core region, all thermodynamic quantities and the meridional velocity component are assumed to be spherically symmetric. However, the 1D core is allowed to rotate rigidly, and contains a divergence-free magnetic field with a constant component $B_z$ along the grid axis, and a constant
toroidal field $B_\varphi$.

\section{Progenitor Models and Simulation Setup}\label{sec:progmodels}
We simulate 17 different models using two different progenitor stars, namely the $15\, M_{\odot}$ progenitor m15b6 from \citet{heger_2005_rotprofile} and the $39 \, M_{\odot}$ progenitor  m39 from  \citet{2018ApJm39}. The parameters of the 17 models are summarised in 
 Table~\ref{tab:modelparams}. The model
 labels denote the progenitor (m15 or m39), the initial rotation profile, and the initial
 magnetic field strength, e.g., model m15nrB10 uses the $15\,M_\odot$ progenitor
 with no rotation and an initial maximum field strength of $10^{10}\, \mathrm{G}$. The progenitor models, rotation profiles, and
 initial magnetic field configurations are explained in more detail below.

Model m15b6, has been evolved up to collapse
using the stellar evolution code \textsc{Kepler}
with magnetic torques \citep{heger_2005_rotprofile}
based on the Tayler-Spruit dynamo \citep{spruit_differential_1999,spruit_dynamo_2002}, and has a moderate
core spin rate.
Its central rotation rate, of $0.05\,  \mathrm{rad}\, \mathrm{s}^{-1}$
at the onset of collapse translates to a birth spin period of $\mathord{\sim} 11\, \mathrm{ms}$ for a nascent neutron star assuming that angular momentum is not exchanged between the collapsing core and the ejecta during the explosion. This amounts to a rotational energy of $ \mathord{\sim }2 \times 10^{50}\, \mathrm{erg}$, too low to power a CCSN by magnetorotational effects alone.
The default progenitor rotation
profile of m15b6 is denoted `ps' 
(in the model labels)
for `progenitor, slow rotation'  in Table~\ref{tab:modelparams}.

We also simulate model m15b6 without rotation
(model label `nr' for no rotation) and with a
artificial fast rotation profile, labelled `af' in the model names. The `af' profile is adapted from the `artrot' profile used previously in the works \citet{summa_rotation-supported_2018,muller_2004_artrot,buras_two-dimensional_buras_artrot,marek_delayed_2009_artrot} and involves a rotation profile that changes from uniform to differential rotation at the edge of the iron core. For model m15b6, it entails a constant angular velocity
 of $\Omega=0.5\, \mathrm{rad}\, \mathrm{s}^{-1}$
within the iron core, $r \, <\,1500 \, \mathrm{km}$. Outside of the core the angular velocity decreases as  $\Omega\propto r^{-3/2}$. 

\begin{figure}
    \centering
    \includegraphics[width=1\linewidth]{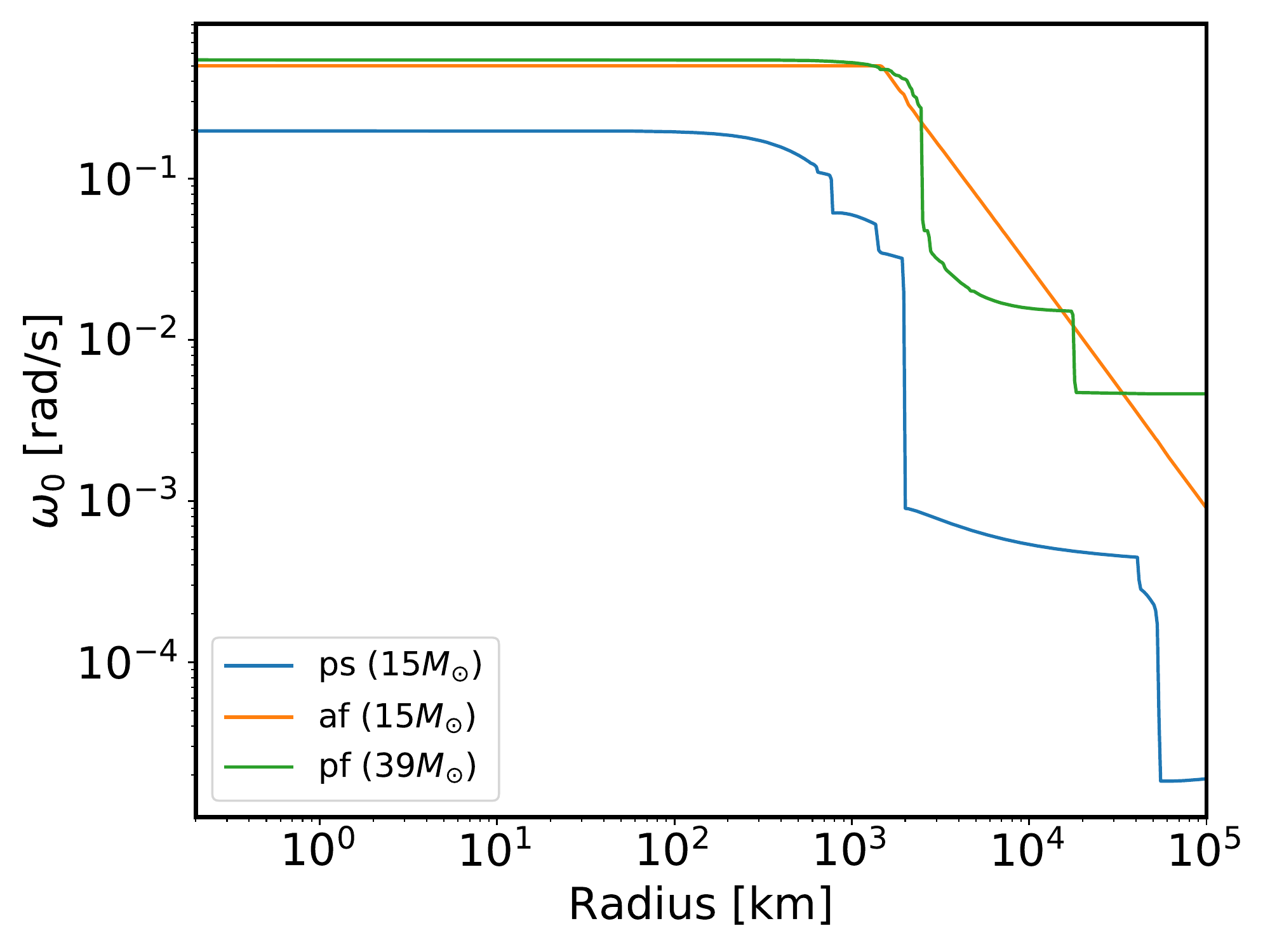}
    \caption{Initial rotation profiles of each of our models at the onset of collapse.}
    \label{fig:rotprofile}
\end{figure}

Model m39 \citep{2018ApJm39} is a rapidly rotating Wolf-Rayet star
with a low metallicity of $1/50\, Z_{\odot}$,
 a pre-collapse mass of
$22\, M_{\odot}$ and an initial helium star mass of $39\, M_{\odot}$. 
The model
has been computed using 
\textsc{Mesa}
(Modules for Experiments in Stellar Astrophysics; \citealp{mesa}), again using magnetic torques following
\citet{spruit_dynamo_2002} and \citet{heger_2005_rotprofile}.
Starting with a fast initial surface rotation velocity of $600 \, \mathrm{km}\, \mathrm{s}^{-1}$, the model retains substantial
angular momentum in the core 
such that it would collapse to a neutron star with a spin period
of $\mathord{\sim} 4.15\, \mathrm{ms}$ under
the assumption of angular momentum conservation, and with a rotational energy of $2.6 \times 10^{51}\, \mathrm{erg}$, sufficient to undergo a magnetorotational explosion
with moderate energy. The high initial mass of the star is thought to be sufficient for the model to form a black hole with a mass of $\sim 5\,M_{\odot}$. Additionally, 
the high specific angular momentum  $j$ in the shells
surrounding the core
($j=9.36 \times 10^{15}\, \mathrm{cm}^2\, \mathrm{s}^{-1}$ at $5\,M_{\odot}$)
implies the progenitor would also be a candidate for a hypernova explosion and long duration gamma-ray burst (lGRB) within the 
collapsar model. This model was investigated both with its default rotation profile from the stellar evolution calculation 
(model label `pf' for `progenitor, fast rotation') 
and also without rotation (model label `nr'). The initial core angular velocities, $\omega_0$, are presented in Table \ref{tab:modelparams} and initial rotation profiles are presented in Figure \ref{fig:rotprofile}.

For both progenitors and every rotation
profile, we investigate models with initial maximum field
strengths $b_{ \mathrm{p},0}$ and $b_{ \mathrm{t},0}$
for the poloidal and toroidal component of the magnetic field of $0\,\mathrm{G}$, $10^{10}\, \mathrm{G}$ and  $10^{12}\, \mathrm{G}$ in the stellar core. In addition, one model, 
m15psB11, with a field strength of  $10^{11}\,\mathrm{G}$ was included
for the $15\, M_\odot$ progenitor
with the default rotation profile. In all cases the initial magnitudes of both the toroidal and poloidal components of the magnetic field were identical. 

We employ a dipole magnetic field geometry similar to the implementation of \citet{obergaulinger_17}, taking the following form:
\begin{align}
    (A_r,A_{\theta},A_{\varphi})=\left(r b_{ \mathrm{t},0} \frac{r_0^3}{(r_0^3+r^3)} \cos \theta,0,\frac{r}{2}b_{ \mathrm{p},0} \frac{r_0^3}{(r_0^3+r^3)} \sin \theta \right),
\end{align}
with $r_0=10^8\, \mathrm{cm}$.

The actual magnetic field strengths in the
core and inner shells of massive stars
are somewhat uncertain. A range of plausible
estimates can be made based on white dwarf
magnetic fields \citep{ferrario_06,ferrario_15}, dynamo models for 
radiative zones \citep{heger_2005_rotprofile},
and equipartition arguments for the field
in convective zones \citep{christensen_09,mueller_20b}.  The field geometry is even more uncertain.
Following previous studies of MHD
effects in non-rotating \citep{obergaulinger_14,matsumoto_20} and rotating
\citep[e.g.,][]{burrows_07,suwa_07,winteler_12,moesta_14b,obergaulinger_17,obergaulinger_20,kuroda_magnetorotational_2020} CCSN models, we use initial fields of $10^{10}$ G and $10^{12}$ G, to study the impact
of moderate and high magnetisation
 on the dynamics and the GW signal,
respectively, and contrast these with  non-magnetised models as controls. 
Pre-collapse fields of $10^{12}\,\mathrm{G}$ are sufficiently high to reach dynamically relevant field strengths (for powering
magnetically-dominated explosions) via field compression during collapse alone, whilst secondary amplification mechanisms
like the magnetorotational instability \citep{balbus_91,akiyama_03}
or an $\alpha$-$\Omega$ dynamo \citep{thompson_93,raynaud_20}
would be required to reach this regime
from an initial field strength of $10^{10}\,\mathrm{G}$. Random density fluctuations of $10^{-4}$ were applied in all cells at the onset of collapse in the non-rotating and non-magnetic models 
m15nrB0 and m39nrB0
in order to break spherical symmetry. No explicit perturbations were added in the other models.

\begin{table*}
\centering
 \begin{tabular}{||l c c c c c c c c c c||} 
 \hline
  Name & $M$  & $B$  &  Rotation  & $\omega_0$ & $t_{\text{exp}}$ & $E_{\text{expl}}$  & 
  Explosion 
  &
  $E_{\text{GW}} $ &  $f_{\text{p}}$   &  $A^\mathrm{E2}_{20,\mathrm{max}}$   \\
     &  ($M_{\odot}$)  &  (G) &  profile & ($\mathrm{rad}\, \mathrm{s}^{-1}$) & (s) &  ($10^{50}$ erg) & 
   mechanism
  &
  ($10^{46}$ erg) &   (Hz)  &    (cm) \\
  [0.5ex]
 \hline\hline
  m15psB0 & 15 & $0$ & progenitor, slow  & $2.0\times10^{-1}$ & DNE & --- & --- & \textgreater \, 2.28 & 700 & 47\\
 \hline 
  m15psB10 & 15 & $10^{10}$ & progenitor, slow   & $2.0\times10^{-1}$ &  0.28 & 0.34 
  & neutrino-driven &
  \textgreater \, 5.20 & 1240 & 58\\ 
 \hline 
 m15psB11 & 15 & $10^{11}$ & progenitor, slow   & $2.0\times10^{-1}$ &  0.23 & 0.49 
 & neutrino-driven 
 & \textgreater \, 4.87 & 950 & 46\\
 \hline 
 m15psB12 & 15 & $10^{12}$ & progenitor, slow    & $2.0\times10^{-1}$ & 0.20 & \textgreater \, 4.54 
 & neutrino-driven &
 \textgreater \, 6.25 & 700 & 76\\
 \hline 
 m15afB0 & 15 & $0$ & artificial, fast rotation & $5.0 \times 10^{-1}$ &  DNE & ---  & --- & \textgreater \, 1.29 & 1350 & 50\\
 \hline 
 m15afB10 & 15 & $10^{10}$ & artificial, fast & $5.0 \times 10^{-1}$  & DNE& --- & --- & \textgreater \, 0.98 & 730 & 40\\
 \hline 
 m15afB12 & 15 & $10^{12}$ & artificial, fast & $5.0 \times 10^{-1}$  &  0.16   & \textgreater \, 86.48 &
  magnetorotational &
 \textgreater \,1.24 
 & 450  & 56  \\
  \hline 
  m15nrB0 & 15 & $0$ & no rotation & 0 & DNE & --- & --- & \textgreater \, 3.78 & 1160 & 33\\
 \hline 
 m15nrB10 & 15 & $10^{10}$ & no rotation & 0 &  DNE & --- & --- & \textgreater \, 1.99 & 1230 & 38\\
 \hline 
  m15nrB11 & 15 & $10^{11}$ & no rotation & 0 &  DNE & --- & --- & \textgreater \, 1.79 & 1650 & 42\\
 \hline 
 m15nrB12 & 15 & $10^{12}$ & no rotation & 0 & 0.44 & \textgreater \, 1.01 & 
 neutrino-driven &
 \textgreater \, 2.49 & 1370 & 67\\
 \hline 
 m39pfB0 & 39 & $0$ & progenitor, fast & $5.4 \times 10^{-1}$ & 0.33 & \textgreater \, 2.36 & neutrino-driven &\textgreater \,39.07 & 1140 & 99\\
 \hline 
 m39pfB10 & 39 & $10^{10}$ & progenitor, fast & $5.4 \times 10^{-1}$ & 0.40 & \textgreater \, 3.70 & neutrino-driven & \textgreater \,45.23 & 1460 & 76\\
 \hline 
 m39pfB12 & 39 & $10^{12}$ & progenitor, fast & $5.4 \times 10^{-1}$ & 0.16 & \textgreater \, 6.25  & magnetorotational & \textgreater \, 9.03 & 990 & 62\\
 \hline 
m39nrB0 & 39 & $0$ & no rotation & 0 & DNE & --- & --- & \textgreater \, 23.55 & 1440 & 98\\
 \hline 
m39nrB10 & 39 & $10^{10}$ & no rotation & 0 & 0.26 & \textgreater \, 5.67  & neutrino-driven & \textgreater \, 27.35       & 1310 & 108\\
 \hline 
m39nrB12 & 39 & $10^{12}$ & no rotation & 0 & $\sim 0.5^\dagger$ & \textgreater \, 0.25  &
neutrino-driven & \textgreater \, 11.36 & 1790 & 57\\ 
 \hline
\end{tabular}
\caption{Initial conditions and key outcomes
for the  CCSN simulations performed in this study. 
$M$ is the progenitor mass; simulations with $15\,M_{\odot}$ use progenitor model m15b6, whilst $39\, M_{\odot}$ models use m39. 
$B$ denotes the initial toroidal and poloidal magnetic field strengths at the centre of the star. The rotation profiles are described in detail in Section~\ref{sec:progmodels}. $\omega_0$
is the central angular velocity $\omega_0$; for detailed rotation
profiles see Figure~\ref{fig:rotprofile}.
$t_{\mathrm{exp}}$ denotes the time of explosion,  defined as the time at which the average shock radius exceeds 
$500\, \mathrm{km}$, to the nearest $10\, \mathrm{ms}$. Entries `DNE' (`does not explode') denote models without shock revival. 
$E_{\text{exp}}$ is the diagnostic explosion energy at the end of the simulation, and ``explosion mechanism''
specifies whether the explosion (if one occurs)
is neutrino-driven or magnetorotational.
 $E_{\text{GW}}$ gives the energy emitted in GWs, 
 $f_{\text{p}}$ the frequency corresponding to the peak of the GW energy spectrum,
 and $A^\mathrm{E2}_{20,\mathrm{max}}$ is the peak amplitude of GWs prior to the tail phase.
\\
$^\dagger$ As seen in Figure~\ref{fig:shock}e, while the shock 
has not yet reached a radius of $500\, \mathrm{km}$ within the time of the simulation, model 
m39nrB12 is clearly in the process of exploding.
}
\label{tab:modelparams}
\end{table*}

\section{Results}
\label{sec:results}

\subsection{Impact of Rotation and Magnetic Fields on Post-Bounce Dynamics}\label{sec:expdyn}
A summary of the simulation outcomes for all models, and
if applicable, the explosion times and energies can be found in Table~\ref{tab:modelparams} which illustrates
the sensitivities of the the dynamical evolution
to the initial rotation rate and field strength. 
The explosion energies given in the table are
``diagnostic energies'' at the end of the simulations
\citep{buras_two-dimensional_buras_artrot}, i.e.,
they are computed as the integral of the net
total (kinetic, internal, magnetic, and potential) energy
over the material that is nominally unbound.
Angle-averaged shock radii for all models are presented in
Figure~\ref{fig:shock}.
Our models substantially conform to the trends
and effects seen in previous systematic studies of
rotation in non-magnetised supernova
simulations \citep{summa_rotation-supported_2018}, and
magnetic fields in non-rotating models \citep{obergaulinger_14,matsumoto_20}.
We therefore confine ourselves to a brief descriptive discussion
of the post-bounce dynamics of our models, which mainly serves
to set the ground for the subsequent analysis of the GW signals.

We find a general trend towards earlier shock revival
with rotation and with higher initial magnetic field strength.
Trends in explosion energy
$E_\mathrm{expl}$ are more difficult to discern because
$E_\mathrm{expl}$ generally has not asymptoted to its final values at the end of the simulations yet.
However, the hierarchy of
the explosion energies at the end of the simulations
in Table~\ref{tab:modelparams} generally reflects the hierarchy during the entire rise phase and is not just a momentary snapshot.
The rapidly rotating and strongly
magnetised model 
m15afB12 is the only one to exceed
$10^{51}\, \mathrm{erg}$, even though it is still
not in the hypernova regime with
$E_\mathrm{expl}=1.4 \times 10^{51}\, \mathrm{erg}$
at the end of the simulation. There are, however,
significant exceptions to this trend that are in line
with effects observed in the literature and are best
illustrated by discussing the $15 M_\odot$
and $39 M_\odot$ progenitors separately.

\subsubsection{$15\, M_\odot$ models}
\label{sec:s15}
For the $15\,M_{\odot}$ models, we invariably find that higher magnetisation leads to an earlier explosion, irrespective
of the progenitor's rotation profile. This is consistent with the findings of \citet{obergaulinger_14} for non-rotating
progenitors. Specificially,
the $10^{12}\, \mathrm{G}$  case is the only one among the non-rotating model to develop an explosion because the initial field is strong enough
to roughly reach equipartition between magnetic and kinetic energy
in the gain region after collapse so that the fields substantially
shape the post-shock flow, similar to the strong-field
($10^{12}\, \mathrm{G}$) model of \citet{obergaulinger_14}.

\begin{figure}
    \centering
    \includegraphics[width=\linewidth]{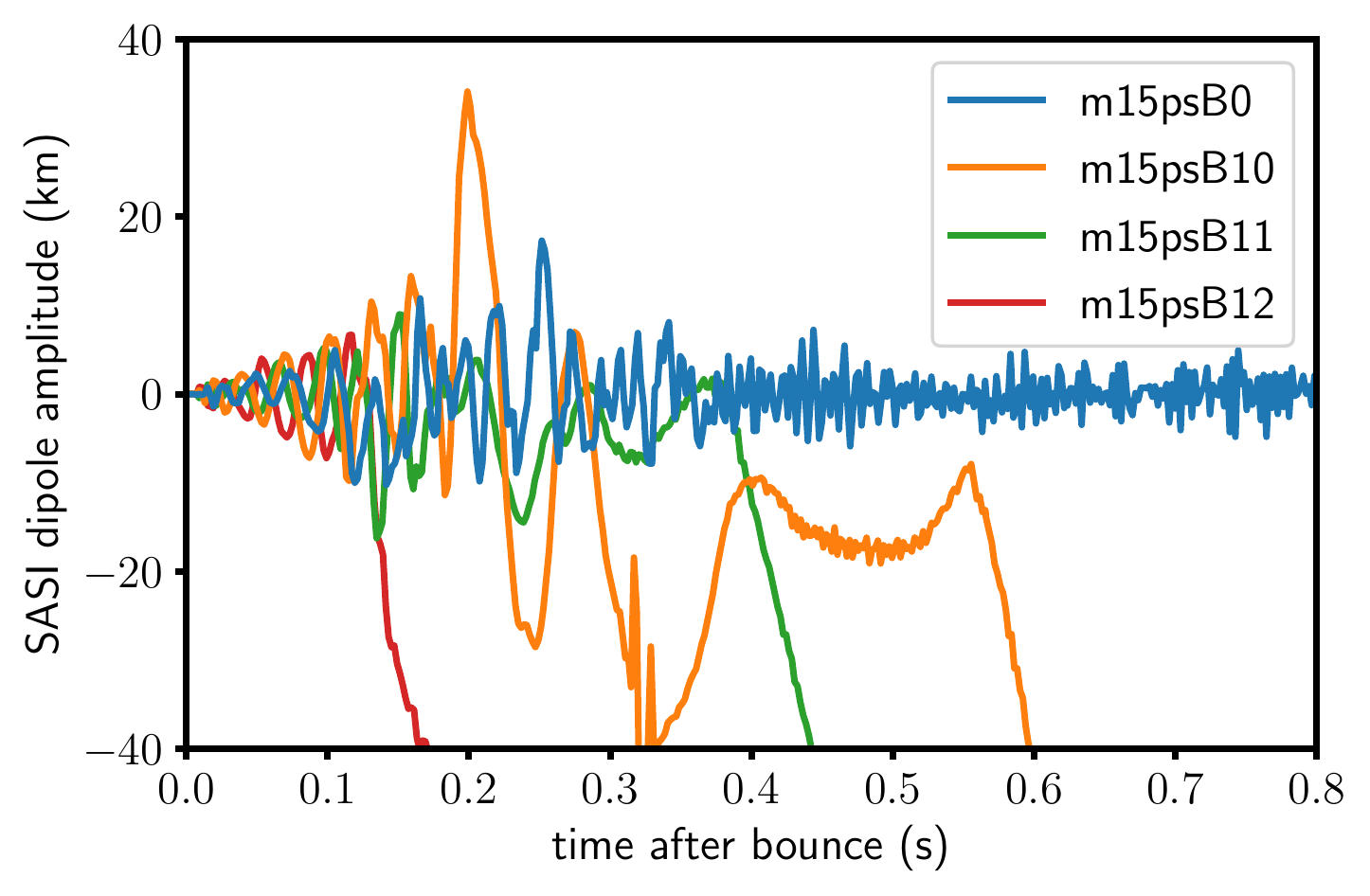}
    \caption{SASI diople amplitude
    for the m15ps models. Despite
    some fluctuations, models m15psB0 starts
    out with markedly smaller SASI amplitudes than
    the other three models prior to the infall
    of the Si/O shell interface, and therefore
    fails to start runaway shock expansion shortly
    afterwards, unlike models m15psB10, m15psB11, and
    m15psB12.
    }
    \label{fig:m15_sasi}
\end{figure}

Interestingly, for the default rotation profile, even a ``weak'' initial magnetic field of $10^{10}\, \mathrm{G}$ makes a difference between explosion and failure (Figure~\ref{fig:shock}a). However, this does not necessarily indicate a robust effect of such weak fields on shock revival. A comparison with the literature shows that the
$15\,M_{\odot}$ model is on the margin between explosion and failure. For example, regardless of rotation rate, none of our $15\,M_{\odot}$ models  explode \emph{without} including magnetic fields, nor do the models of \citet{pajkos_support}
for the same progenitor 
(even though they did not evolve their
simulations far beyond $300\, \mathrm{ms}$ post-bounce).
By contrast, the 2D models with no rotation and
default rotation in \cite{summa_rotation-supported_2018} develop an explosion (whereas their 3D counterparts do not). Under such circumstances, even a minor impact of the (dipole) magnetic field on the growth of instabilities in the post-shock region
in the m15ps models
can tilt the balance towards shock revival by slightly facilitating the emergence of an $\ell=1$ mode in the flow (Figure~\ref{fig:m15_sasi}) so that the shock is already more extended when the Si/O shell
interface reaches the shock 
around $0.2\, \mathrm{s}$ after bounce (Figure~\ref{fig:shock}a). It must be borne in mind that
such subtle effects may generally be swamped by stochastic model variations even though they can be diagnosed for precisely controlled seed perturbation for non-radial instabilities as in our magnetic models.

While magnetic fields always prove beneficial for shock revival in the m15 series, the effect of rotation is non-monotonic, i.e., 
a higher rotation rate in 2D does not guarantee a greater likelihood of explosion, which is consistent with \cite{summa_rotation-supported_2018}. 
The 
m15nr and m15af
models do not explode except in the
presence of strong initial fields of $10^{12}\, \mathrm{G}$. 
As pointed out in the literature, such non-monotonicities arise due to
competing effects in rotating models. On the one hand greater centrifugal support provided by rotation is expected to lead to an extended shock front and, in turn, larger gain region and thus stronger
neutrino heating \citep{summa_rotation-supported_2018}, which  can also be seen in the comparison of the late-time shock trajectory of model m15afB0 vis \'a vis  m15psB0 and  m15nrB0. The resulting increase
of the mass in the gain region can be compensated by lower neutrino
luminosities and mean energies \citep{summa_rotation-supported_2018} due
to the larger radius of a PNS with substantial centrifugal
support. Furthermore, as pointed out by \citep{pajkos_support}, rotation can also hurt shock revival because a positive angular momentum inherited from the progenitor can inhibit convection in the gain region according to the 
Solberg–H\o{}iland criterion \citep[e.g.,][]{kippenhahn,maeder_solberg-hoiland_convection_2013}.\footnote{
As noted by \citet{andresen_19}, rotation can also inhibit convection in the proto-neutron star convection zone, which has implications for gravitational wave emission.}
The analytic rotation profile chosen for the m15af series
indeed exhibits a positive angular momentum gradient. The specific angular momentum $j_z$ scales as $j_z\propto r^2$ inside the core
and $j_z\propto r^{1/2}$ outside the core, and hence the angular
momentum gradient may contribute to the lower explodability
of the m15af series compared to the m15ps models.
We stress, however, that the impact of rotation can be qualitatively different in 3D, where the beneficial effects of a strong spiral mode of the SASI \citep{summa_rotation-supported_2018}, or the low-$|T|/W$ instability \citep{takiwaki_16} for rapidly spinning progenitors outweigh the adverse effects of rotation. As expected, adverse effects of rotation are also overcome
in 2D in the presence of strong magnetic fields. Model 
m15afB12
develops
a classical early bipolar explosion as familiar from 2D simulations
of magnetorotational explosions
\citep{burrows_07,sawai_08,obergaulinger_17,obergaulinger_20} with a rapid growth of the explosion that has reached $8.6 \times 10^{51}\, \mathrm{erg}$ and is still increasing at this point.

\subsubsection{$39\,M_\odot$ models}
\begin{figure*}
\centering
\begin{subfigure}{0.48\textwidth}
  \centering
  \includegraphics[width=1\linewidth]{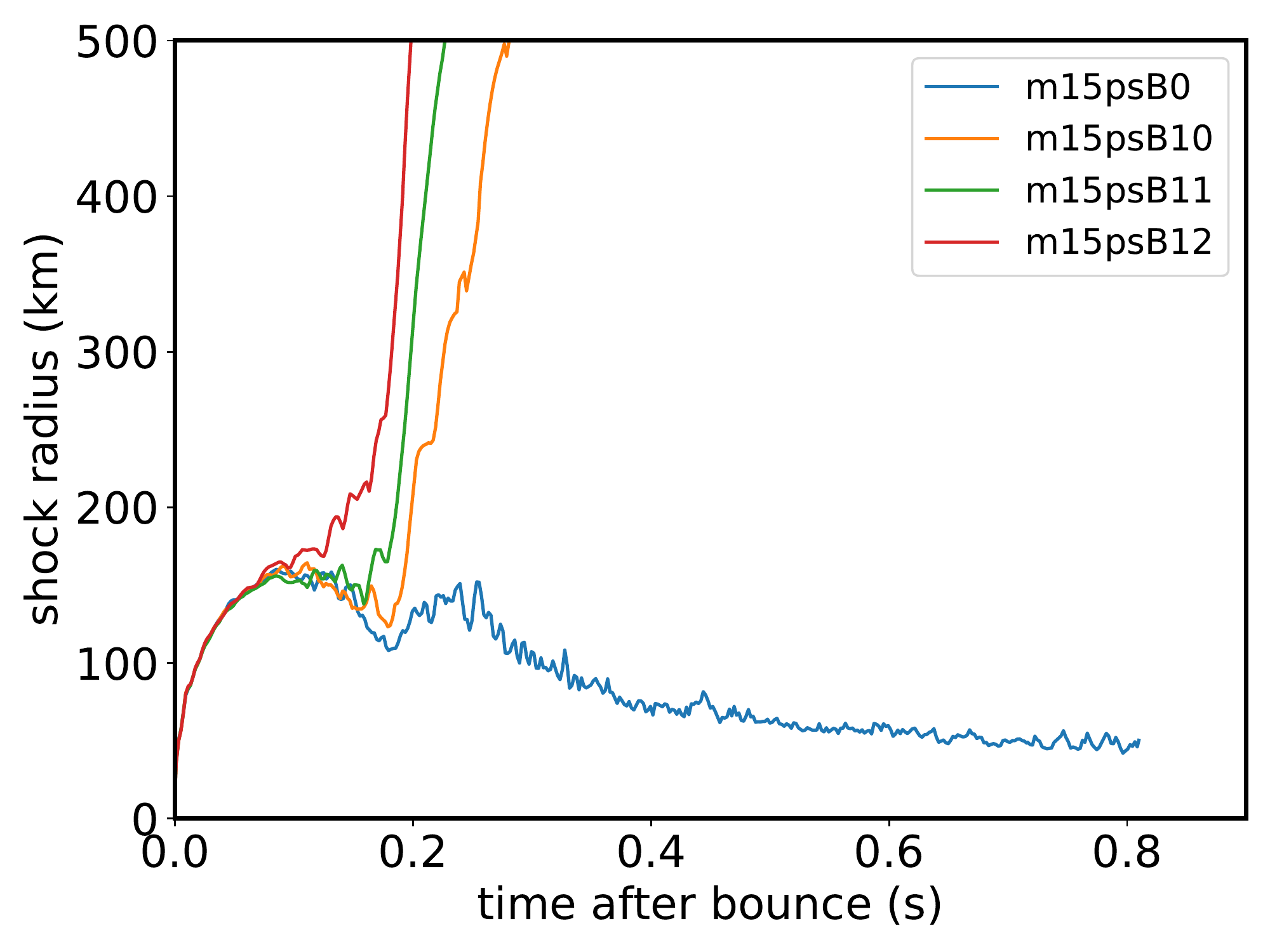}
  \caption{}
\end{subfigure}
\begin{subfigure}{0.48\textwidth}
  \centering
  \includegraphics[width=1\linewidth]{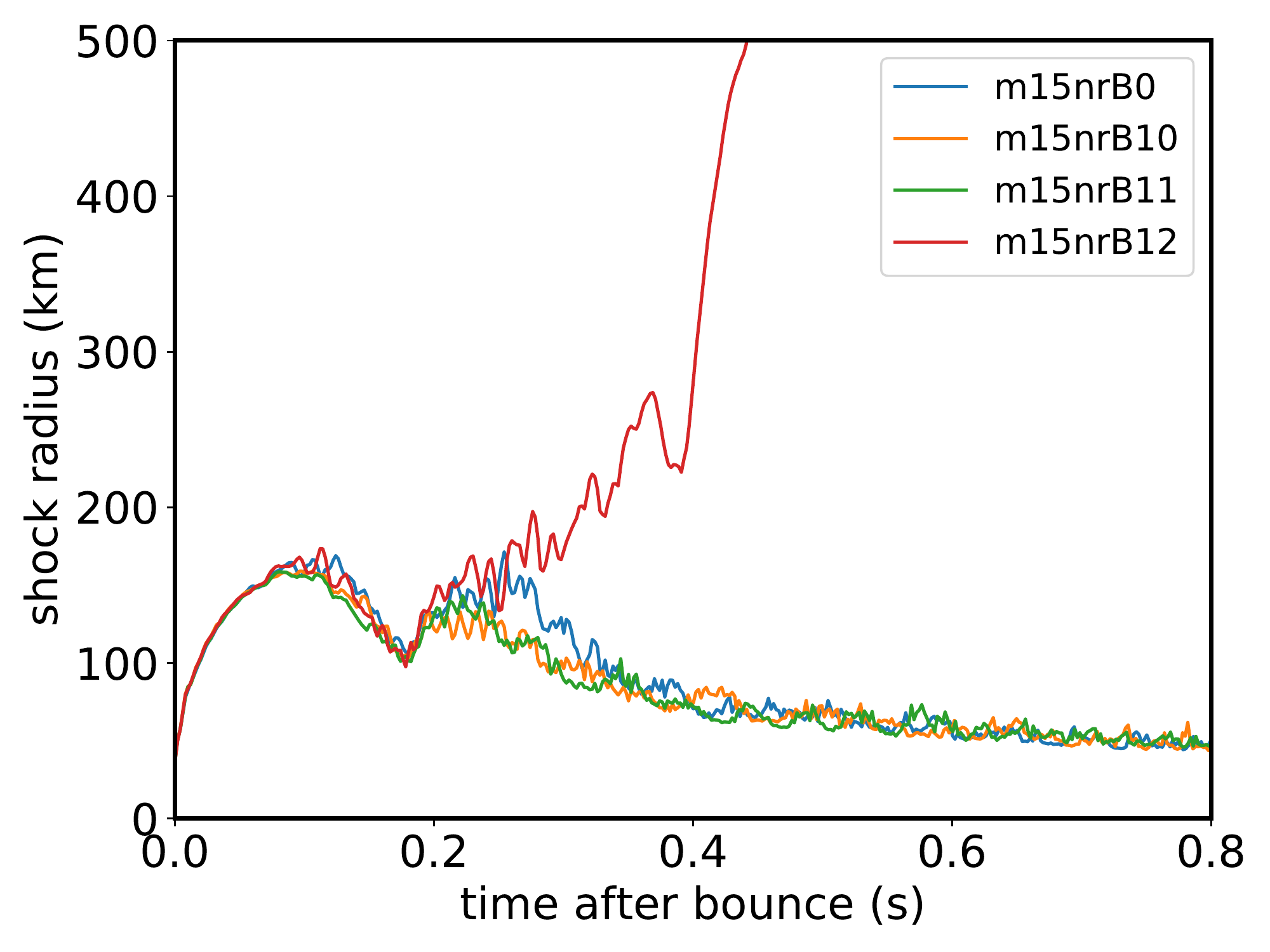}
  \caption{}
\end{subfigure}
\begin{subfigure}{0.48\textwidth}
  \centering
  \includegraphics[width=1\linewidth]{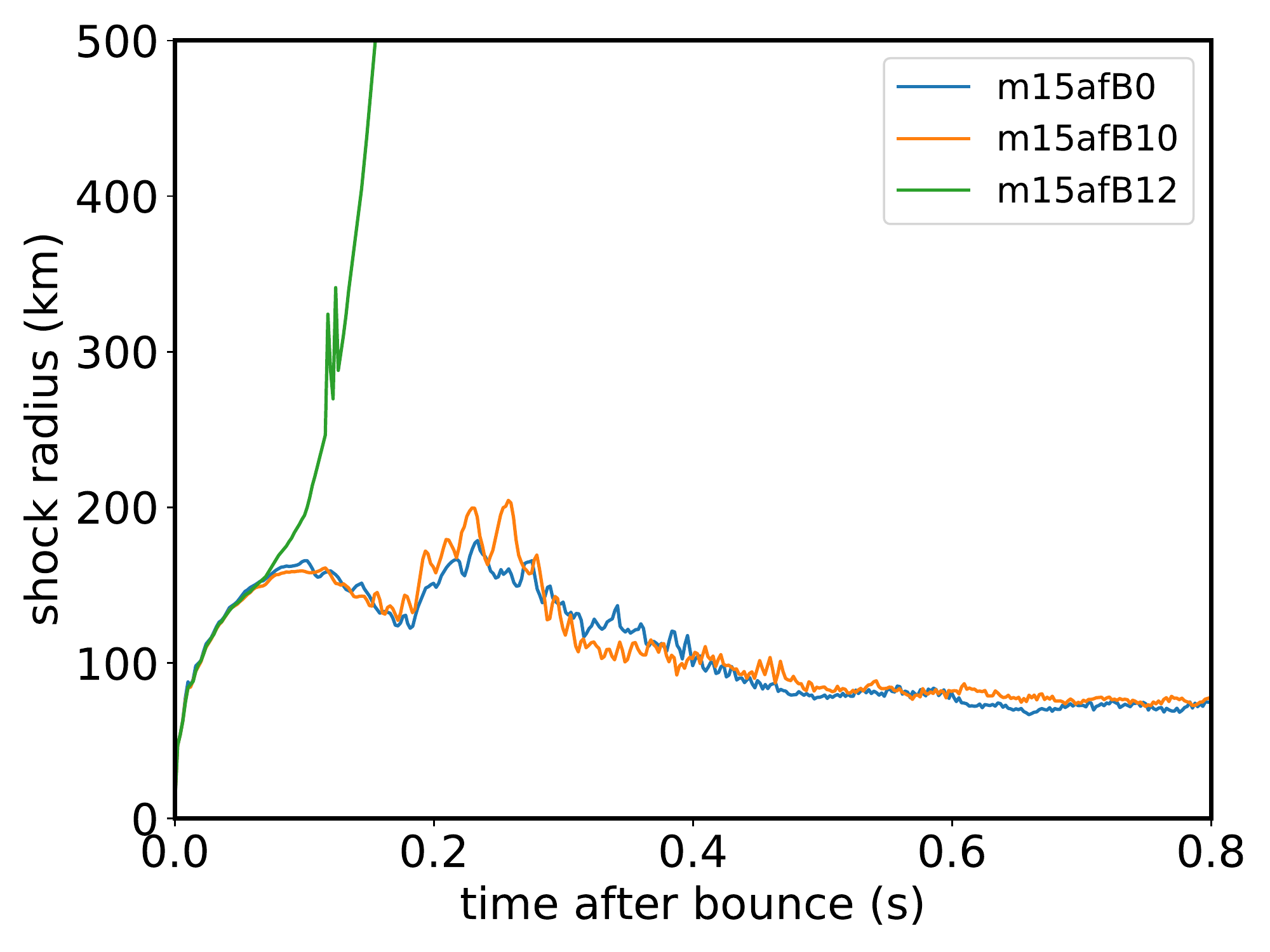}
  \caption{}
  \end{subfigure}
  \begin{subfigure}{0.48\textwidth}
  \centering
  \includegraphics[width=1\linewidth]{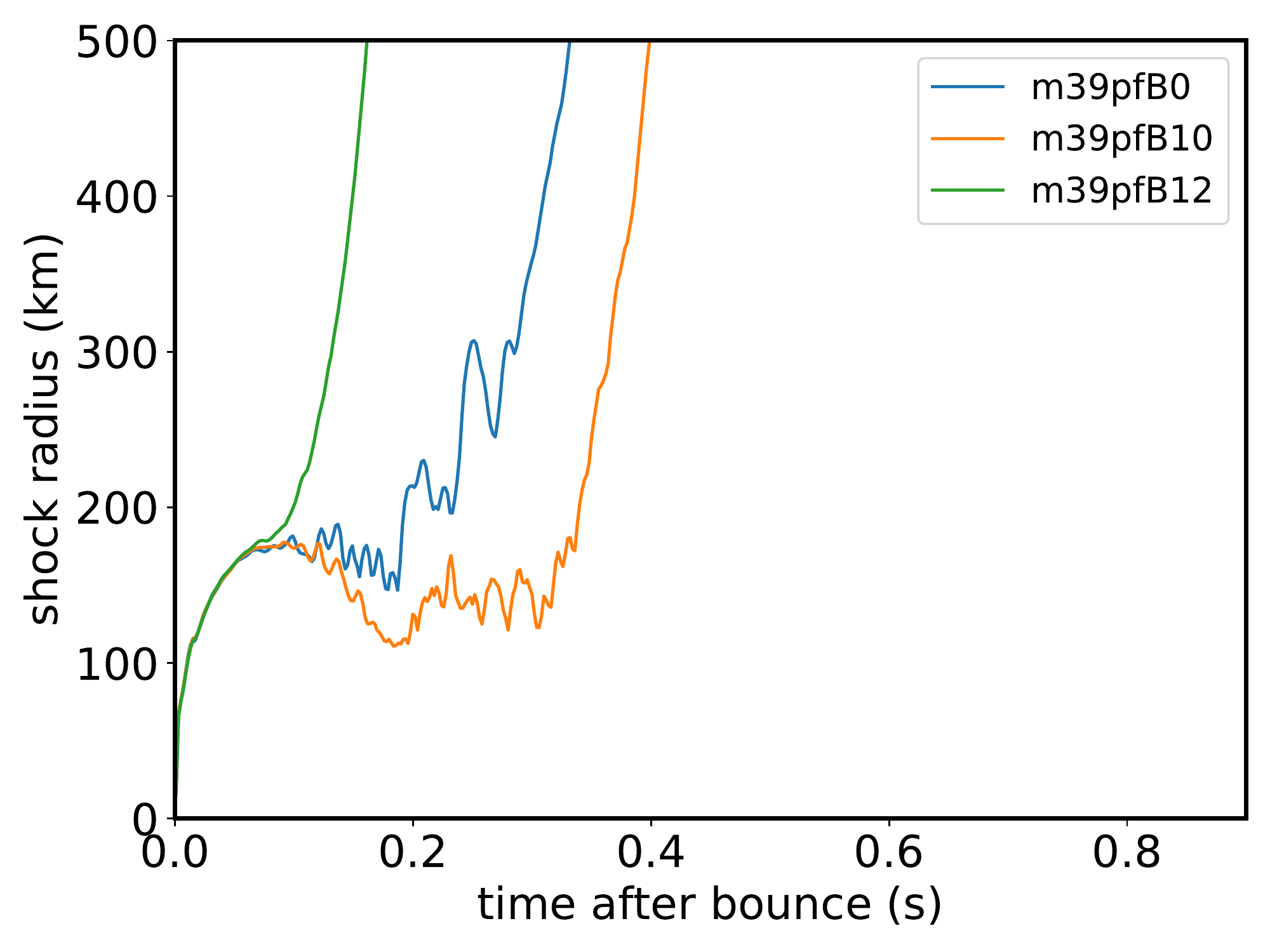}
  \caption{}
\end{subfigure}
\begin{subfigure}{0.48\textwidth}
  \centering
  \includegraphics[width=1\linewidth]{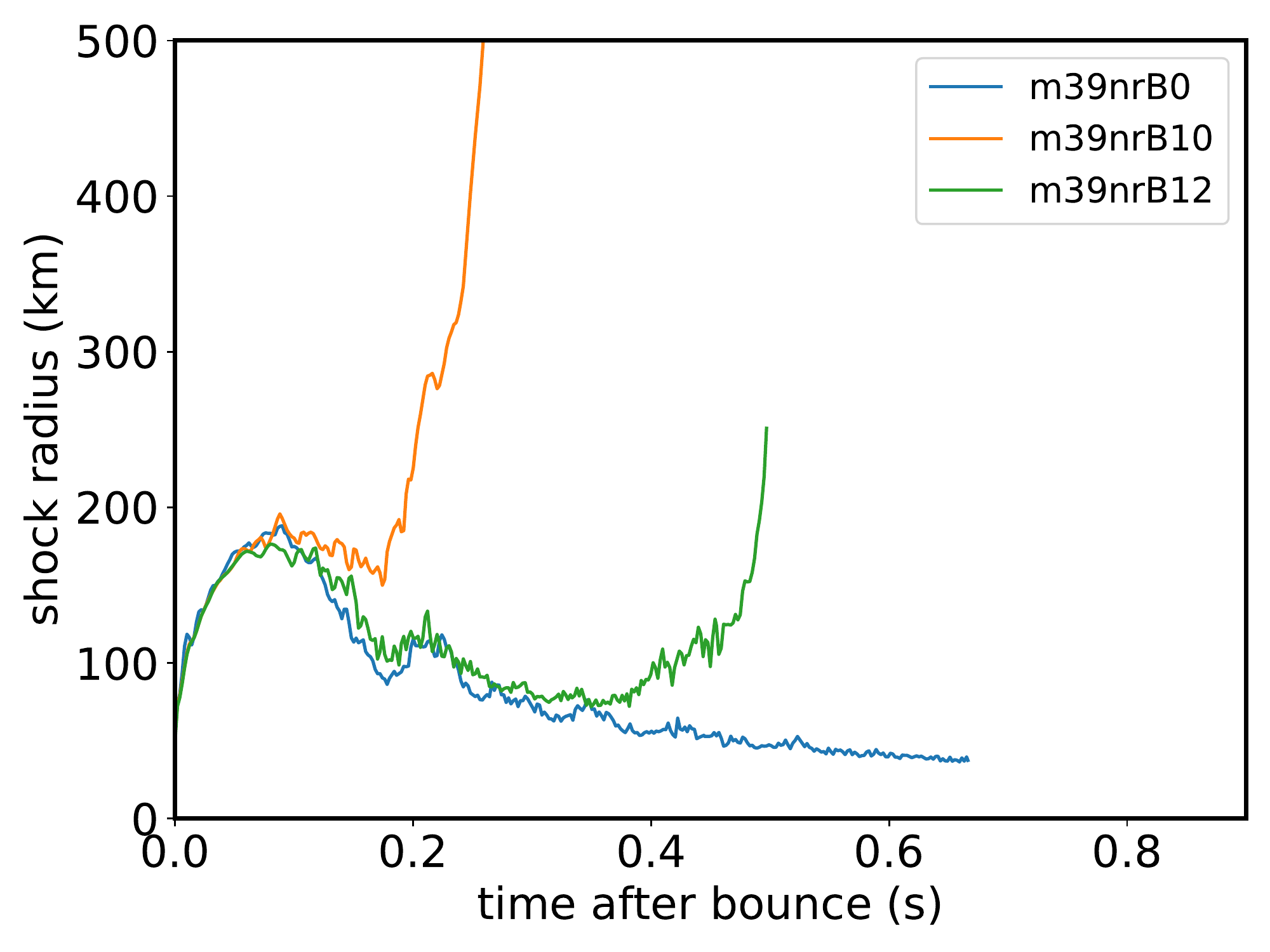}
  \caption{}
\end{subfigure}
 \caption{Angle-averaged shock radii as a function of post-bounce time for our $15\,M_\odot$ and $39\,M_{\odot}$ models. Each panel shows results for one specific rotation profile, but different initial magnetic fields.
 For $15\,M_{\odot}$ progenitors, models generally tend to explode more readily with increased magnetic field strength, while the impact of rotation is not monotonic. For the $39\,M_{\odot}$ progenitor, the impact of increased magnetic field strength is not monotonic. 
  }\label{fig:shock}
\end{figure*}

\begin{figure}
    \centering
    \includegraphics[width=\linewidth]{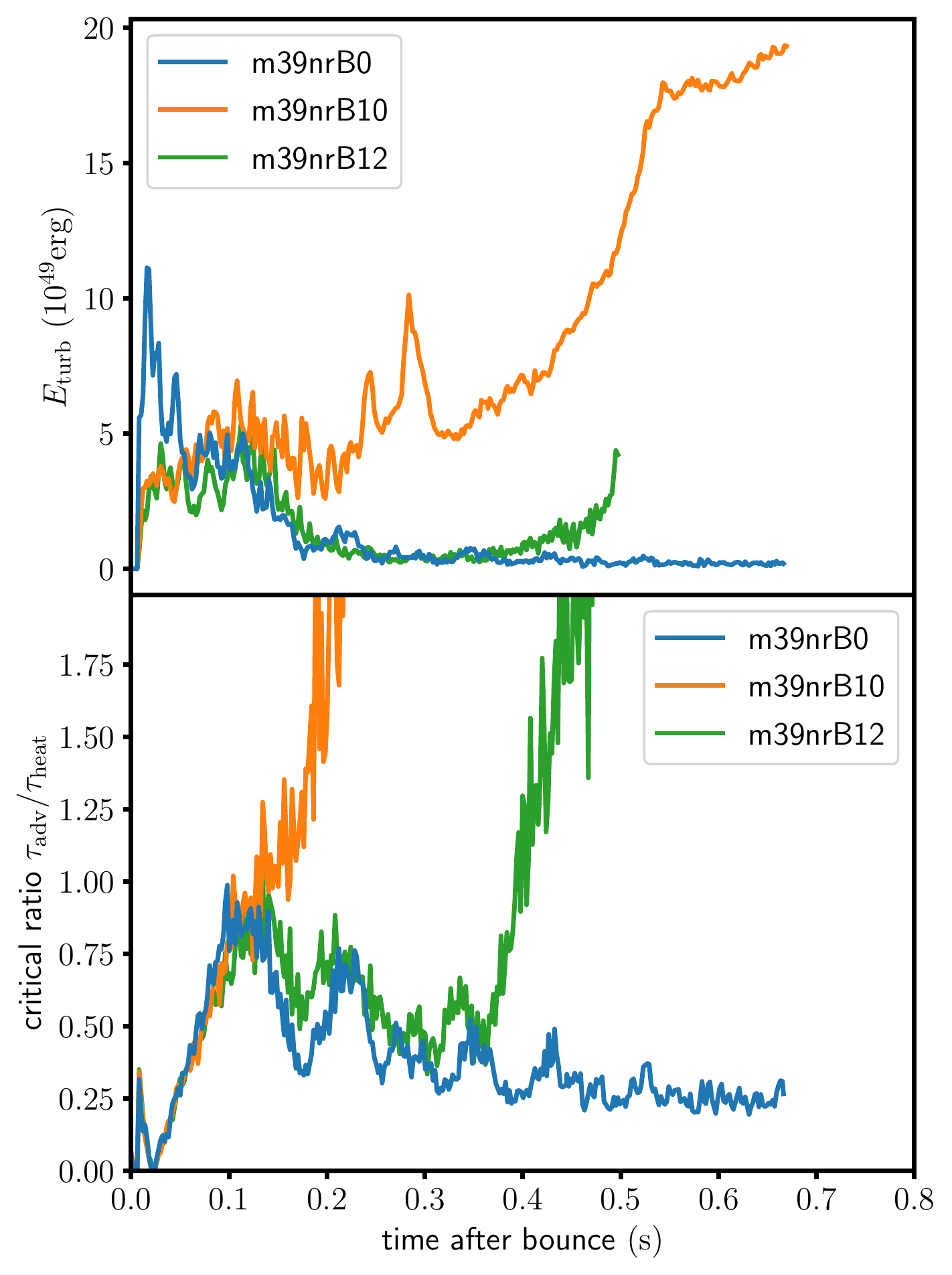}
    \caption{The turbulent kinetic energy
    $E_\mathrm{turb}$ in the gain region and the
    critical time scale ratio $\tau_\mathrm{adv}/\tau_\mathrm{heat}$
    for the m39nr models. Model m39nrB12
    maintains a lower turbulent kinetic energy in
    the gain region than model m39nrB10
    by the time when
    $\tau_\mathrm{adv}/\tau_\mathrm{heat}$  approaches unity and falls slightly short
    of an explosive runaway despite similar heating conditions. In model 
    m39nrB0, $E_\mathrm{turb}$  evolves
    very differently from the other two models with an early peak and a fast decline because of the different seed perturbations from which non-radial instabilities grow (random
    in m39nrB0 vs.\ dipolar perturbations
    from the magnetic field in m39nrB10
    and m39nrB12).
    }
    \label{fig:m39_turb}
\end{figure}

The $39\,M_\odot$ models present an even more intricate picture. Different from
the m15 series, all the rotating $39\,M_\odot$ models develop explosions. Rotation precipitates shock revival in the non-magnetised models and the $10^{12}\, \mathrm{G}$ models, but delays shock revival slightly for an initial field
of $10^{10}\, \mathrm{G}.$
It is noteworthy that the $39\,M_\odot$ progenitor exhibits a negative
gradient in specific angular momentum at the Si/O shell interface
\citep{powell_20}, which may explain why rapid progenitor rotation
is not hurtful in the non-magnetised case, different from
the $15 M_\odot$ model.
However, in contrast to the $15\,M_\odot$ models, we also find non-monotonic behaviour with respect to the initial field strength. For the unmodified progenitor rotation profile (which is tantamount to rapid rotation
in this case), the weak-field model 
m39pfB10 explodes considerably later than the non-magnetised and strongly magnetised models (Figure~\ref{fig:shock}d). The non-rotating series m39nr
is even more interesting (Figure~\ref{fig:shock}e). Here the weak-field model m39nrB10
explodes early about $0.26 \, \mathrm{s}$ after bounce, 
m39nrB0 does not explode at all, and the strong-field model 
m39nrB12 explodes with a considerable delay around half a second after bounce. 
The significant delay of shock revival in model m39nrB12 compared to 
m39nrB10 is unexpected and to be contrasted with the findings of
\citet{obergaulinger_14}, where strong initial fields help bring about an earlier explosion. Close inspection reveals that model 
m39nrB12 just narrowly misses an explosion at the same time as 
m39nrB10; the critical ratio between the advection time scale
$\tau_\mathrm{adv}$
and the heating time scale $\tau_\mathrm{heat}$
 \citep{buras_two-dimensional_buras_artrot}
comes close to unity before turning around and only reaching the explosion threshold about $250 \, \mathrm{ms}$ later (Figure~\ref{fig:m39_turb}, bottom). This can be explained by slightly
smaller turbulent kinetic energy
$E_\mathrm{turb}$ in the gain region, which can be computed as
\begin{equation}
    E_\mathrm{turb}
    =\frac{1}{2}\int \rho[ (v_r -\langle v_r \rangle)^2+v_\theta^2)\,\ud V,
\end{equation}
where $v_r$ and $v_\theta$ are the velocity components
in the radial and meridional direction, 
and $\langle v_r \rangle$ is the mass-weighted
spherical average of the radial velocity (Figure~\ref{fig:m39_turb}, top).
The adverse effects of strong magnetic fields on the explosion conditions is reminiscent of 
the convection-dominated
$15\,M_\odot$ model of \citet{matsumoto_20}, who found that strong initial fields can suppress the growth of medium- to small-scale turbulence, and thereby reduce the overall turbulent kinetic energy stored in the gain region.
We note, however, that that the situation was reversed for
the $15 M_\odot$ model (Section~\ref{sec:s15}), where a strong seed field of $10^{12}\, \mathrm{G}$ resulted in an explosion whereas
fields of $10^{10}\, \mathrm{G}$ and $10^{11}\, \mathrm{G}$ did not.
Hence we cannot yet robustly reproduce a suppression effect that 
is detrimental to shock revival.
Different seeds for the growth of neutrino-driven convection  (dipole perturbation vs.\ random perturbations) may account for the divergence between the weak-field runs ($10^{10}\, \mathrm{G}$) and the corresponding non-magnetised runs.

\begin{figure*}
\centering
\begin{subfigure}{1\textwidth}
  \centering
  \includegraphics[width=1\linewidth]{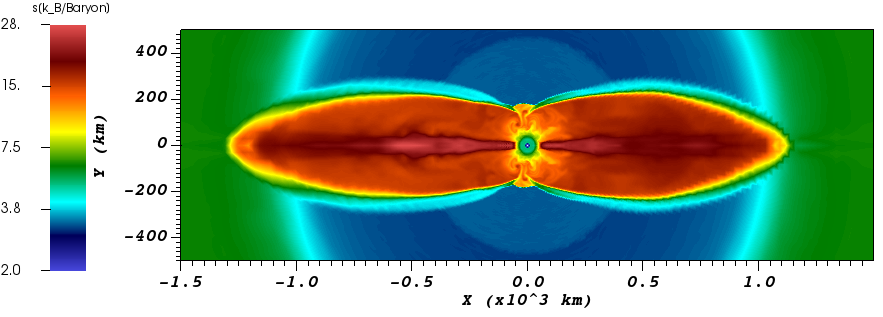}
  \caption{Model m39pfB12 at a time of $140\, \mathrm{ms}$ after bounce. }
\end{subfigure}
\begin{subfigure}{1\textwidth}
  \centering
  \includegraphics[width=1\linewidth]{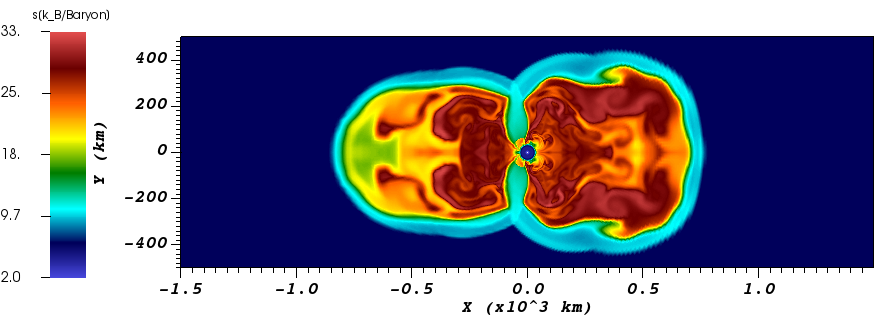}
  \caption{Model m39nrB10 at a time of $256\, \mathrm{ms}$ after bounce.}
\end{subfigure}
 \caption{2D entropy snapshots illustrating the different flow structure of 
model m39pfB12  as an example of a magnetorotational explosion (Panel~a), and m39nrB10 as an example of a neutrino-driven explosion (Panel~b) at a stage of shock expansion with a similar angle-averaged shock radius.
}\label{fig:entro}
\end{figure*}

Again, the combination of rapid rotation and strong magnetic fields 
in model m39pfB12 leads to a characteristic bipolar explosion geometry in which magnetic fields clearly shape the flow into collimated outflows. Compared to the $15\,M_\odot$, the one case that develops a true magnetorotational explosion among the $39\,M_\odot$ models sticks out less prominently in terms
of shock propagation and explosion energetics; the explosion energy has only
reached $6.25 \times 10^{50}\, \mathrm{erg}$ by the end of the simulation.
However, the explosion morphology of this model remains very distinct.
Compared to model m39pfB10,
which develops a neutrino-driven explosion, m39pfB12
exhibits a much more prolate explosion geometry, no volume-filling
turbulence in the post-shock region, and early quenching of accretion downflows
in the equatorial plane (Figure~\ref{fig:entro}).

\begin{figure*}
\centering
\begin{subfigure}{0.33\linewidth}
  \centering
  \includegraphics[width=1\linewidth]{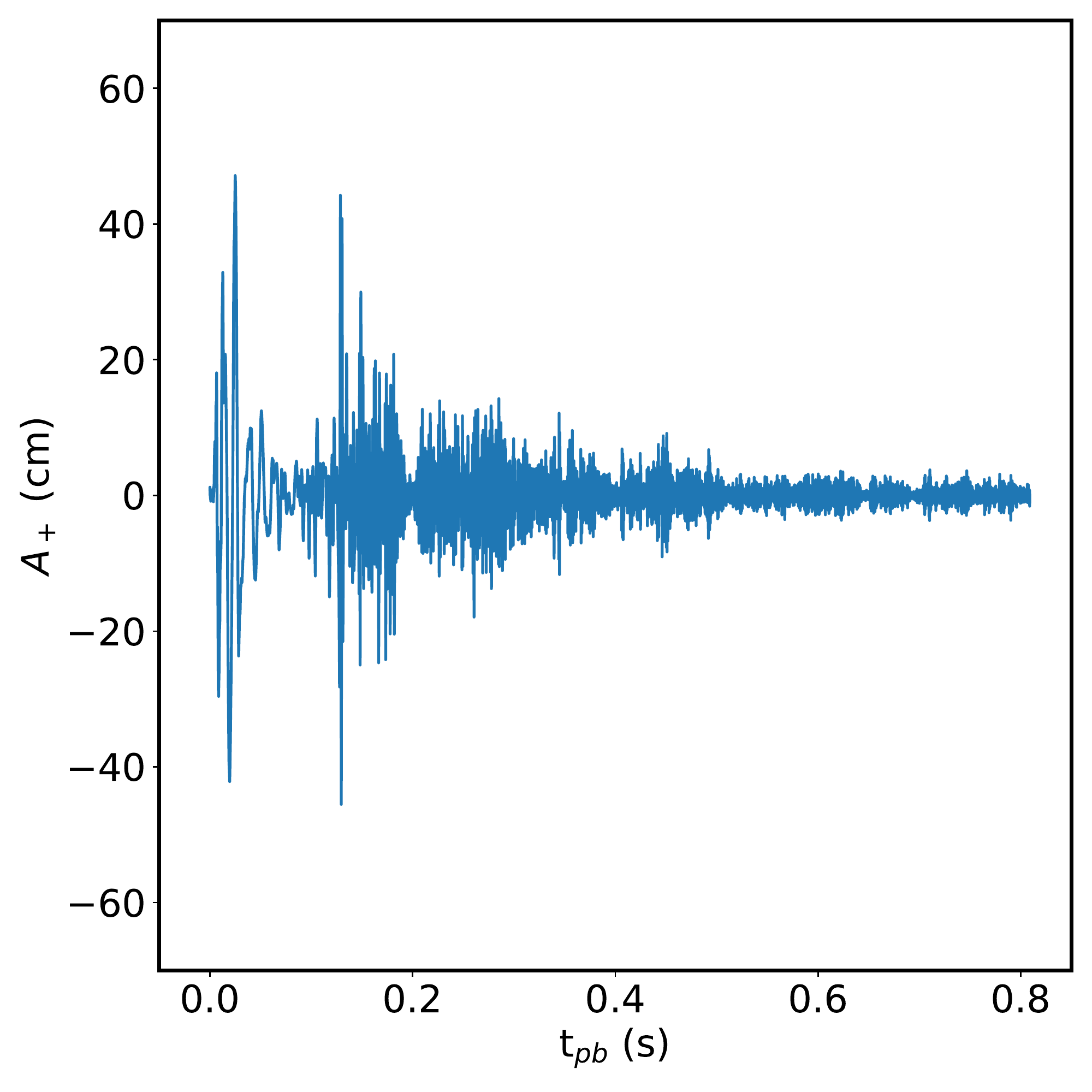}
  \caption{m15psB0}
\end{subfigure}%
\begin{subfigure}{0.33\textwidth}
  \centering
  \includegraphics[width=1\linewidth]{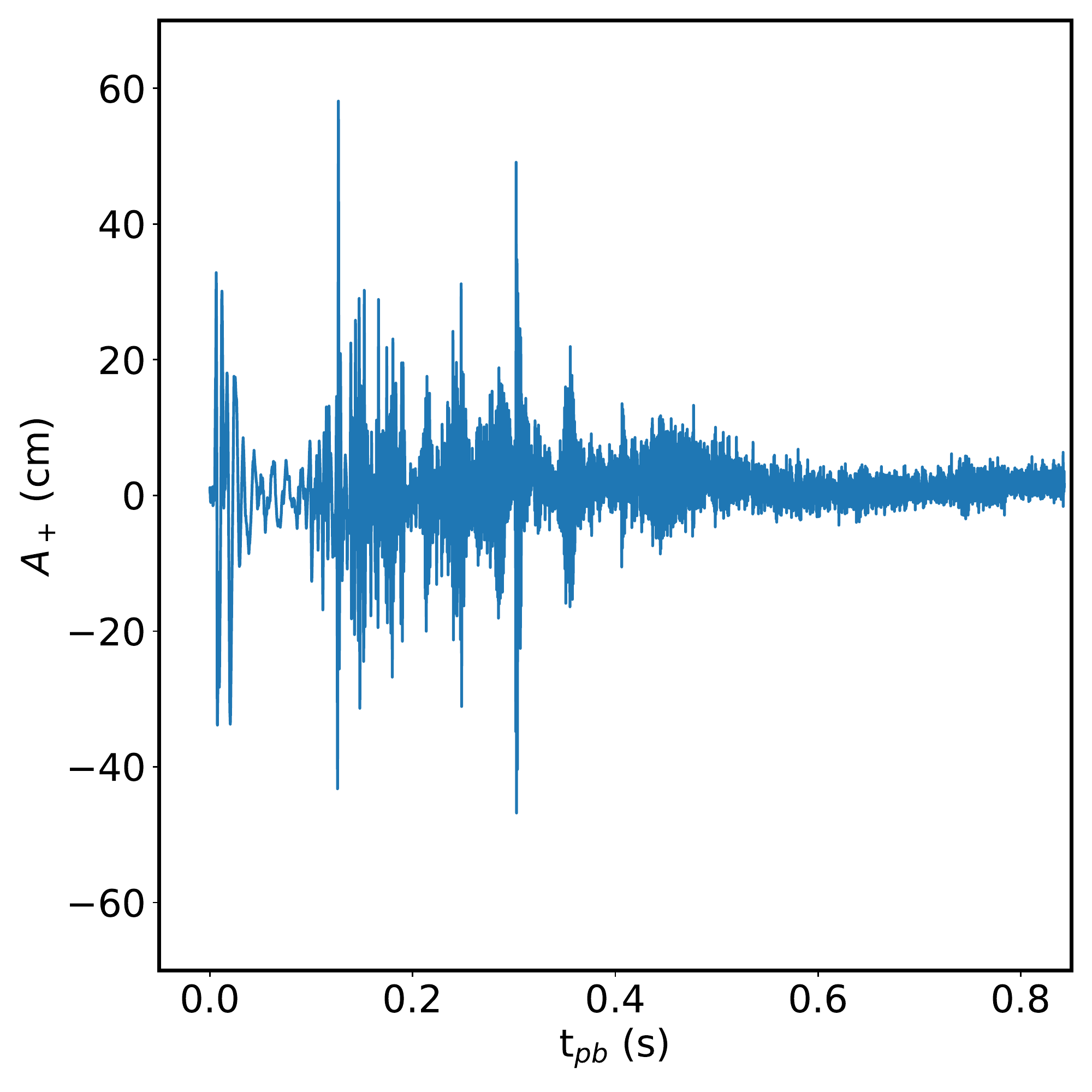}
  \caption{m15psB10}
\end{subfigure}%
\begin{subfigure}{0.33\textwidth}
  \centering
  \includegraphics[width=1\linewidth]{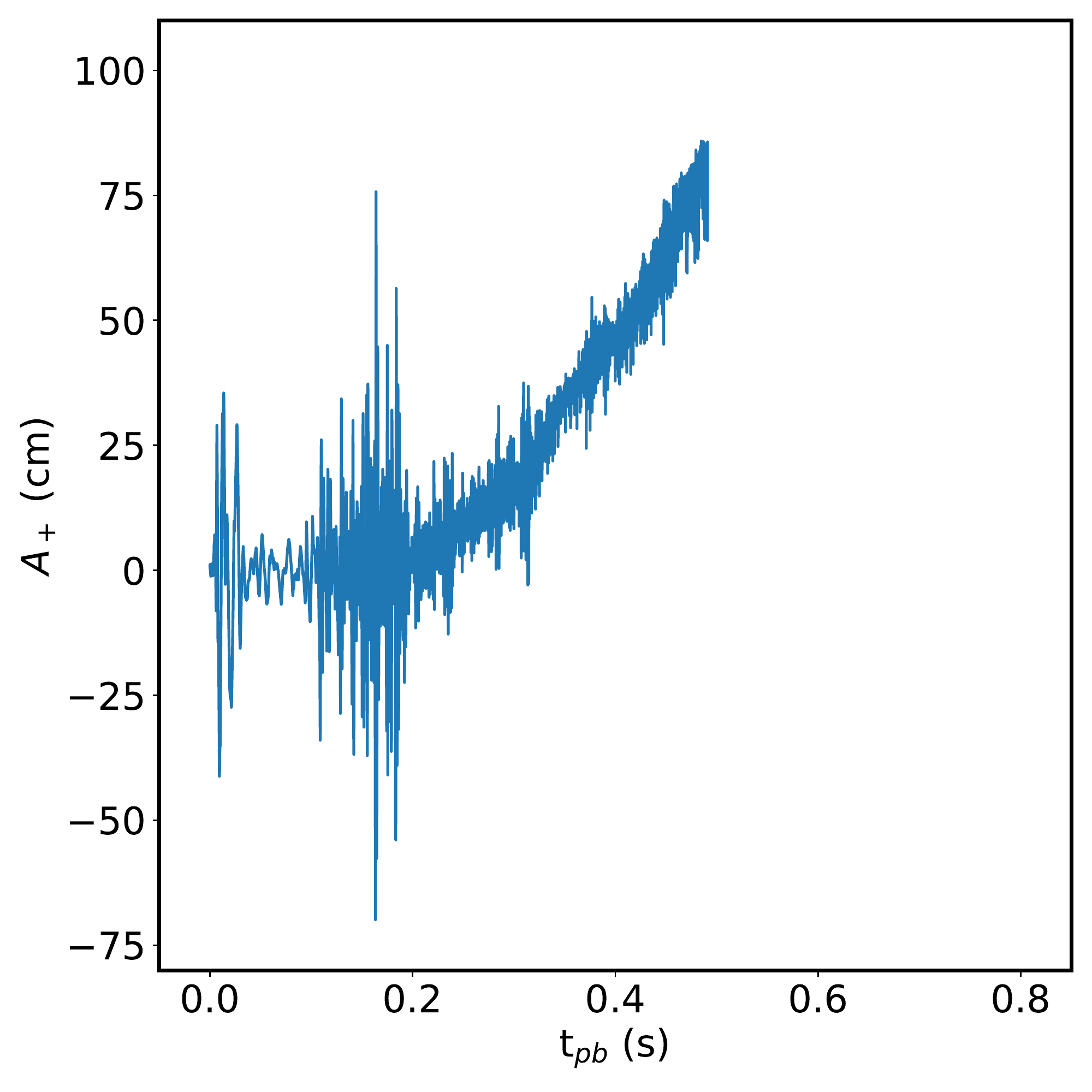}
  \caption{m15psB12}
\end{subfigure}
\begin{subfigure}{0.33\textwidth}
  \centering
  \includegraphics[width=1\linewidth]{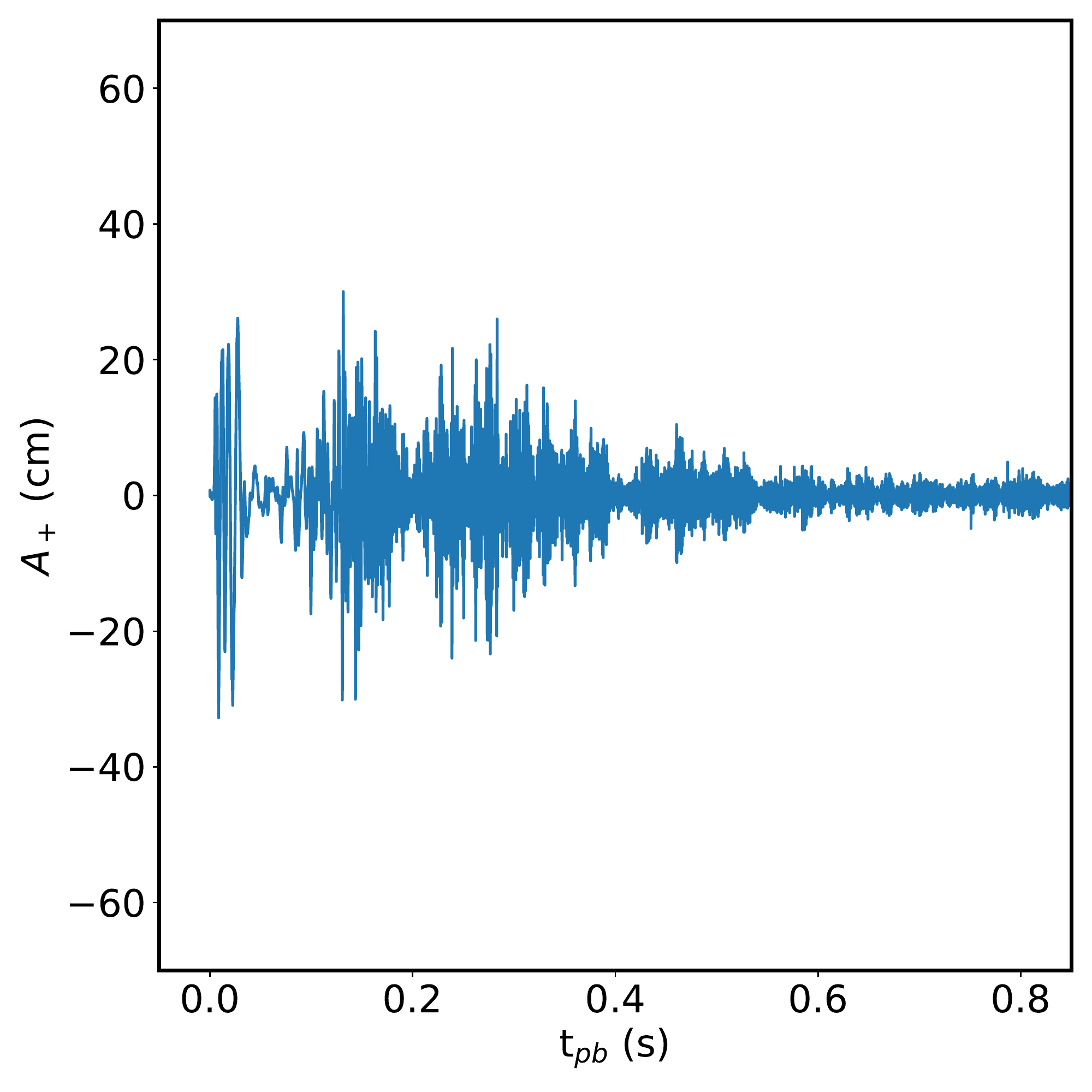}
  \caption{m15nrB0}
\end{subfigure}%
\begin{subfigure}{0.33\textwidth}
  \centering
  \includegraphics[width=1\linewidth]{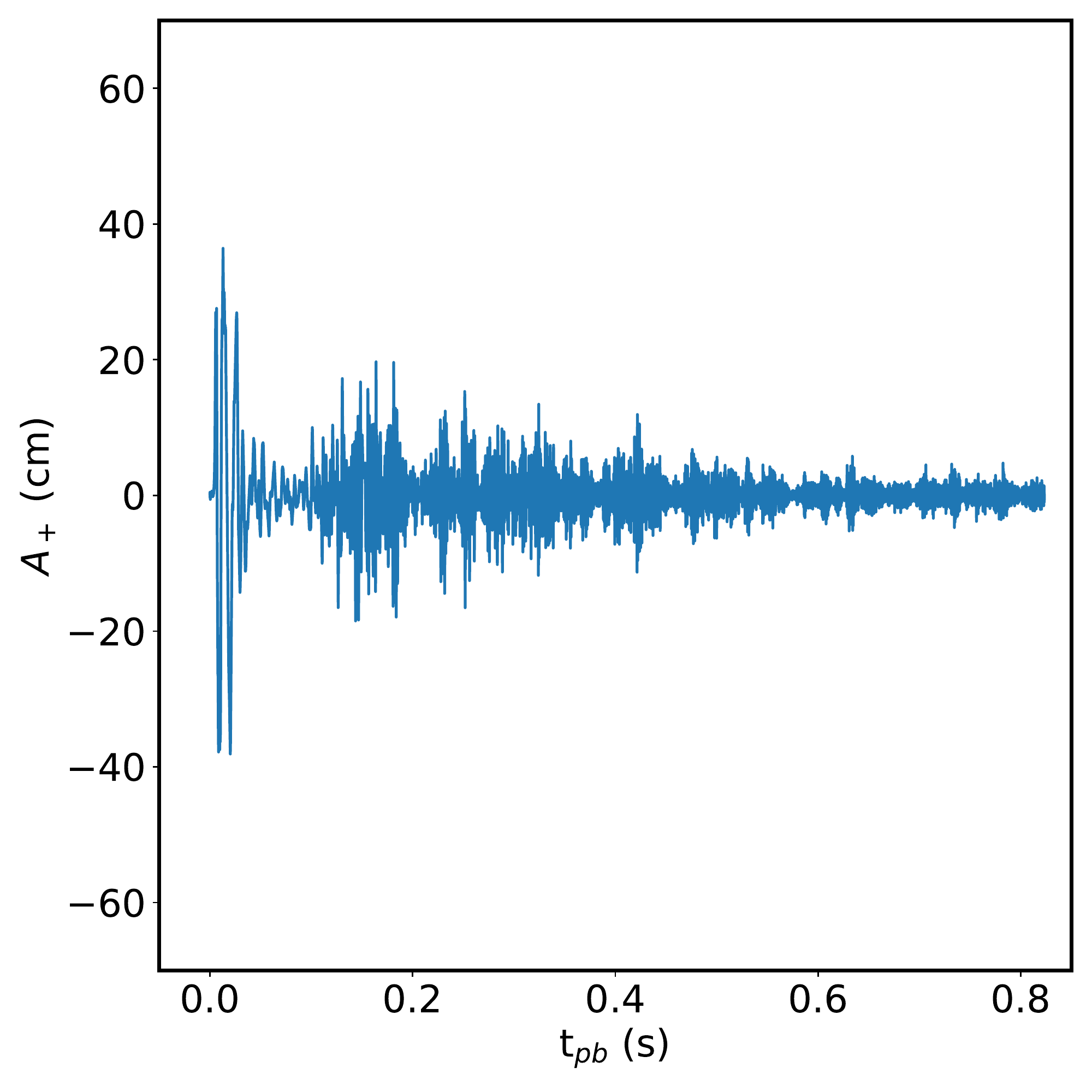}
  \caption{m15nrB10}
\end{subfigure}%
\begin{subfigure}{0.33\textwidth}
  \centering
  \includegraphics[width=1\linewidth]{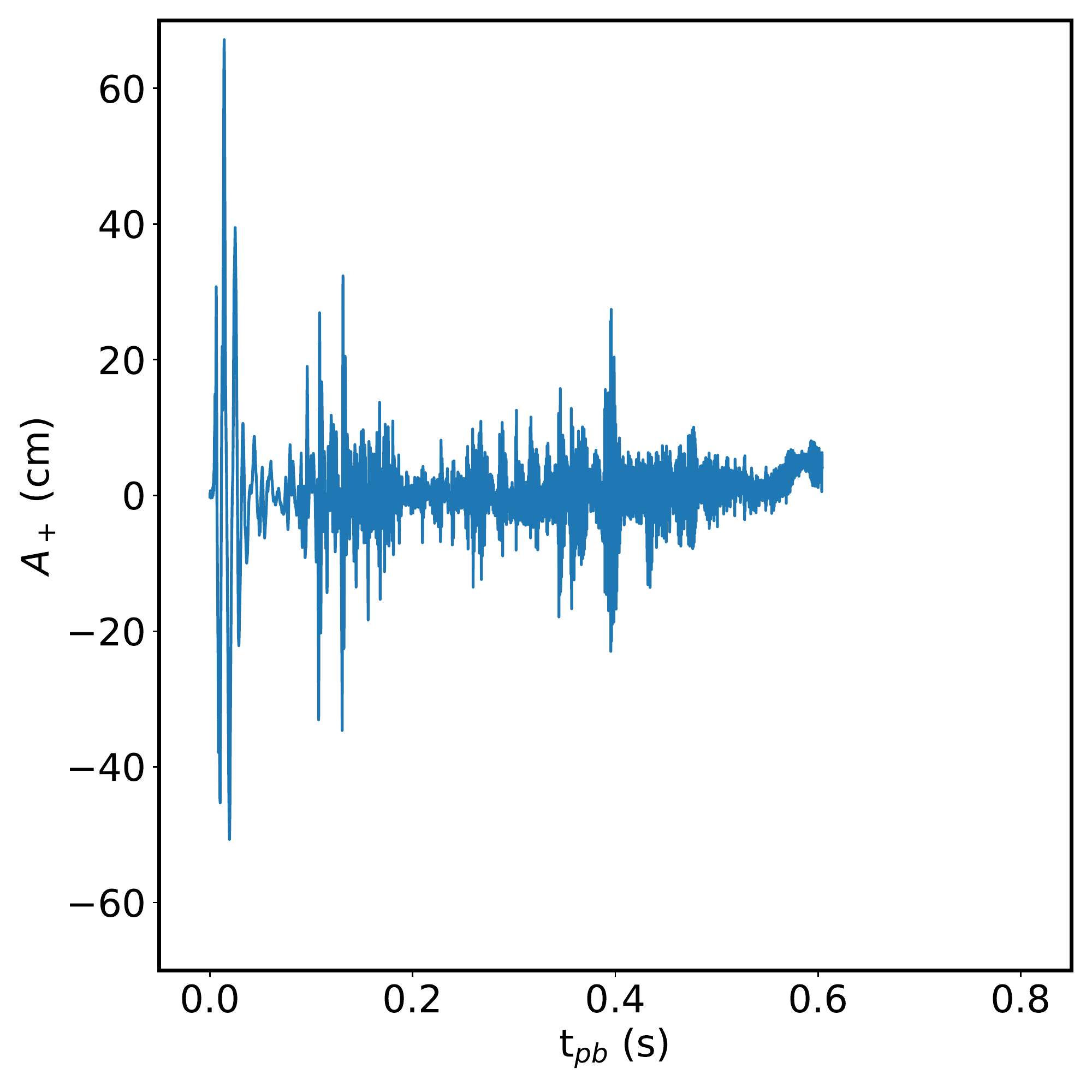}
  \caption{m15nrB12}
\end{subfigure}
\begin{subfigure}{0.33\textwidth}
  \centering
  \includegraphics[width=1\linewidth]{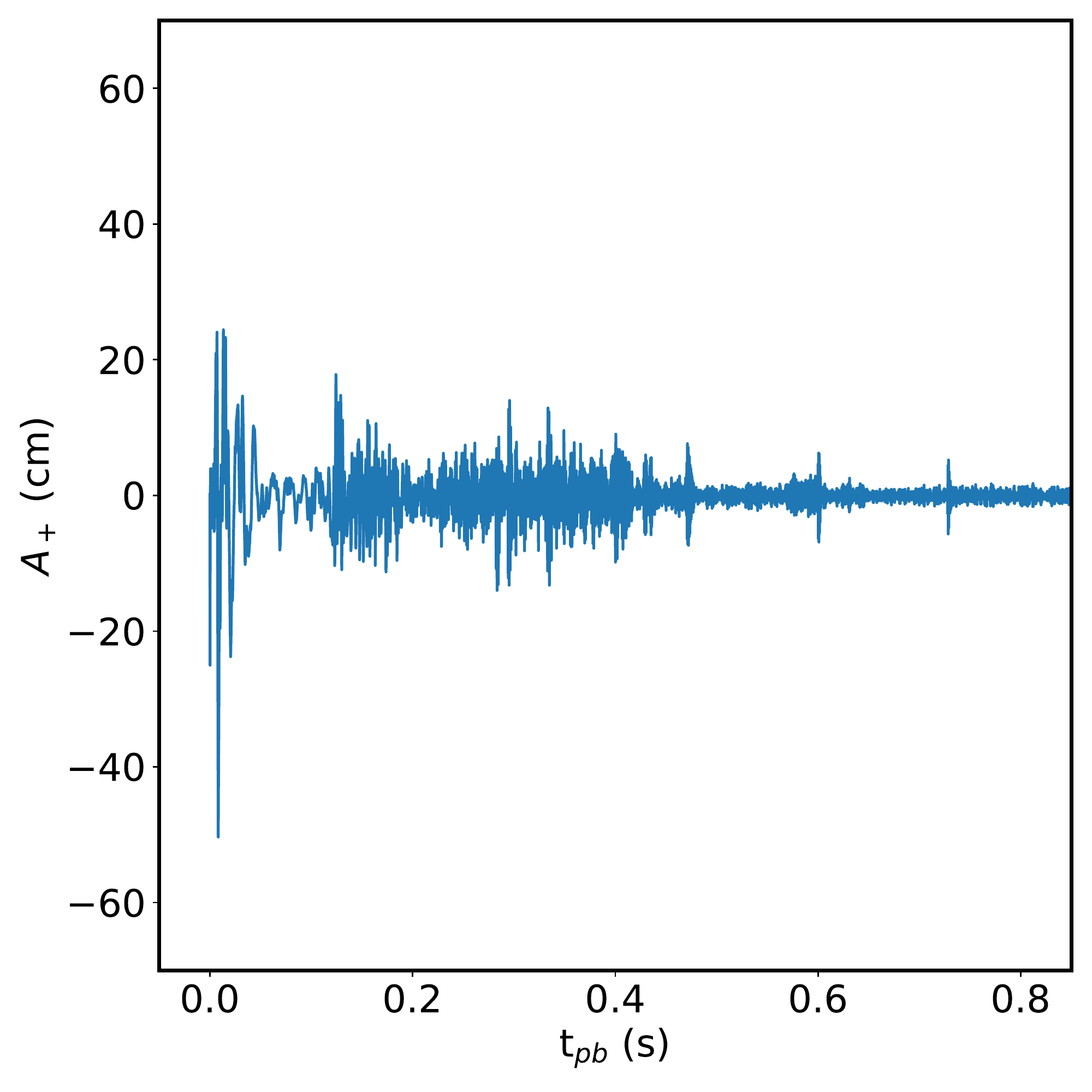}
  \caption{m15afB0}
\end{subfigure}%
\begin{subfigure}{0.33\textwidth}
  \centering
  \includegraphics[width=1\linewidth]{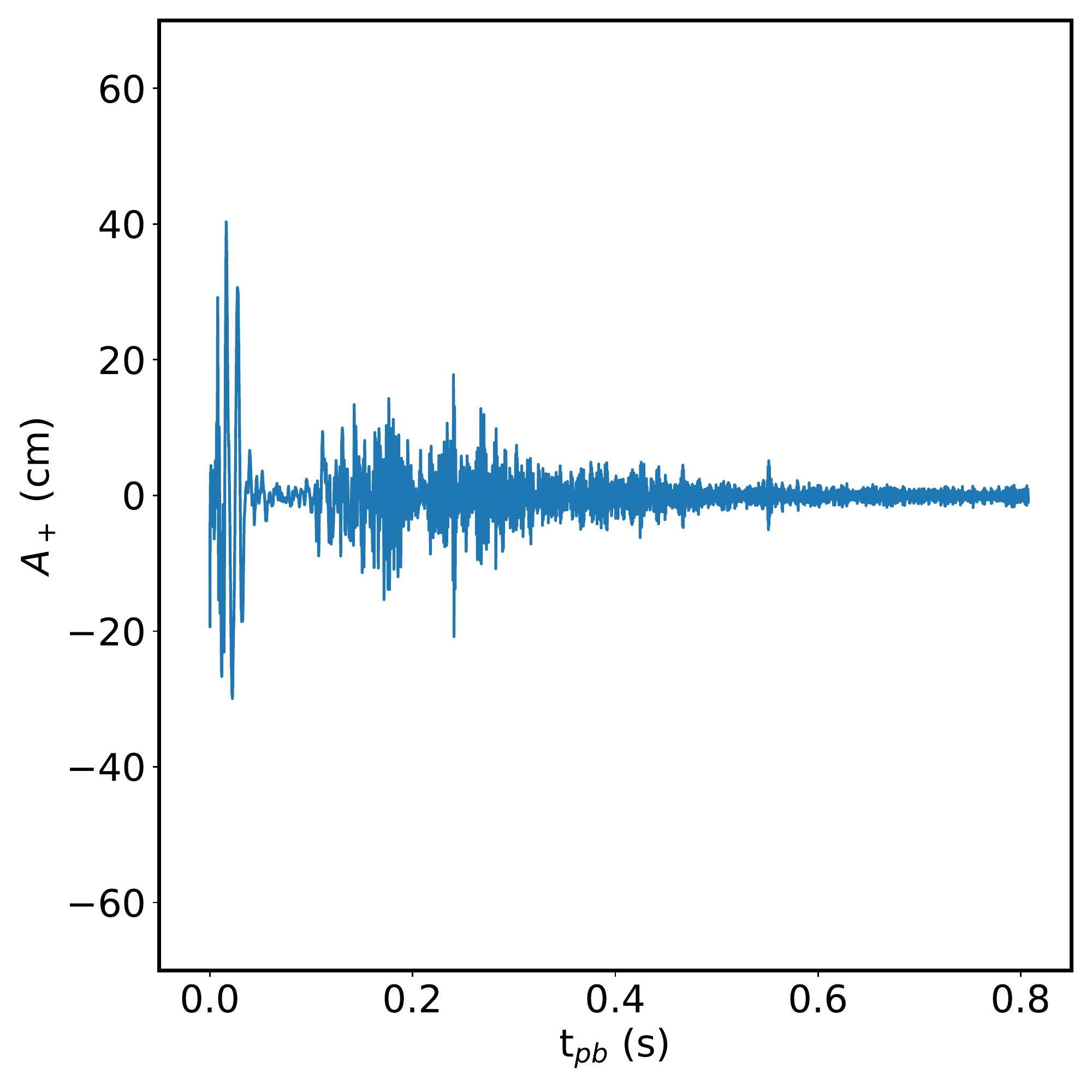}
  \caption{m15afB10}
\end{subfigure}%
\begin{subfigure}{0.33\textwidth}
  \centering
  \includegraphics[width=1\linewidth]{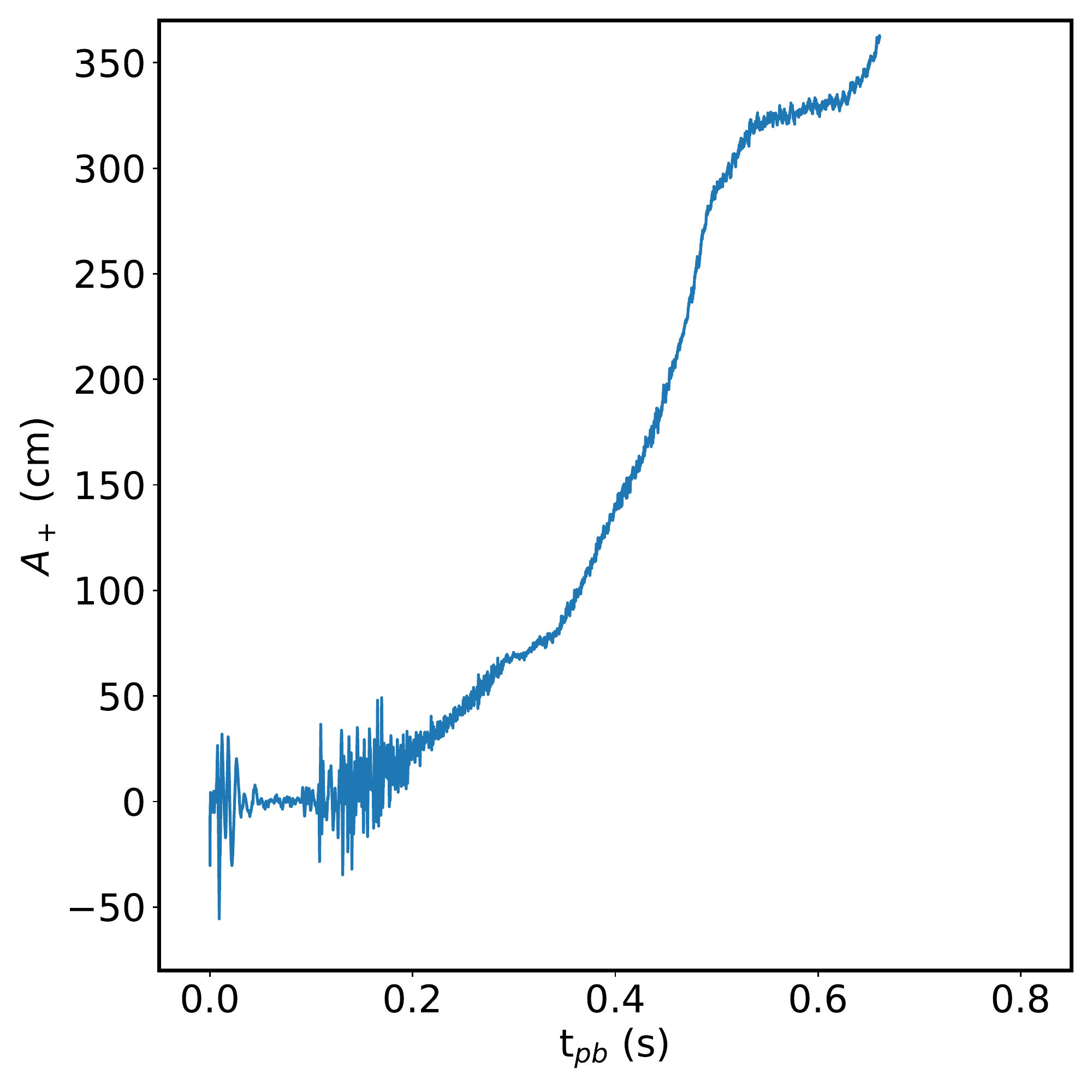}
  \caption{m15afB12}
\end{subfigure}%
 \caption{Gravitational wave amplitudes $A_+$ 
 of the plus polarisation mode in
 the equatorial plane for the $15\,M_{\odot}$ models.
  The strong tails present in Panels (c) and (i), with steadily increasing gravitational wave amplitude are a result of a strongly prolate shock expansion in the explosion phase. 
  Note the conspicuous resurgence of high-frequency emission in late times for model m15afB12 in Panel (i).
  As much as possible, the same   scale for $A_+$ is used for different panels  to allow for better comparison, except in Panels~(c) and (i) because of the pronounced tail signal.}
  \label{fig:amp}
\end{figure*}

\begin{figure*}
\centering
\begin{subfigure}{0.33\linewidth}
  \centering
  \includegraphics[width=1\linewidth]{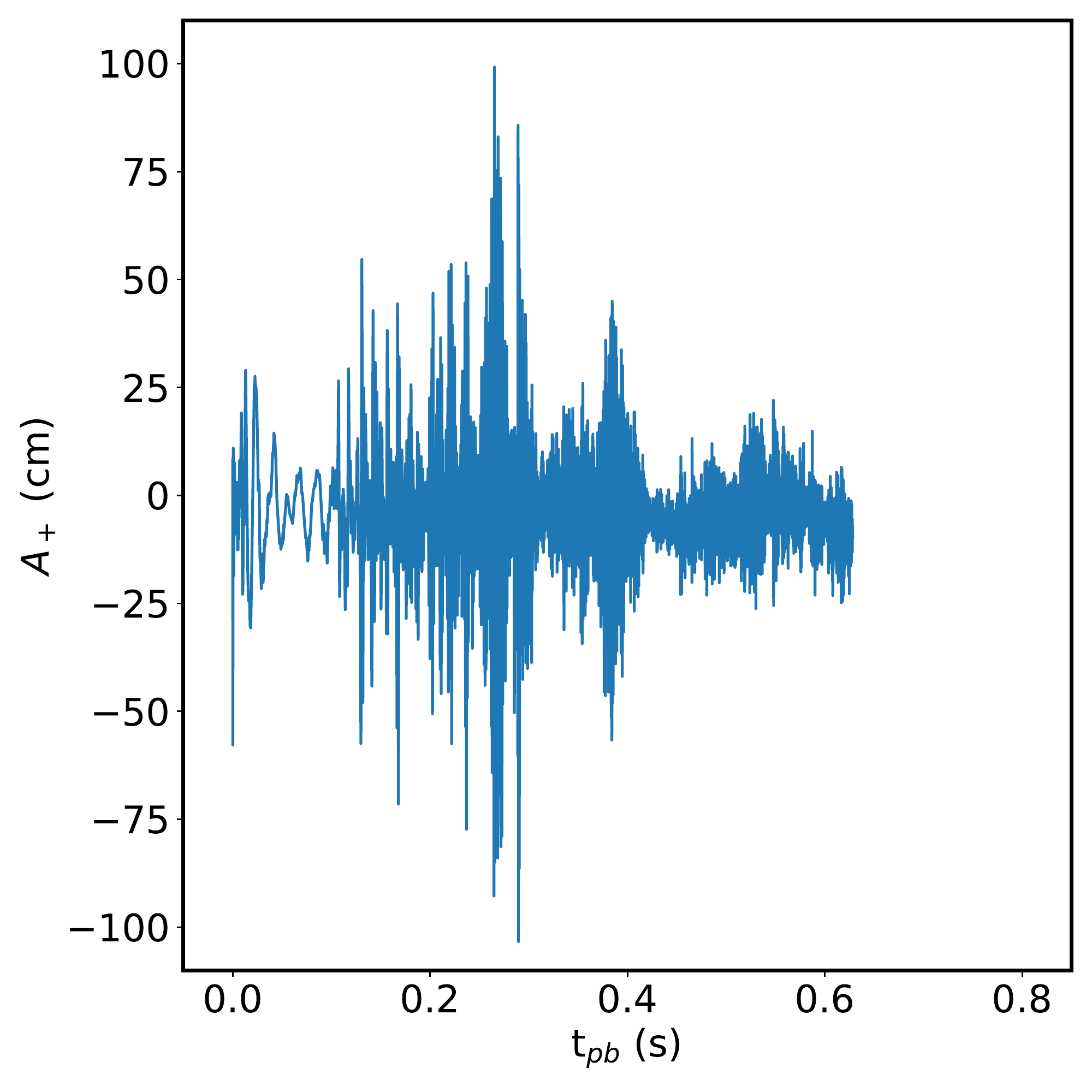}
  \caption{m39pfB0}
\end{subfigure}%
\begin{subfigure}{0.33\textwidth}
  \centering
  \includegraphics[width=1\linewidth]{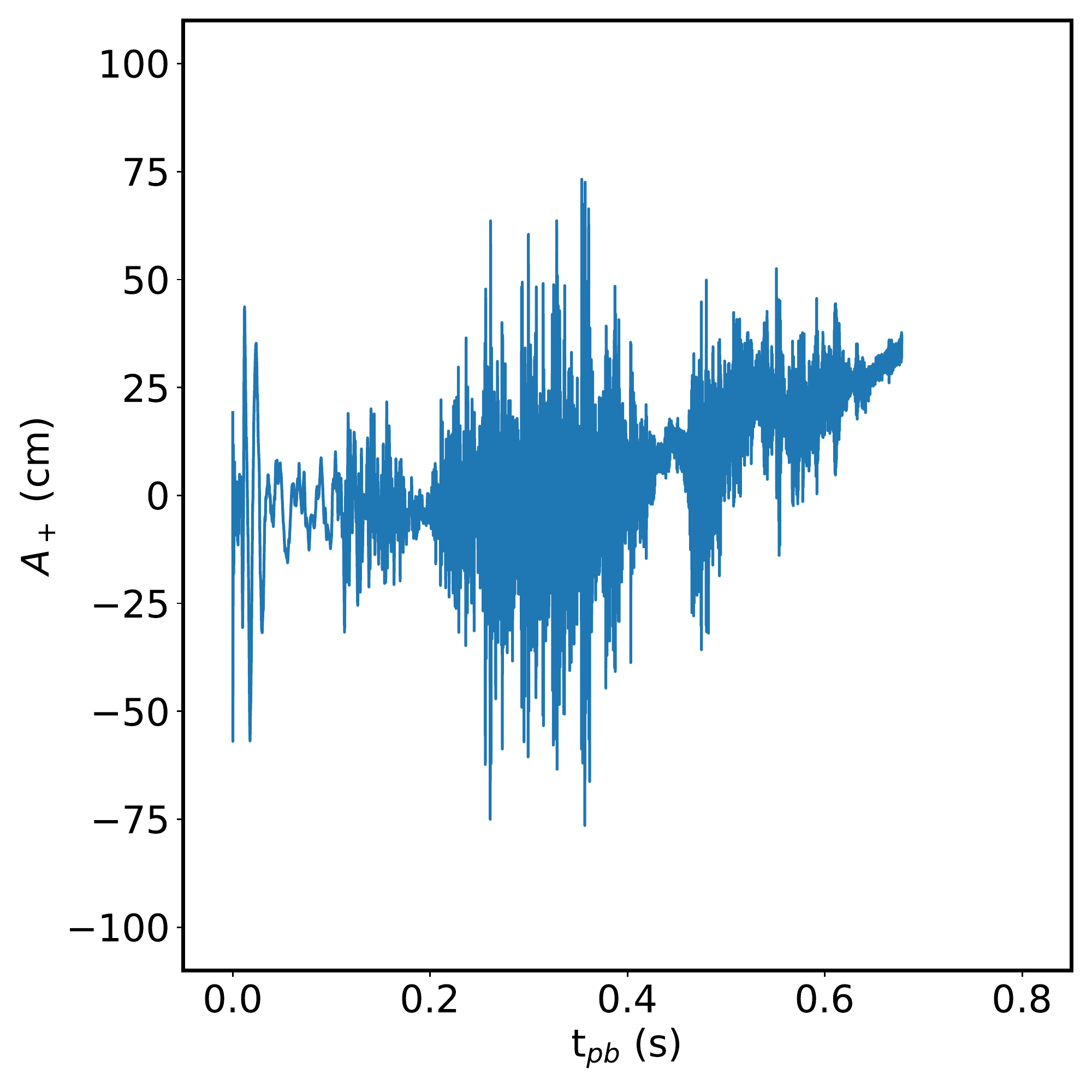}
  \caption{m39pfB10}
\end{subfigure}%
\begin{subfigure}{0.33\textwidth}
  \centering
  \includegraphics[width=1\linewidth]{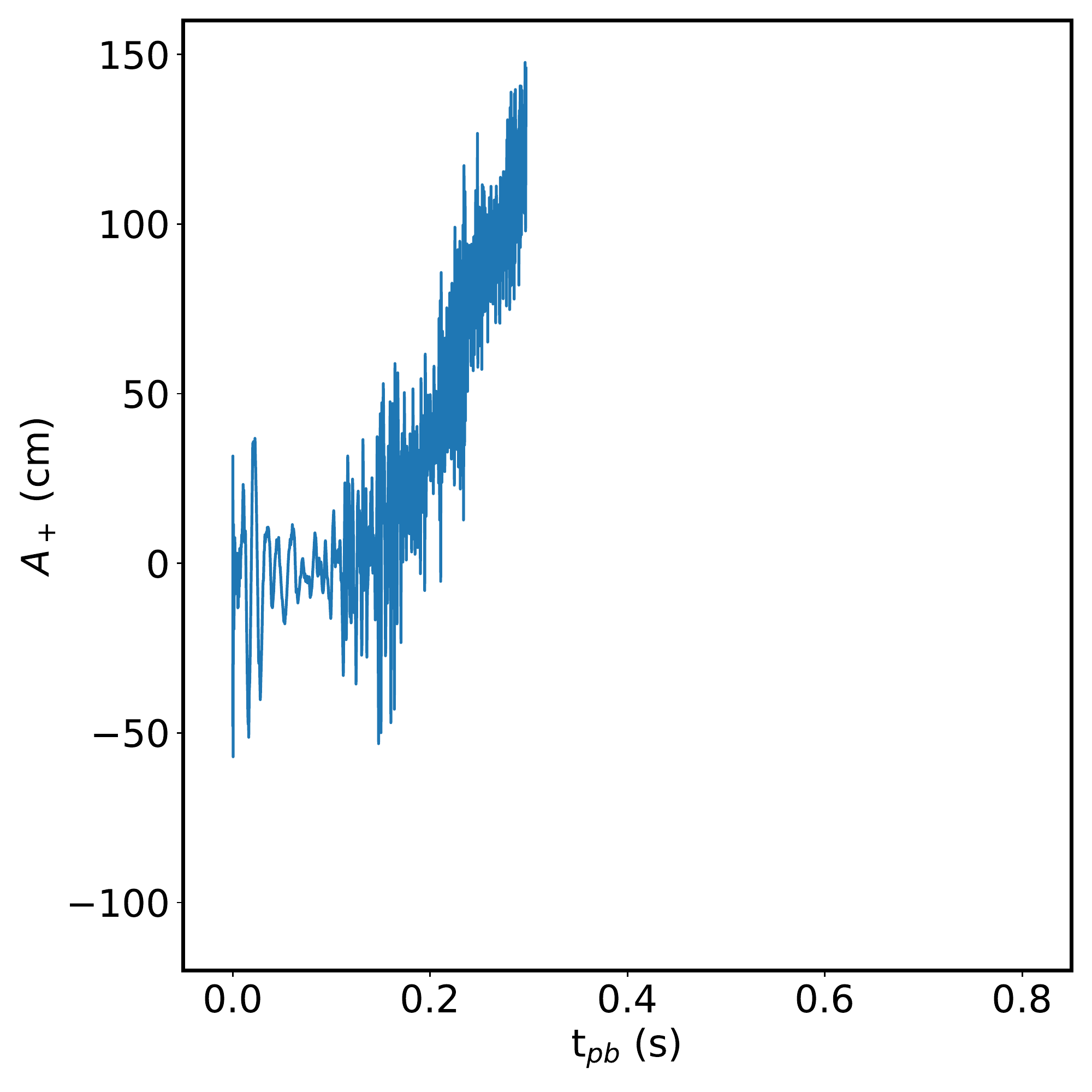}
  \caption{m39pfB12}
\end{subfigure}
\begin{subfigure}{0.33\textwidth}
  \centering
  \includegraphics[width=1\linewidth]{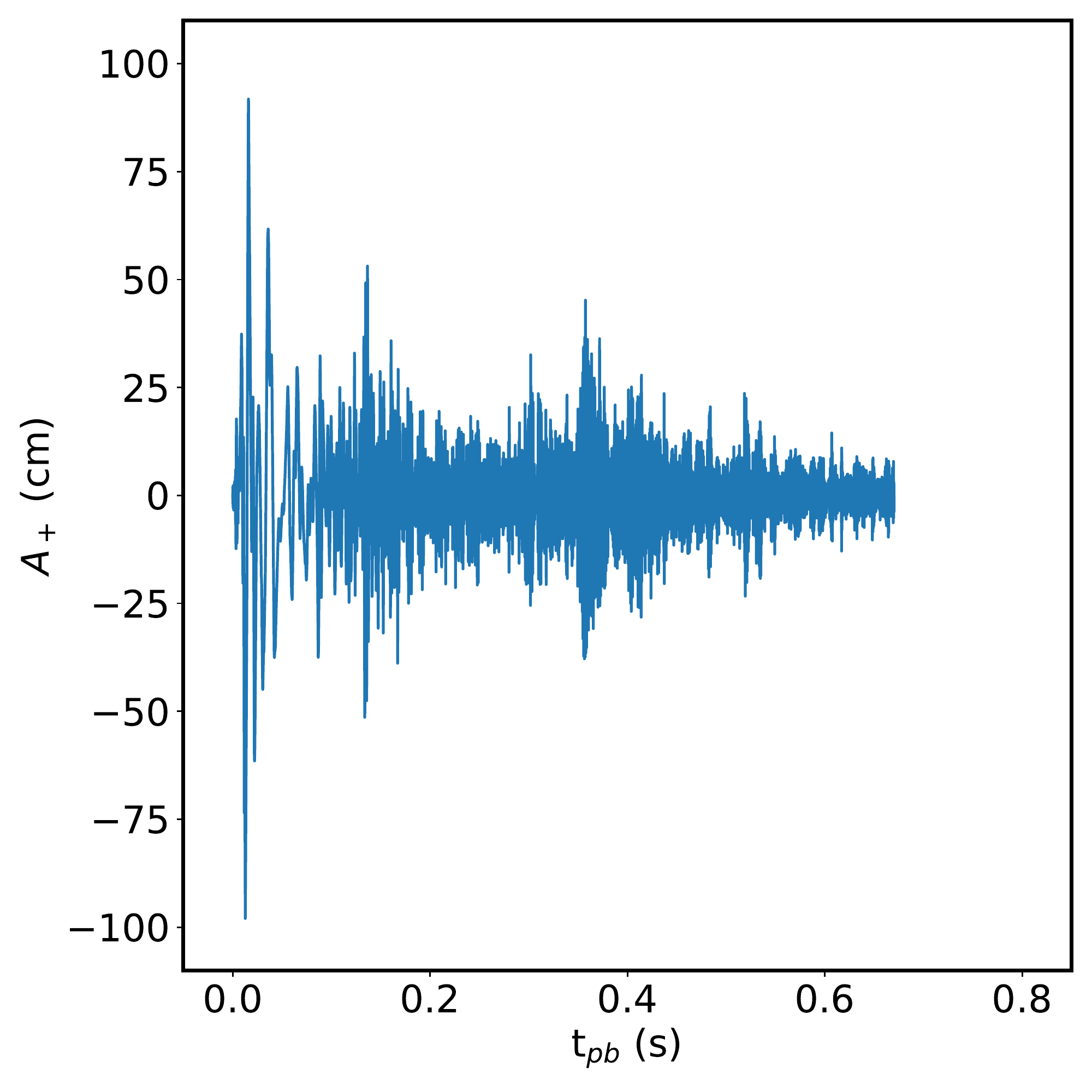}
  \caption{m39nrB0}
\end{subfigure}%
\begin{subfigure}{0.33\textwidth}
  \centering
  \includegraphics[width=1\linewidth]{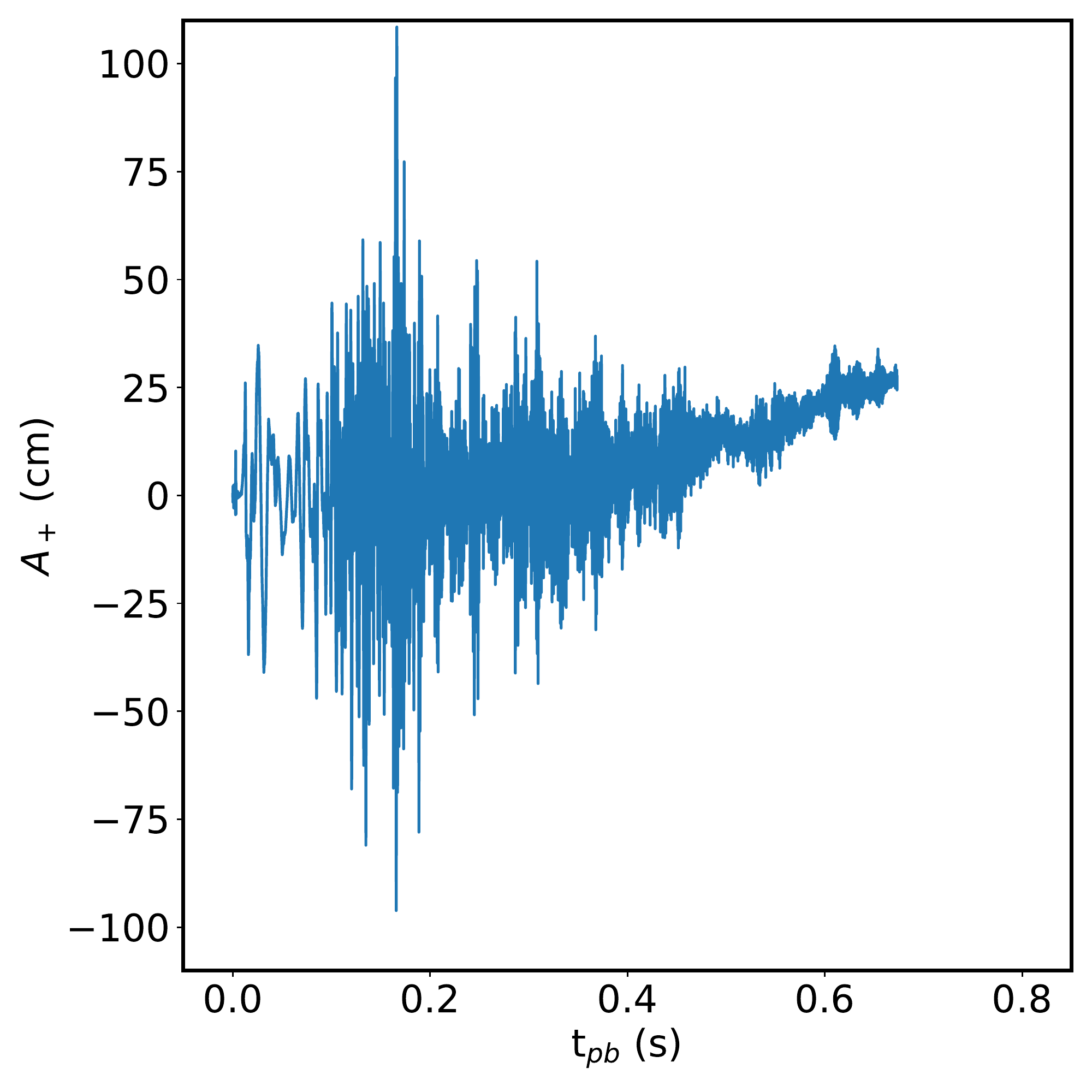}
  \caption{m39nrB10}
\end{subfigure}%
\begin{subfigure}{0.33\textwidth}
  \centering
  \includegraphics[width=1\linewidth]{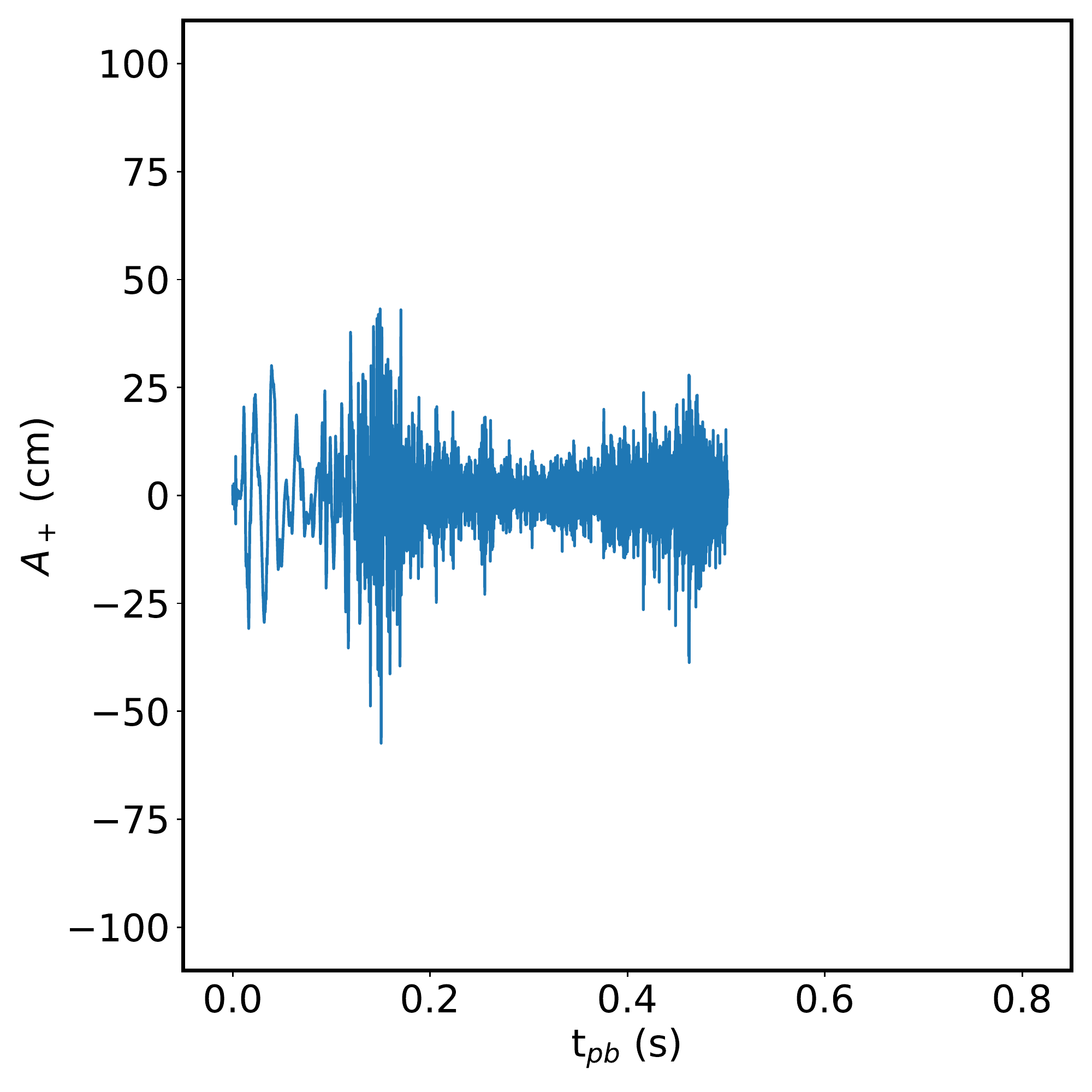}
  \caption{m39nrB12}
\end{subfigure}
 \caption{
 Gravitational wave amplitudes $A_+$ 
 of the plus polarisation mode in
 the equatorial plane for the $39\,M_{\odot}$ models.
 Note again the pronounced tail for the magnetorotational explosion model m39pfB12, similar to model m15afB12 in Figure~\ref{fig:amp}.}\label{fig:ampbig}
\end{figure*}

\subsection{Gravitational Wave Signals}
The main focus of this paper is investigating the impact of magnetisation and rotation on the GW signal. 
To this end, we compute the GW amplitude
$A^\mathrm{E2}_{20}$, which is the only non-vanishing quadrupole component in the decomposition of the far-field metric
perturbations into pure-spin tensor harmonics \citep{kip1980}
in axisymmetry.
$A^\mathrm{E2}_{20}$  can be calculated in spherical polar coordinates in axisymmetry using the time-integrated quadrupole formula
\citep{finn_89,blanchet_90,Kotake_Sawai_Yamada_Sato_2004},
\begin{align}
A_{20}^{\mathrm{E} 2}(t)=& \frac{G}{c^{4}} \frac{16 \pi^{3 / 2}}{\sqrt{15}} 
\frac{\ud}{\ud t}
\int\limits_{-1}^{1} \int\limits_{0}^{\infty} \rho(r, z, t)
v_r\left(\frac{3}{2}z^2-1\right)
-3v_\theta z \sqrt{1-z^2}
\,\mathrm{d} r
\,\mathrm{d} z \nonumber,
\end{align}
where $z=\cos(\theta)$ and $\rho$ is the rest mass density. The velocity, $v_i$, is expressed in terms of unit vectors in the $r$, $\theta$ and $\phi$ directions. 
Note that the magnetic field does not enter in the time-integrated Newtonian quadrupole formula, different from the stress formula \citep{obergaulinger_06}.
For an observer located at an angle $\theta$ from the symmetry axis of the source, at distance $R$,  the dimensionless GW strain, $h$ can be computed from
$A^{E2}_{20}$ as
 \citep[see, e.g.,][]{mueller_gravitational_1997}
 \begin{align}
h=\frac{1}{8} \sqrt{\frac{15}{\pi}} \sin ^{2} \theta \frac{A_{20}^{\mathrm{E} 2}(t)}{R}.
\end{align}
We do not include the GW signal from anisotropic neutrino emission
\citep{epstein_78} in this work, which only adds a low-frequency contribution and usually does not significantly increase the detectability of the signal in currently operating detectors.

We show gravitational waveforms for all our models
(except m15psB11 and m15nrB11, which do not fit into the
$3\times 3$-tableau) in
Figures~\ref{fig:amp} and \ref{fig:ampbig}, and
spectrograms in Figures~\ref{fig:freq} and \ref{fig:freqbig}. The spectrograms are constructed using a discrete short-time Fourier transform with a Blackman window function \citep{blackman1958measurement}
of width 
$31.25\, \mathrm{ms}$.

\subsubsection{Quasi-periodic Early Signal and Quiescent Period}
During the first $\mathord{\sim}100\, \mathrm{ms}$ after bounce, the characteristics of the GW emission in our models conform to the usual behaviour observed in 2D models \citep{murphy_gw_2009,marek_gw_2009,yakunin_gravitational_2010,yakunin_16,muller_2013_gw,mezzacappa_2020}.
There is no prominent signal from rotational bounce, even in the 
m15af and m39 models with relatively fast core rotation, where it is artificially suppressed because we treat the density and velocity field as spherically symmetric in the innermost 
$10\, \mathrm{km}$ of the computational domain.
During the first   $\mathord{\sim}30 \texttt{-} 50\, \mathrm{ms}$, all models show the characteristic quasi-periodic signal  with amplitudes of a few $10\, \mathrm{cm}$ as typical for 2D models, which is due to the ``ringing'' of the shock and acoustic wave propagation in the wake of prompt convection
\citep{yakunin_gravitational_2010,muller_2013_gw}.
The spectrograms show broadband power from 0 to $\mathord{\sim}1000\texttt{-}2000\, \mathrm{Hz}$, with a peak around $100\texttt{-}200\, \mathrm{Hz}$, although it is difficult to see the high frequency emission due to its transient nature.
The quasi-periodic signal is followed by a relatively quiescent
period of GW emission until $\mathord{\sim}100\ \mathrm{ms}$
after bounce; some models show some recognisable activity
at $100\texttt{-}200\, \mathrm{Hz}$ due to early SASI activity during this phase. The periods of early SASI oscillations tend
to be of order $20\, \mathrm{ms}$ (cp.\ Figure~\ref{fig:m15_sasi}),
which is consistent with the low-frequency features in 
the waveforms if frequency doubling is taken into account \citep{abjgw2017}. A precise matching of the gravitational
wave frequency with the time-frequency structure of the
shock oscillations is difficult because in addition to
frequency doubling, (weak) quadrupole modes of the SASI may
also contribute to gravitational wave emission \citep{abjgw2017}.

The amplitudes of the quasi-periodic signal differ by a factor of several for different models
of a given progenitor. For example, model m39pfB0 sticks out with
a peak amplitude of almost $100\, \mathrm{cm}$ compared
to about $30\, \mathrm{cm}$ in m39nrB10 and m39nrB12.
However, the amplitudes of the early quasi-periodic signal are known to be affected strongly by stochasticity. We therefore cannot identify any clear trends with rotation and initial magnetic field strengths in our models; all features of the GW signal are well within the confines of non-rotating, non-magnetised 2D models during this phase.

\subsubsection{GW Signal from Convection and the SASI}
Subsequently, the models still largely conform to the typical
GW emission patterns known from 2D and 3D models during
the pre-explosion and early explosion phase, although
important differences to the canonical picture  emerge especially in case of the magnetorational explosion models m15afB12 and m39pfB12 upon closer inspection.

Thus, from  $\mathord{\sim}100\,\mathrm{ms}$ after bounce neutrino-driven  convection and PNS convection start to shape the GW signal by exciting various oscillation modes,
specifically a dominant high-frequency emission band from
a quadrupolar f/g-mode \citep{muller_2013_gw,Morogw2018,torres_18}. In some cases, there is strong SASI activity that directly translates into a low-frequency GW emission band.  
The amplitudes of the models during the pre-explosion phase are quite similar for a given progenitor. Rotation and magnetic field primarily exert an indirect influence on GW emission caused by convection and the SASI by influencing the conditions for explosion; in models that undergo shock revival, GW emission characteristically ramps up for a few
$100\, \mathrm{ms}$ \citep{muller_2013_gw}, whereas
the rapidly rotating models m15afB0 and m15afB10 betray a more direct effect of rotation on GW emission because they have significantly lower GW amplitudes than their counterparts with no or slow rotation.  This is in line with the findings of \citet{andresen_19}, who found that high-frequency GW emission is suppressed for strong rotation because of the stabilising influence
of positive angular moment gradients on convection. Different from
the 3D model m15fr of \citet{andresen_19} which showed strong GW emission from the spiral mode of the SASI, there is nothing to compensate for the reduced high-frequency emisssion, and the overall effect of rapid rotation is a reduction of the GW amplitudes.

\begin{figure*}
\centering
\begin{subfigure}{0.33\linewidth}
  \centering
  \includegraphics[width=1\linewidth]{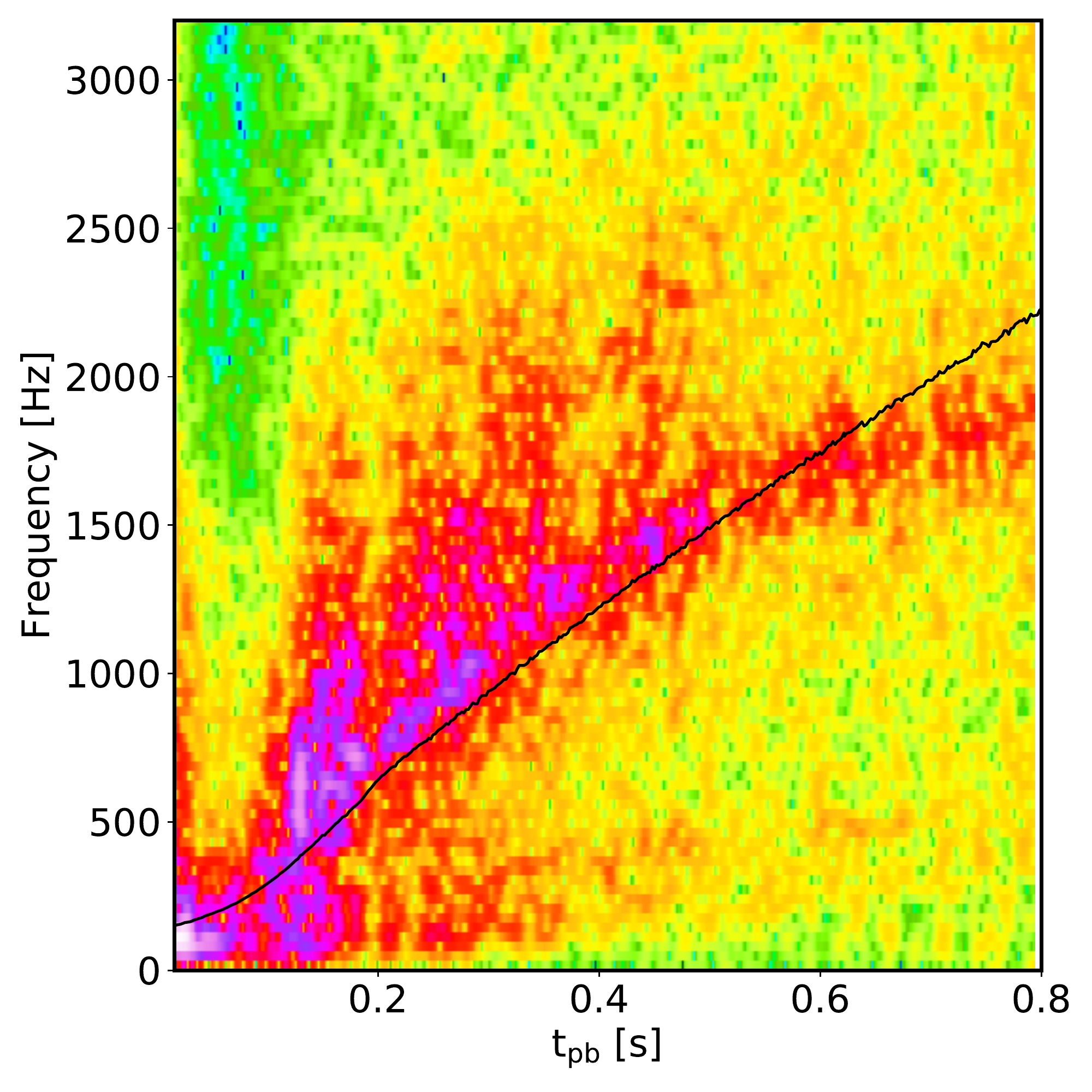}
  \caption{m15psB0}
\end{subfigure}%
\begin{subfigure}{0.33\textwidth}
  \centering
  \includegraphics[width=1\linewidth]{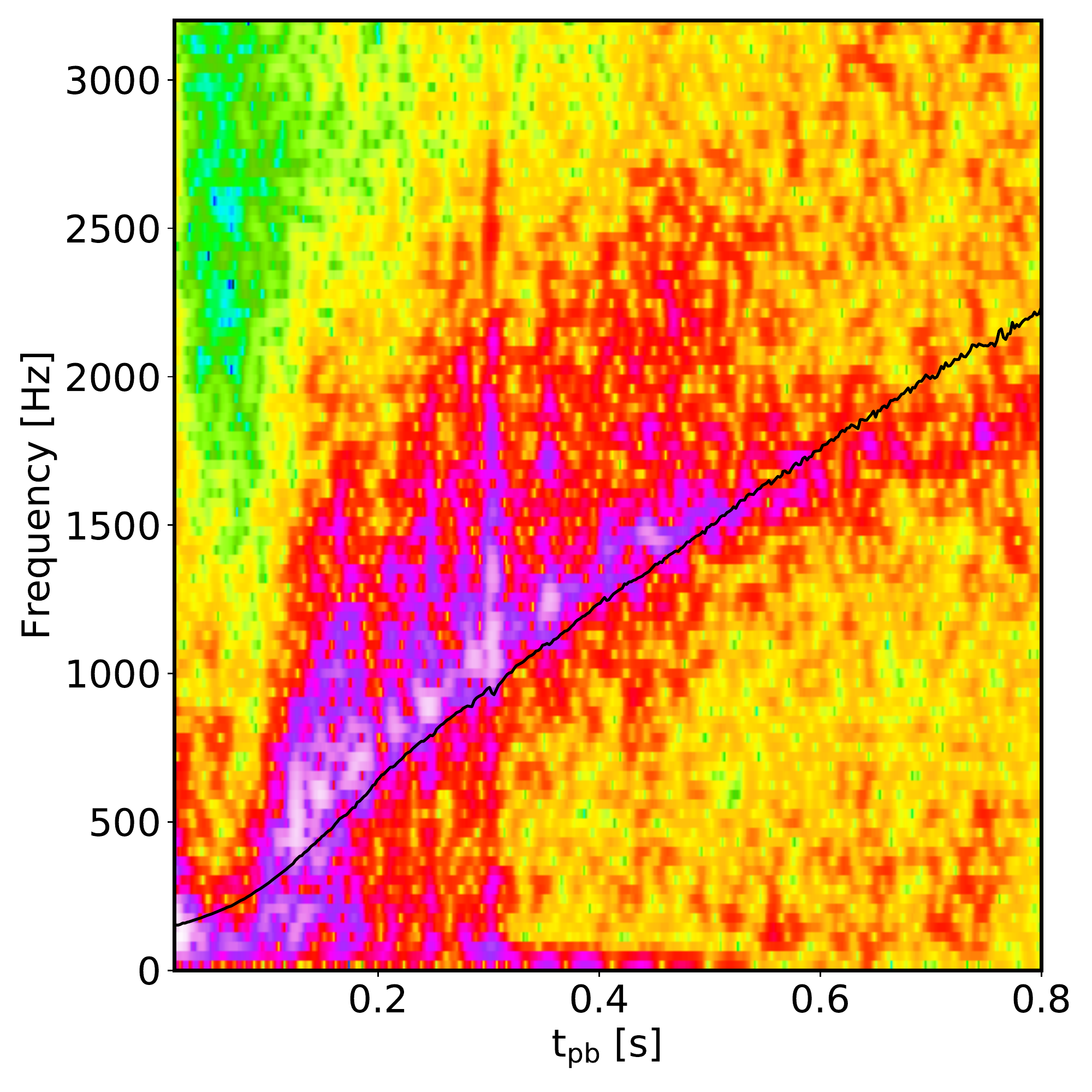}
  \caption{m15psB10}
\end{subfigure}%
\begin{subfigure}{0.33\textwidth}
  \centering
  \includegraphics[width=1\linewidth]{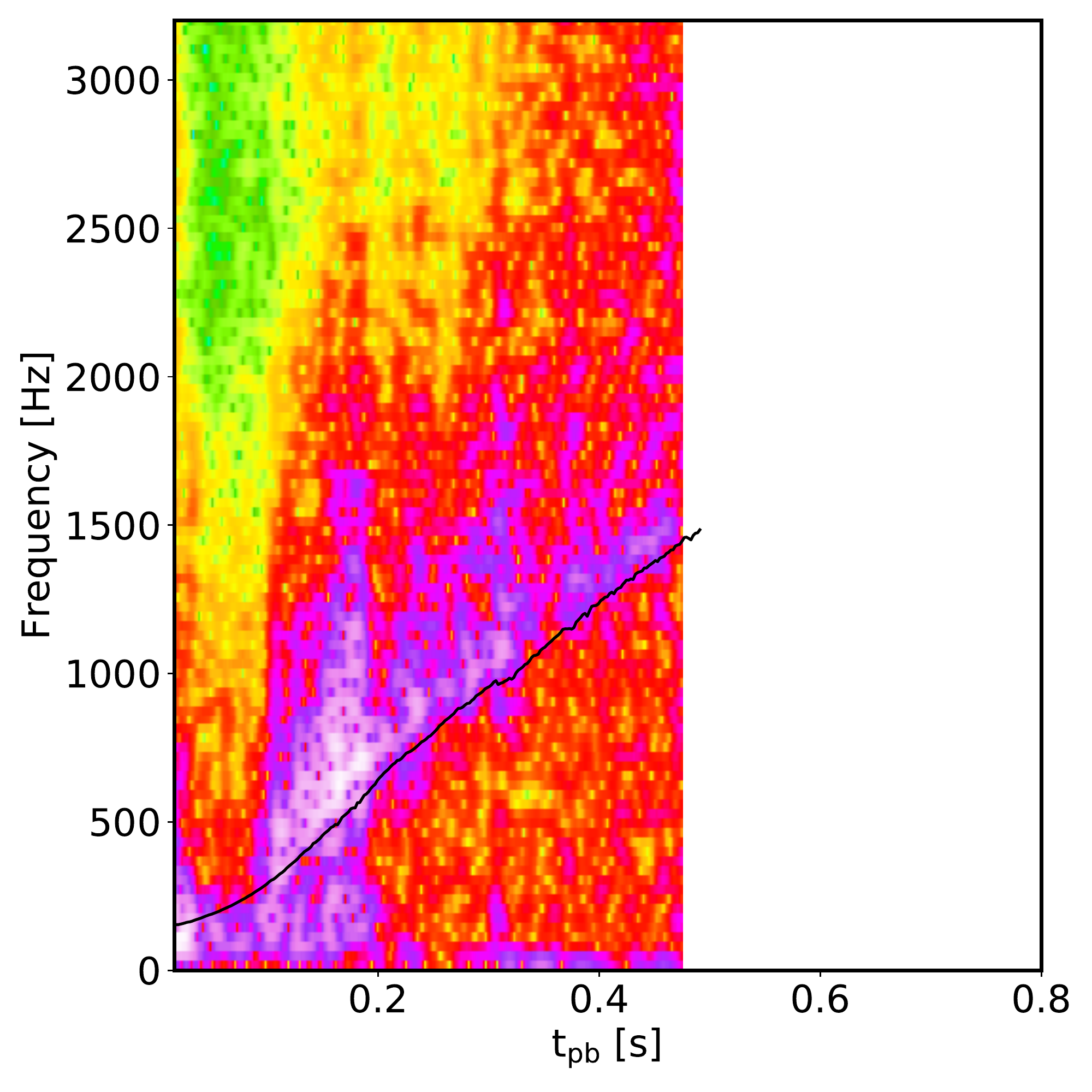}
  \caption{m15psB12}
\end{subfigure}
\begin{subfigure}{0.33\textwidth}
  \centering
  \includegraphics[width=1\linewidth]{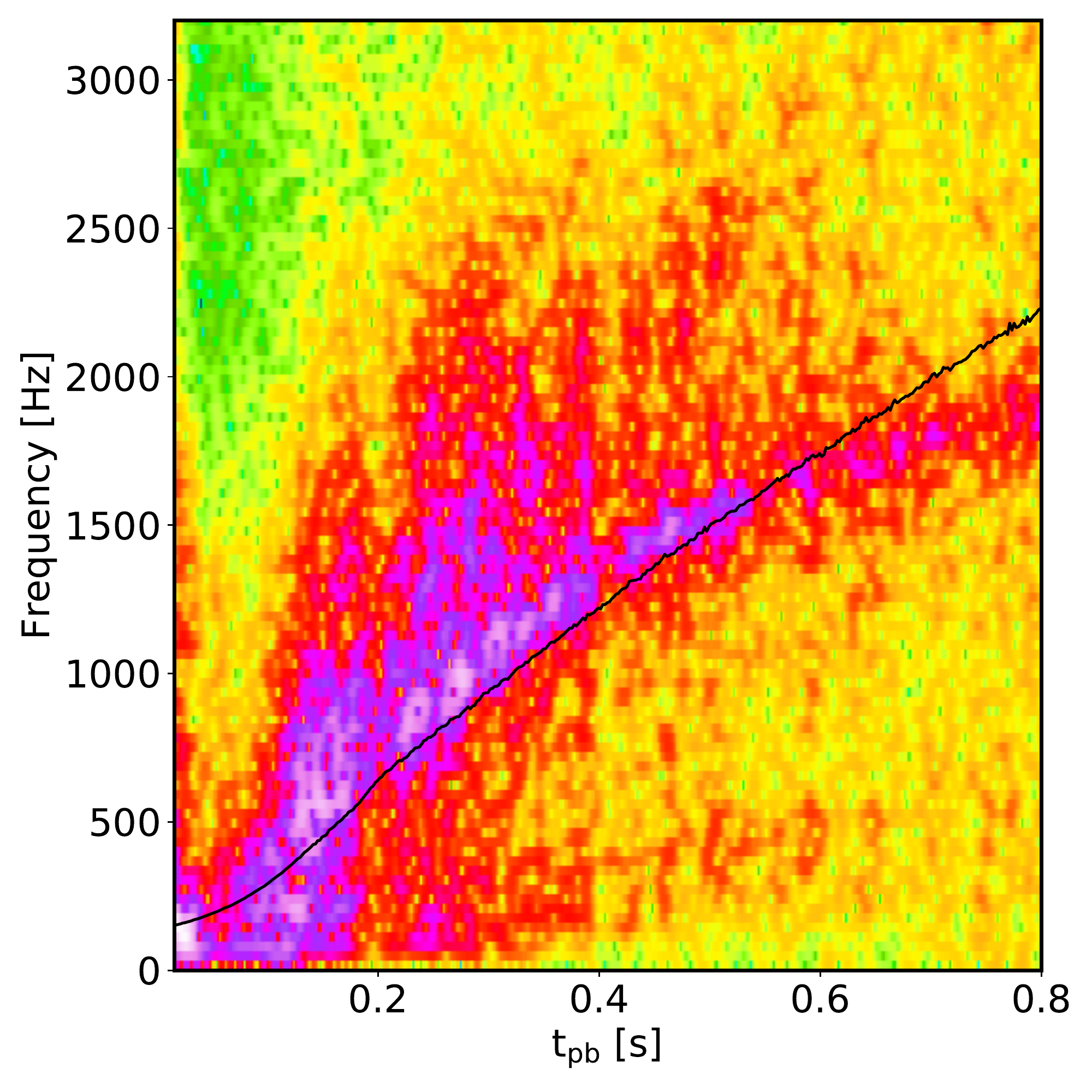}
  \caption{m15nrB0}
\end{subfigure}%
\begin{subfigure}{0.33\textwidth}
  \centering
  \includegraphics[width=1\linewidth]{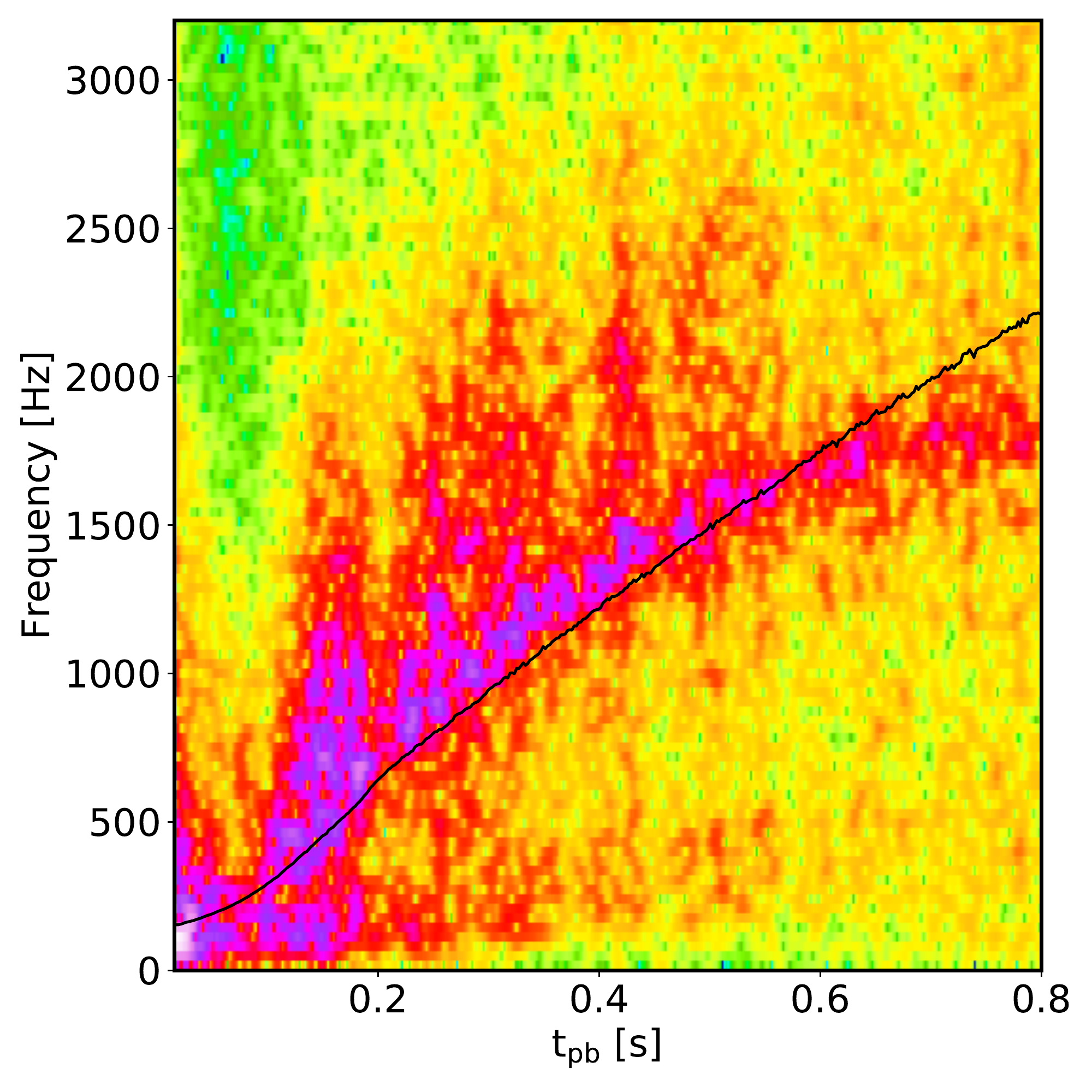}
  \caption{m15nrB10}
\end{subfigure}%
\begin{subfigure}{0.33\textwidth}
  \centering
  \includegraphics[width=1\linewidth]{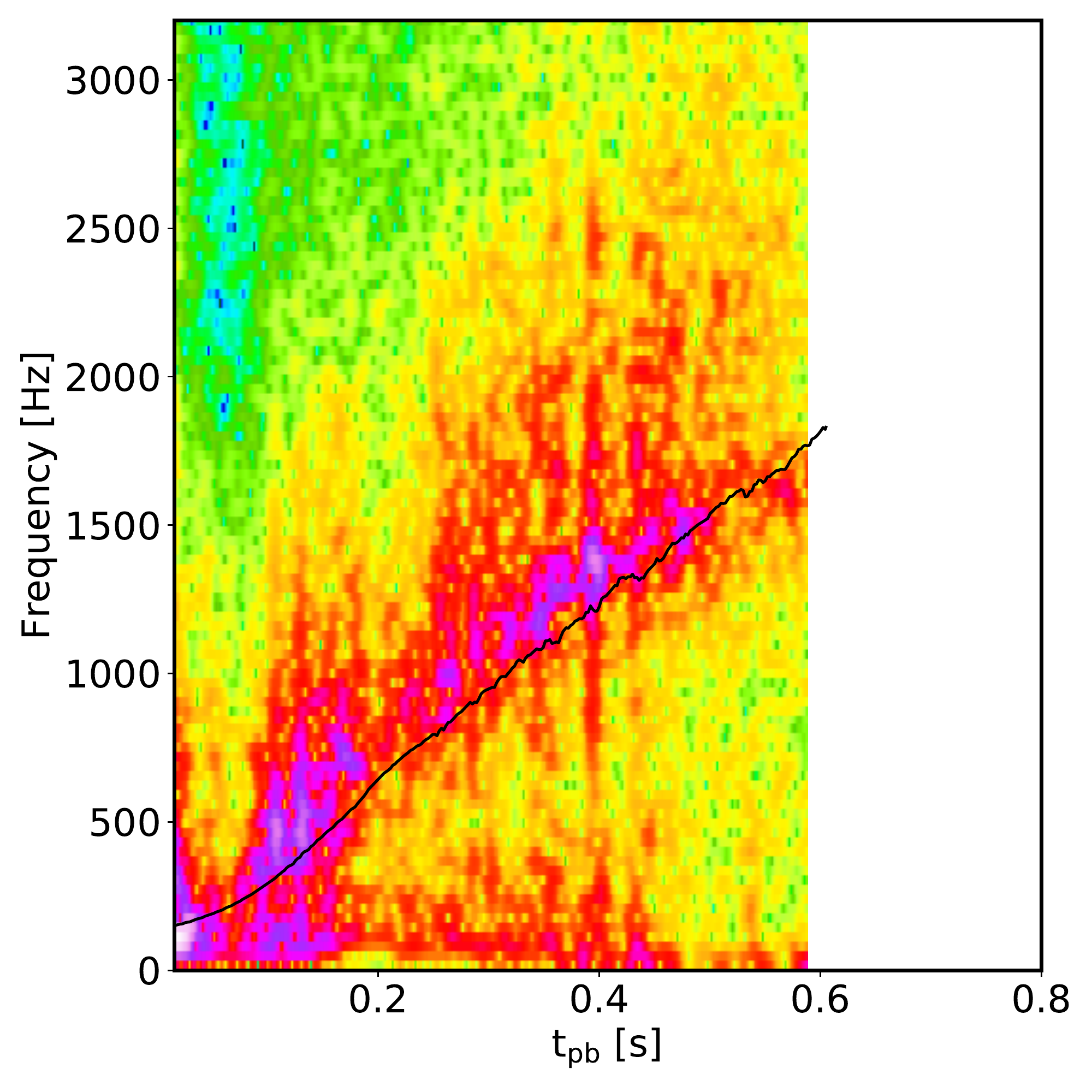}
  \caption{m15nrB12}
\end{subfigure}
\begin{subfigure}{0.33\textwidth}
  \centering
  \includegraphics[width=1\linewidth]{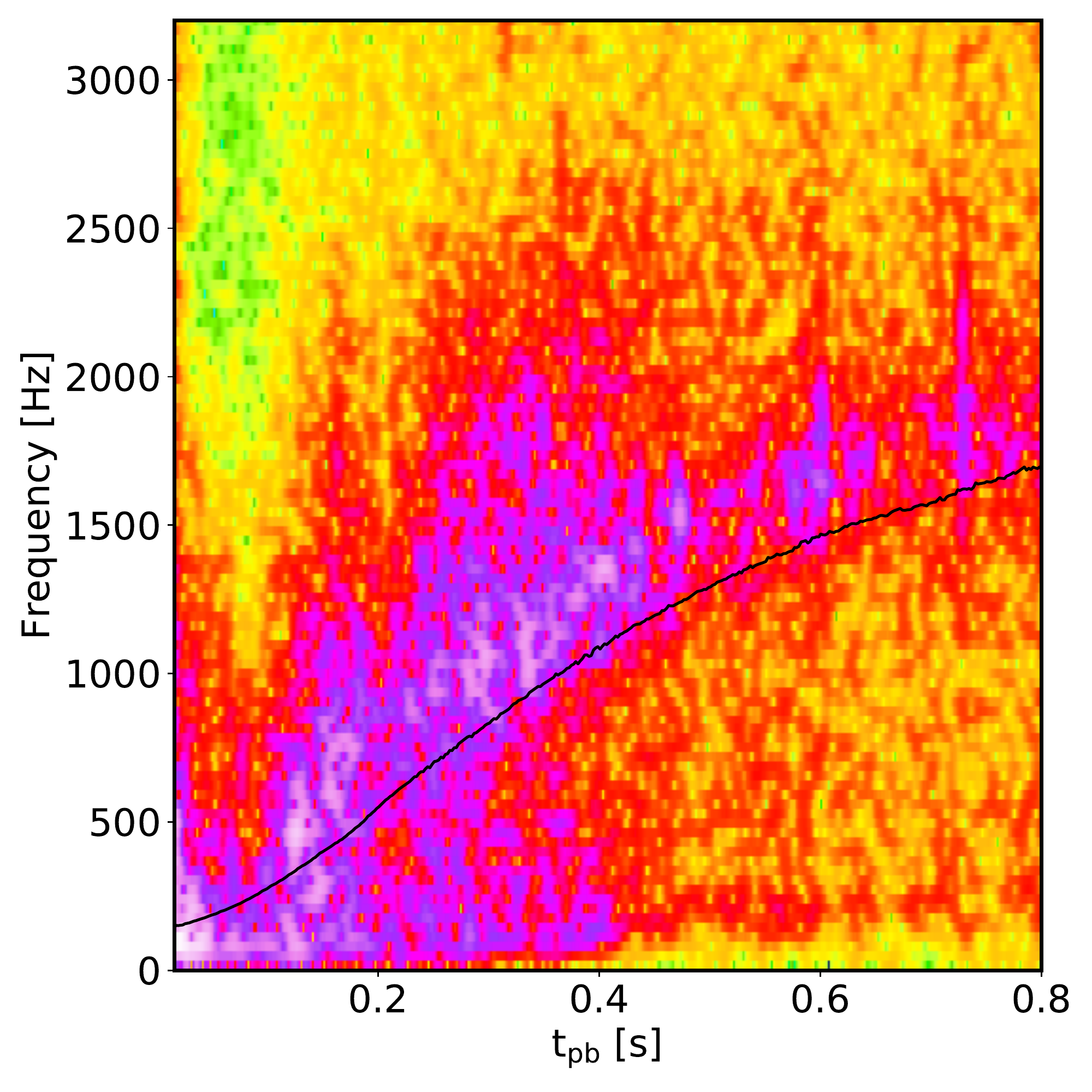}
  \caption{m15afB0}
\end{subfigure}%
\begin{subfigure}{0.33\textwidth}
  \centering
  \includegraphics[width=1\linewidth]{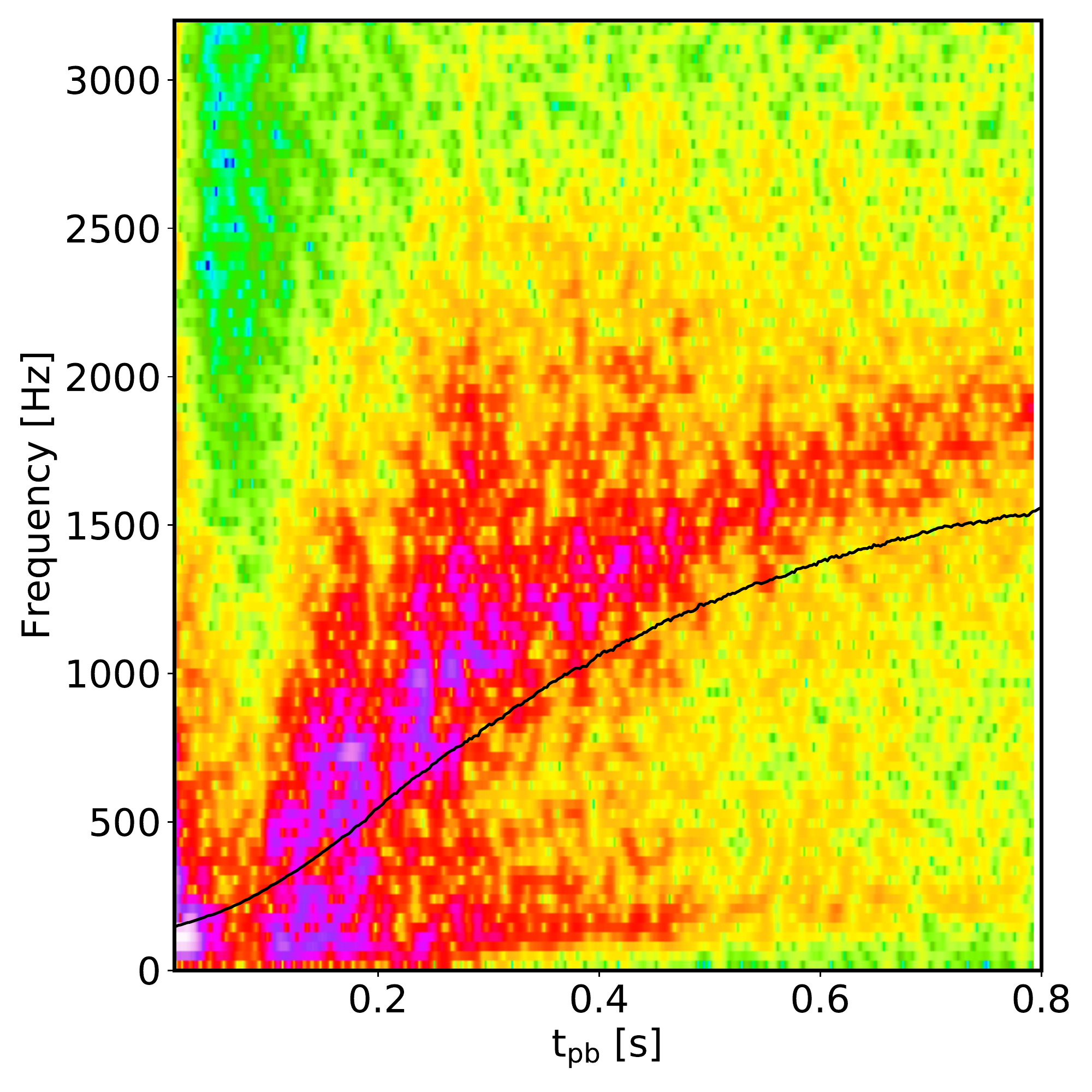}
  \caption{m15afB10}
\end{subfigure}%
\begin{subfigure}{0.33\textwidth}
  \centering
  \includegraphics[width=1\linewidth]{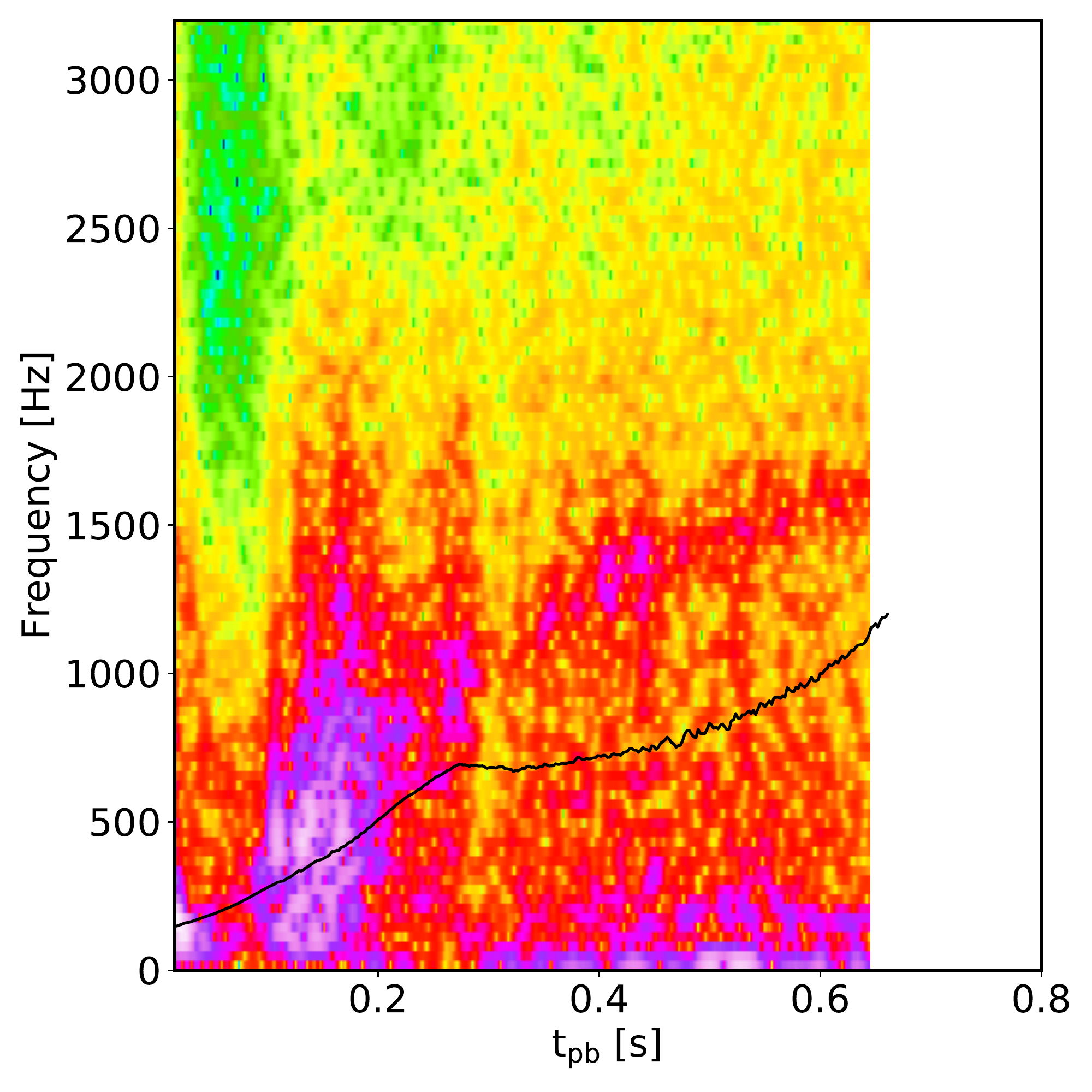}
  \caption{m15afB12}
\end{subfigure}%
 \caption{GW spectrograms for the $15\,M_{\odot}$ models. Power increases from green to yellow, red, purple, and white. 
 Note the colours of the spectrogram are represented on a logarithmic scale, whose range is not identical for all models.
 Black curves shows the predicted dominant f/g-mode frequency according to Equation~(\ref{eq:fpeaks}). For the non-rotating and moderately rotating models, the analytic relation
 provides a close fit to the dominant f/g-mode frequency except at late times. For the rapidly rotating models,  Equation~(\ref{eq:fpeaks}) underestimate the dominant emission frequency in the case of no or moderately strong initial magnetic fields (Panels g and h). For the magnetorational explosion model m15afB12
 (Panel i), one can no longer discern a narrow emission band at high frequencies after 
 a post-bounce time of $0.2\, \mathrm{s}$, and broad-band high-frequency emission appears instead. }\label{fig:freq}
\end{figure*}

\begin{figure*}
\centering
\begin{subfigure}{0.33\linewidth}
  \centering
  \includegraphics[width=1\linewidth]{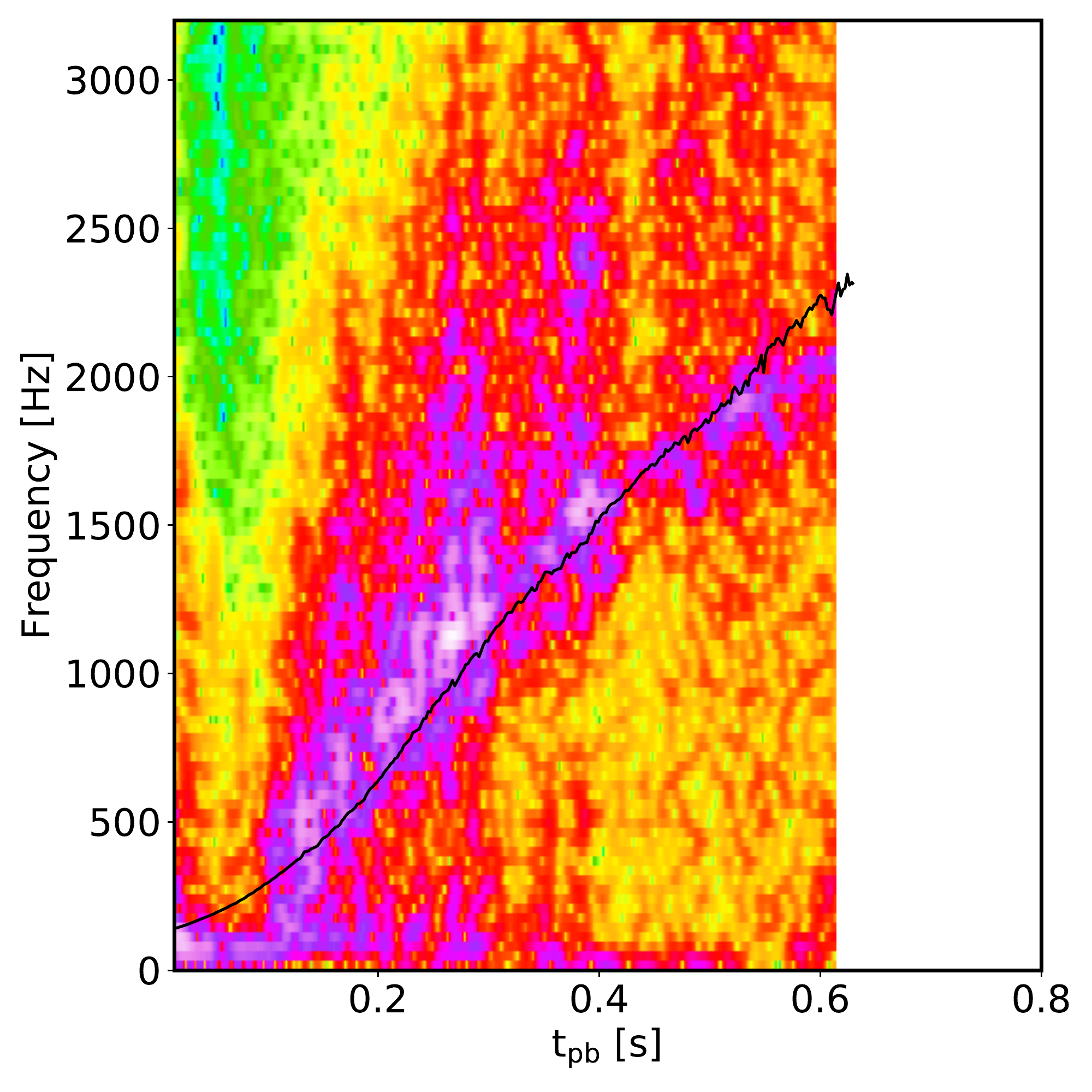}
  \caption{m39pfB0}
\end{subfigure}%
\begin{subfigure}{0.33\textwidth}
  \centering
  \includegraphics[width=1\linewidth]{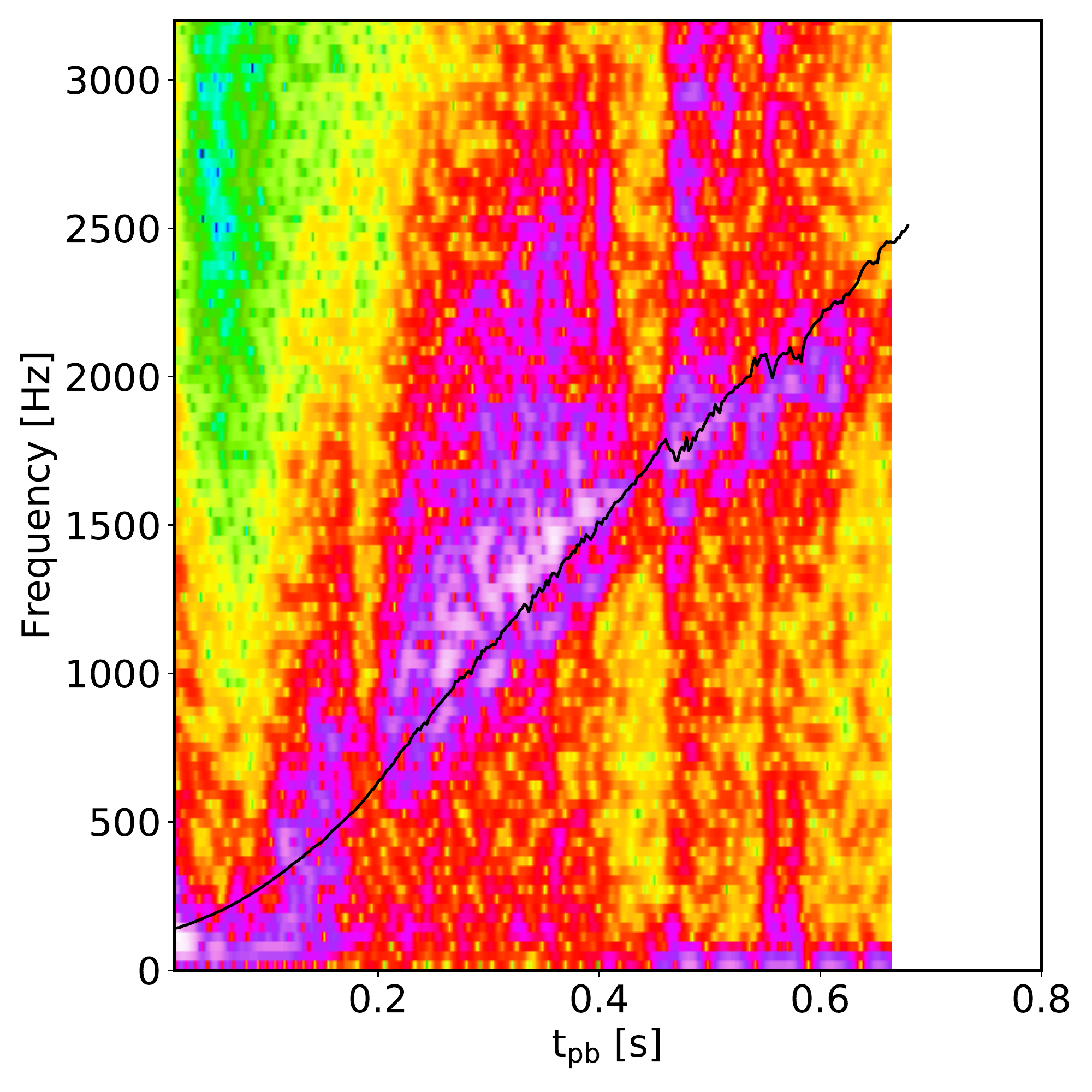}
  \caption{m39pfB10}
\end{subfigure}%
\begin{subfigure}{0.33\textwidth}
  \centering
  \includegraphics[width=1\linewidth]{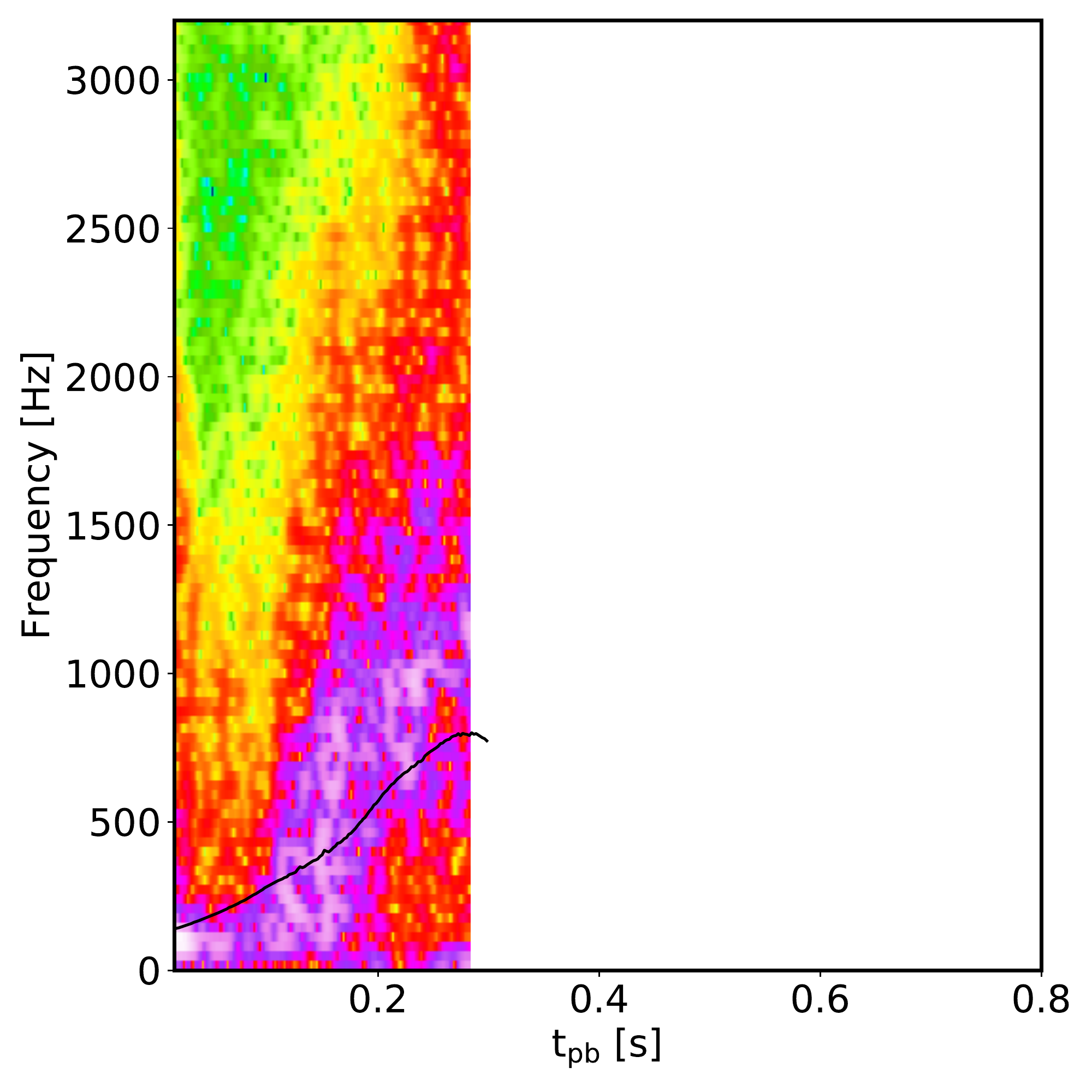}
  \caption{m39pfB12}
\end{subfigure}
\begin{subfigure}{0.33\textwidth}
  \centering
  \includegraphics[width=1\linewidth]{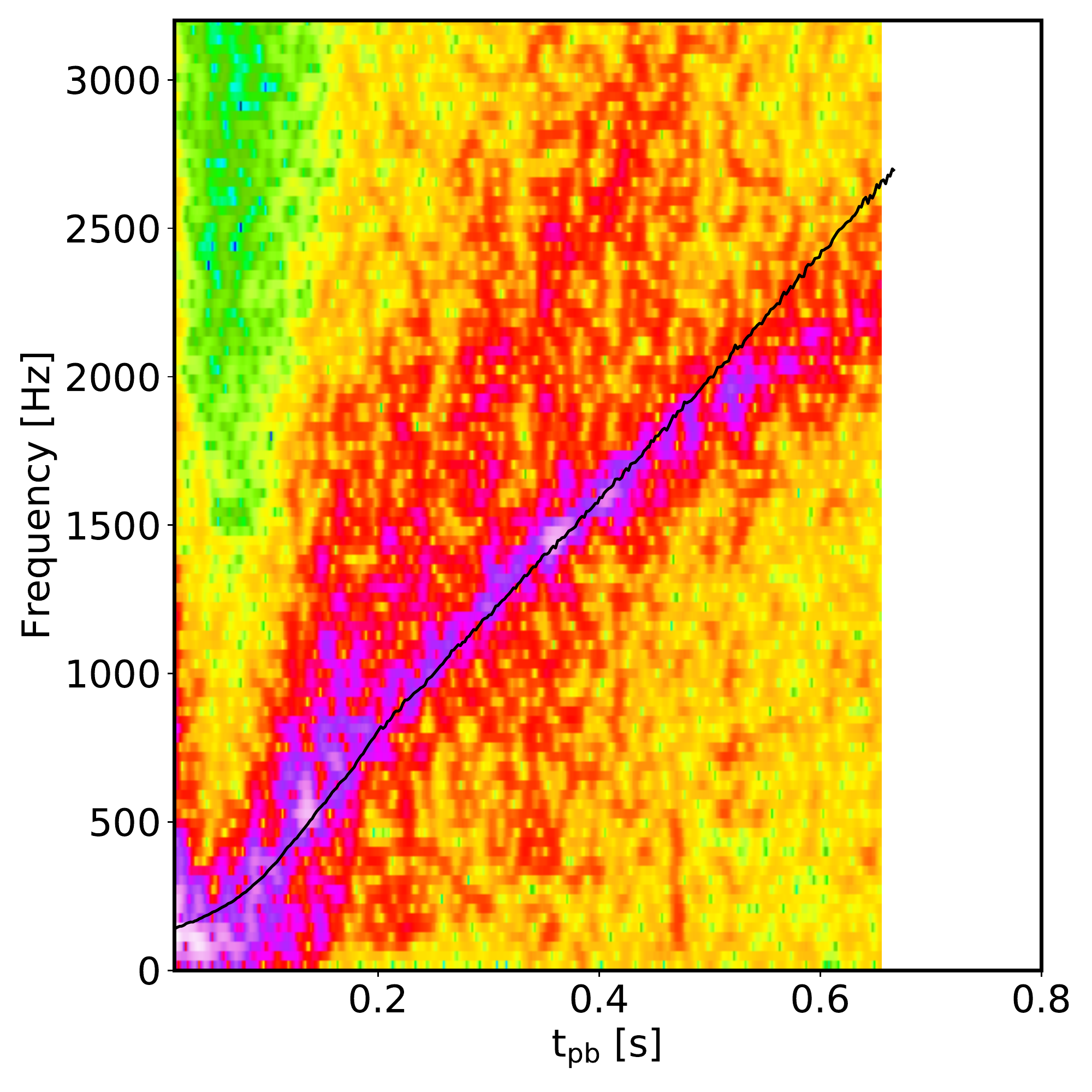}
  \caption{m39nrB0}
\end{subfigure}%
\begin{subfigure}{0.33\textwidth}
  \centering
  \includegraphics[width=1\linewidth]{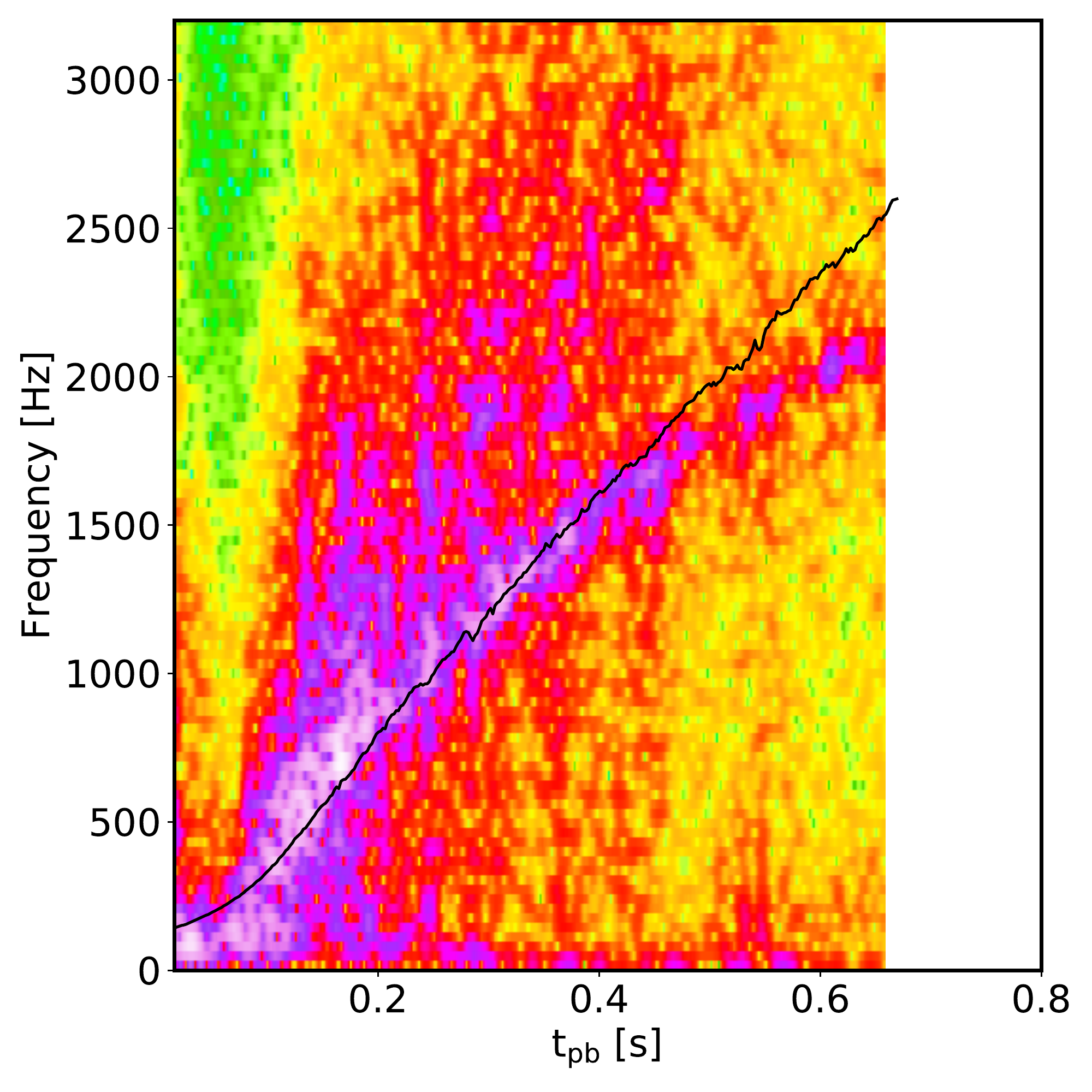}
  \caption{m39nrB10}
\end{subfigure}%
\begin{subfigure}{0.33\textwidth}
  \centering
  \includegraphics[width=1\linewidth]{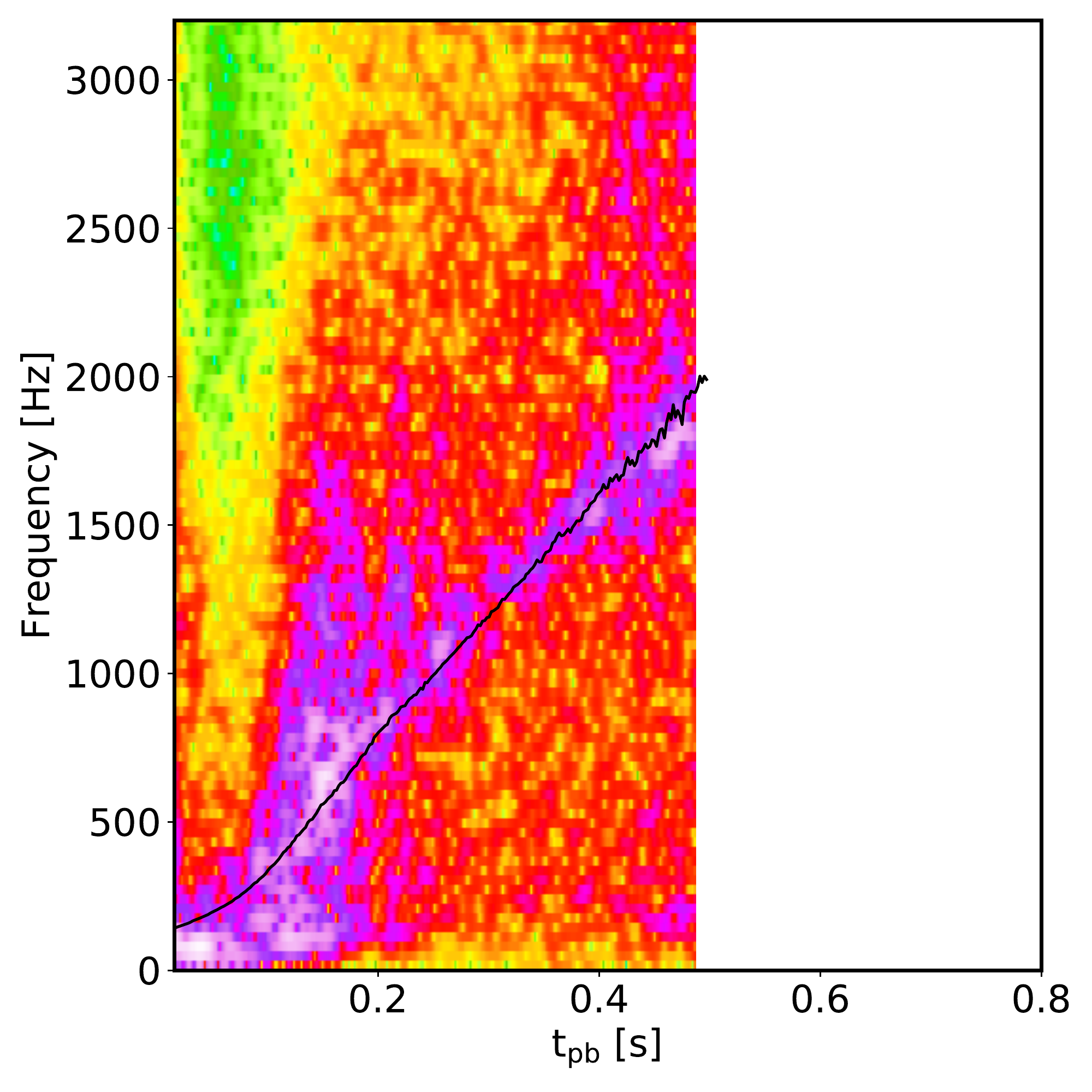}
  \caption{m39nrB12}
\end{subfigure}
 \caption{
 GW spectrograms for the $39\,M_{\odot}$ models. Power increases from green to yellow, red, purple, and white. 
 Note the colours of the spectrogram are represented on a logarithmic scale. 
 Black curves shows the predicted dominant f/g-mode frequency according to Equation~(\ref{eq:fpeaks}), which fits the high-frequency emission band well except for the magnetorotational explosion model
 m39pfB12 (Panel c).}\label{fig:freqbig}
\end{figure*}

Many (but not all) of the differences in the spectrogram can also be explained by the presence or absence of shock revival.
For example, for the non-exploding 
 $15\,M_{\odot}$ models in Figure~\ref{fig:freq} a), d), e), g) and h), we see low-frequency SASI emission in contrast to the exploding models in panels b), c), f), and i), which afterwards dies off. The f/g-mode high-frequency emission band is weaker than other models in
 non-exploding models and sometimes entirely vanishes by $500\texttt{-}600\, \mathrm{ms}$, and there is less power 
 in the p-modes \citep{Morogw2018} above the dominant high-frequency emission bands.  The exception is the m15afB0 case, where the high-frequency emission appears more spread out and longer lasting, although the reason for this is unclear. 
The non-exploding model m39nrB0 stands apart from the other $39\,M_\odot$ in a similar manner (Figure~\ref{fig:freqbig}d), but, unlike the non-exploding $15\,M_\odot$ model, there is no indication of a signal from the SASI.

\begin{figure}
    \centering
    \includegraphics[width=\linewidth]{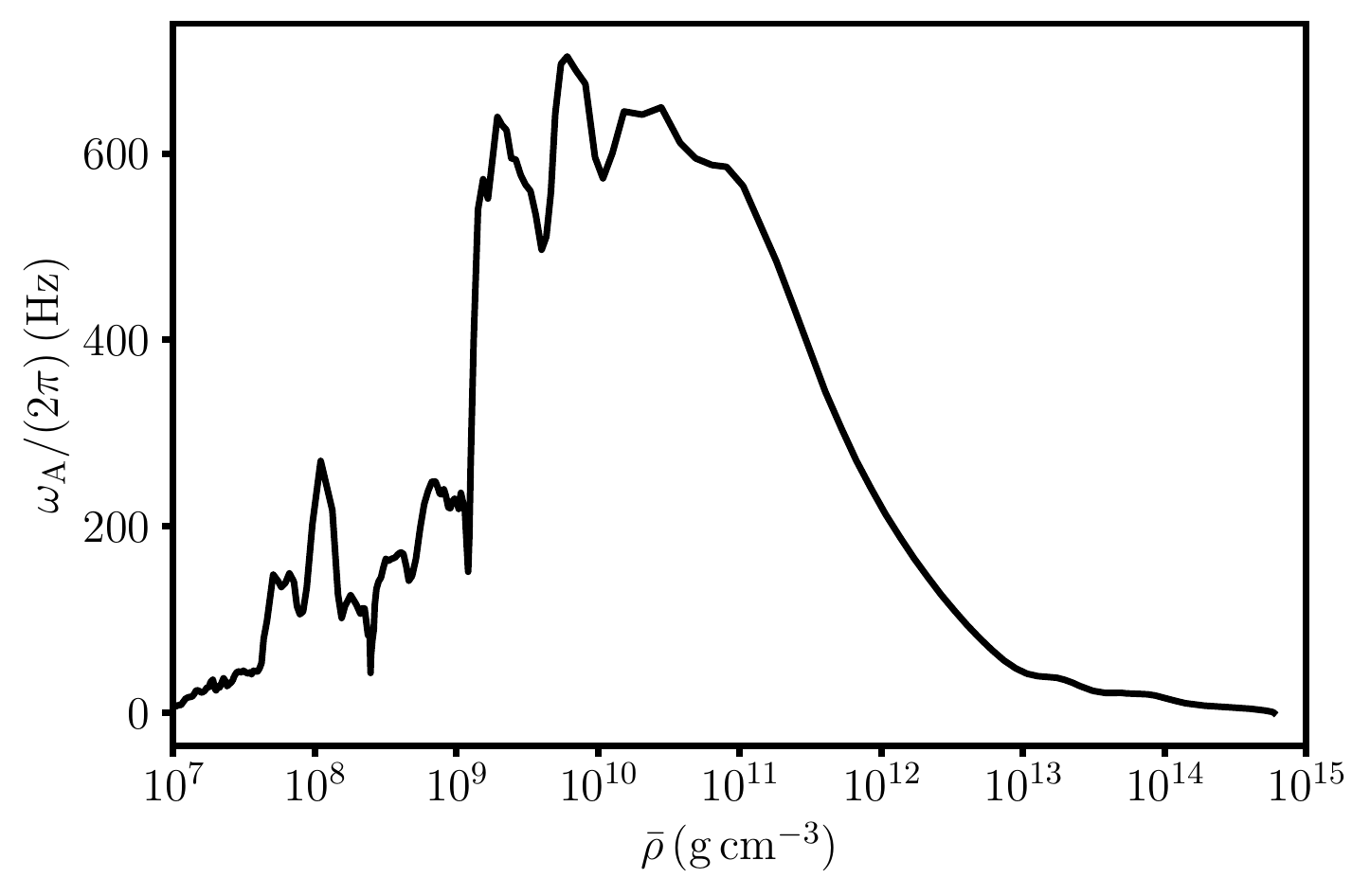}
    \caption{Alfv\'en frequency $\omega_\mathrm{A}$ for modes
    with a radial wave vector and a wavenumber of the order
    of the inverse of the pressure scale height
    (Equation~\ref{eq:alfven}) for model m15psB12 at a time
    of $400\, \mathrm{ms}$ after bounce. $\omega_\mathrm{A}$
    is shown as a function of the angle-averaged
    density $\bar\rho$.}
    \label{fig:alfven}
\end{figure}

Some of the rapidly rotating and highly magnetised models show 
effects in the spectrograms beyond differences
in the strength of the f/g-mode or SASI emission band. The highly 
magnetised models m15psB12 and m39nrB12, still exhibit a clear
f/g-mode emission band, but are noteworthy for showing power around 
this band over a broader range of frequencies compared to
non-magnetised and weakly magnetised models. 
It is possible that the frequency, excitation, coupling of the g- and p-modes above and below the dominant emission band is affected by the presence of strong  magnetic fields. 
For a tentative indication that magnetic fields could be starting
to modify the mode structure and the coupling of fluid motions near
the bottom of the gain region with oscillation modes, it is instructive
to estimate the local Alfv\'en frequency in the vicinity of the PNS
surface. While a detailed analysis of magnetohydrodynamic oscillation
modes is complicated, one can estimate the relevant Alfv\'en frequency
$\omega_\mathrm{A}$ for modes with a high radial wave number of
order of the inverse of the pressure scale height $\Lambda$ \citep[cp.][]{fuller_15}, i.e., for waves mirroring the structure of
f/g-modes with predominantly horizontal displacement and a short
radial wavelength due to the steep density gradient at the PNS surface,
\begin{equation}
\label{eq:alfven}
    \omega_\mathrm{A}\sim \frac{1}{\Lambda}
    \sqrt{\frac{\langle B_r^2\rangle}{4\pi \langle\rho\rangle}}.
\end{equation}
Here angled brackets denote spherical averages. As shown by
Figure~\ref{fig:alfven}, $\omega_\mathrm{A}$ becomes quite large at densities of $10^{9}\texttt{-}10^{11}\, \mathrm{g}\,\mathrm{cm}^{-3}$,
which corresponds to the bottom of the gain region, with $\omega_\mathrm{A}/(2\pi)$ reaching a sizeable fraction of the f/g-mode frequency. In the bulk of the convectively stable PNS surface layer
at densities of $10^{11}\texttt{-}10^{13}\, \mathrm{g}\,\mathrm{cm}^{-3}$, which mostly sets the f/g-mode frequency, $\omega_\mathrm{A}$ remains somewhat lower. The scale of $\omega_\mathrm{A}$ suggests a minor
influence of magnetic fields on the f/g-mode frequency itself, but
a potentially larger role on the interaction of convective motions
in the gain region and the f/g-mode and neighbouring modes.
However, the power outside the dominant frequency band is
still small, so that it is difficult to identify the corresponding
dynamic phenomena in the simulation. More simulations are desirable
to determine whether the broader distribution of power is a robust feature in strongly magnetised models.

Model m15nrB12 does not exhibit such a broad distribution of power, but shows an indirect impact of magnetic fields on the low-frequency emission band, whose frequency begins to decline shortly after $200\, \mathrm{ms}$ simply because
the shock starts to expand, whereas  the low-frequency emission band in the the corresponding models m15nrB0 and m15nrB10 rises monotonically with time due to shock retraction.

\begin{figure}
    \centering
    \includegraphics[width=\linewidth]{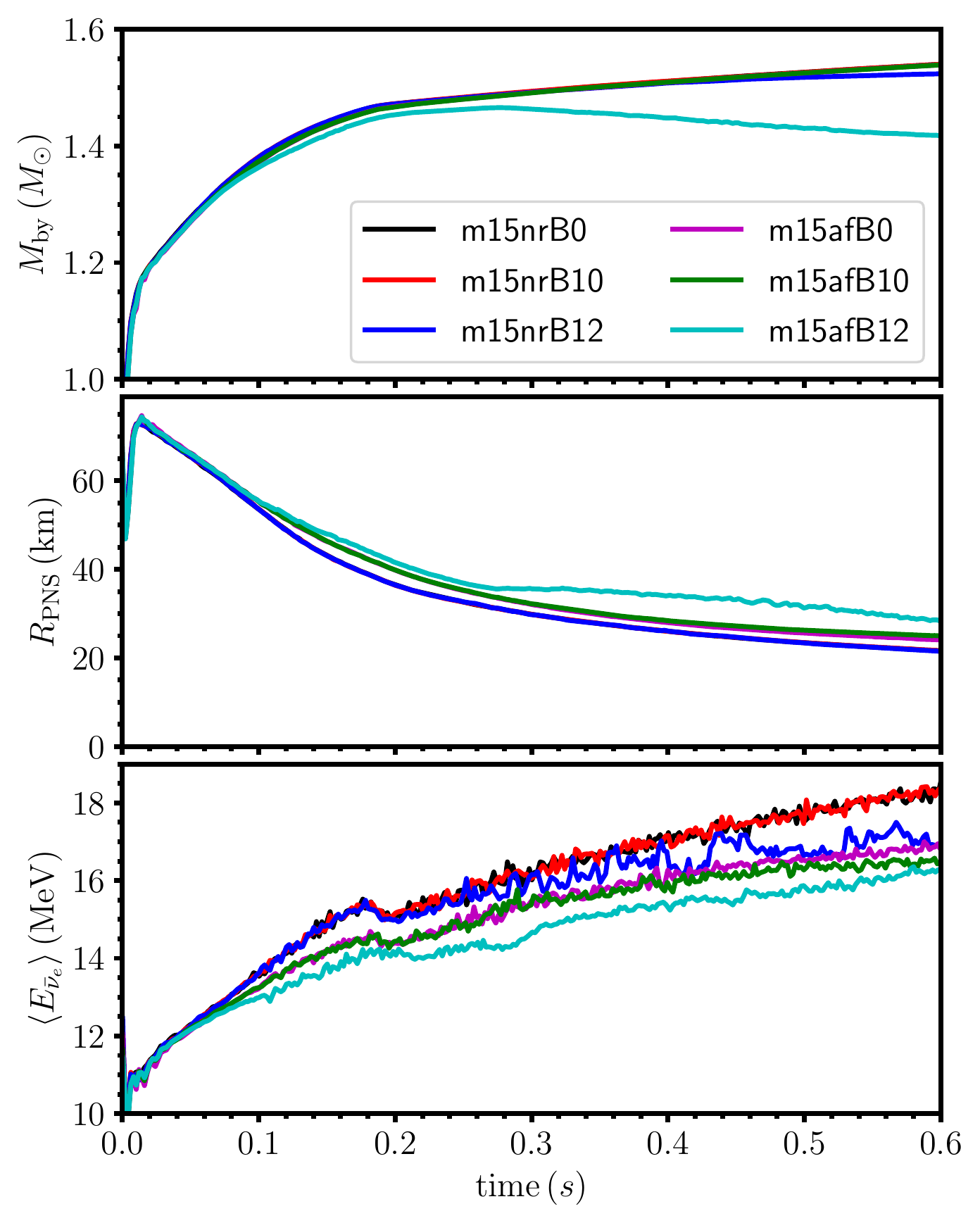}
    \caption{
    PNS parameters entering the analytic relation
    (\ref{eq:fpeaks})} for the f/g-mode frequency for the
    non-rotating $15 M_\odot$ models (m15nr) and the rapidly
    rotating $15 M_\odot$ models (m15af). From top to bottom:
    Baryonic neutron star mass $M_\mathrm{by}$, PNS radius
    $R_\mathrm{PNS}$, and electron antineutrino mean
    energy $\langle E_{\bar{\nu}_e}\rangle$ as a proxy for
    the PNS surface temperature. The rotating models are characterised
    by larger PNS radii and lower $\langle E_{\bar{\nu}_e}\rangle$.
    Model m15afB12 also exhibits a significant decrease in PNAS mass
    at late time due to strong mass outflows.
    \label{fig:gwfit}
\end{figure}

Models m15afB0 and m15afB10 illustrate a noteworthy
effect of rapid rotation on the mode structure. For non-rotating models, the frequency of the dominant f/g-mode emission band in  pseudo-Newtonian simulations is well described by \citep{muller_2013_gw}, 
 \begin{align}\label{eq:fpeaks}
 f_{\text {peak }}  \approx \frac{1}{2 \pi} \frac{G M_{\text {by}}}{R_{\text {PNS}}^{2}} \sqrt{1.1 \frac{m_\mathrm{n}}{\left\langle E_{\bar{v}_{e}}\right\rangle}}, 
 \end{align}
 where $R_{\text{PNS}}$, is the radius of the PNS, $E_{\bar{v}_{e}}$ is the electron antineutrino mean energy, $m_\mathrm{n}$ is the neutron mass and $M_{\text {by}}$ is the baryonic mass of the PNS. In line with previous 2D and 3D simulations, 
 Equation~(\ref{eq:fpeaks}) matches the high-frequency emission band well during the first few hundred milliseconds in almost all models and somewhat overestimates the dominant frequency at late times (black curves in Figures~\ref{fig:freq} and \ref{fig:freqbig}). 
 The m15af models and m39pfB12 are exceptions, however.
 As illustrated by Figure~\ref{fig:gwfit},
 the predicted frequency from Equation~(\ref{eq:fpeaks}) is lowered
 compared to the non-rotating case for these models
 mostly because centrifugal support slightly increases the PNS radii
 and lowers the electron antineutrino mean energy; in case of magnetorotational models, the PNS mass can also be significantly lower
 than in the non-rotating case because of a rapid explosion and
 strong mass outflows. The spectrograms do not show the expected
 decrease in the f/g-mode frequency.
 In m15afB0 and m15afB10, the dominant emission frequency is up to $\mathord{\sim}20\%$ higher than predicted  by Equation~(\ref{eq:fpeaks}). 
 The higher frequency is consistent with the presence of a  positive angular momentum gradient in the region between $10^{12}\, \mathrm{g}\, \mathrm{cm}^{-3}$ and  $10^{13}\, \mathrm{g}\, \mathrm{cm}^{-3}$, which provides additional stabilisation  against convection in the PNS surface region. It is noteworthy, though, that the f/g-mode frequency is not shifted upward to the same degree in model m39 despite a positive angular momentum gradient in this region. Compared to the m39nr models, the dominant frequency still tends to lie slightly
 higher relative to Equation~(\ref{eq:fpeaks}) in the m39 models, but the effect is not as pronounced. It is also interesting that in a 3D model of
 the $39\,M_\odot$ model without magnetic fields,  \citet{powell_20}
 found that rotation \emph{decreases} the dominant emission frequency
 compared to Equation~(\ref{eq:fpeaks}). This, however, can be understood based on the detailed dynamics of the models during the explosion phase.
 In the 3D model of
 \citet{powell_20}, a region around the PNS with retrograde rotation emerges due to stochastic variations in the specific angular momentum of accretion downflows. This leads to a negative, destabilising angular momentum gradient at the neutron star surface that lowers the f/g-mode frequency. No such effect occurs in our 2D models. While it may be possible to understand the impact of rotation in more detail for a given simulation by generalising linear perturbation analysis for eigenmodes \citep{sotani_16,Morogw2018,torres_18} to the rotating case, the disparate results of \citet{powell_20} and our study suggest that the impact of rotation on the mode frequencies may be somewhat stochastic and could introduce a generic uncertainty in interpreting the f/g-mode frequency of the PNS in rapid rotators if it can be measured.

Finally, the spectrograms of the magnetorotational models m15afB12 and m39pfB12 deviate even more strongly from the familiar picture from extant 2D and 3D simulations. During the pre-explosion and early explosion phase, they are not remarkable in terms of amplitudes, but the peak emission clearly does \emph{not}
follow Equation~(\ref{eq:fpeaks}). In model m39pfB12, which has been run for less than $0.3\, \mathrm{s}$ post-bounce, one might discern a well-defined emission band with a rapid rise in frequency that certainly does not follow the familiar
f/g-mode frequency relation. In m15afB12, 
we find a well-defined high-frequency emission band, which is more visibly separated from the predicted trajectory
based on Equation~(\ref{eq:fpeaks}). There is also considerable power at frequencies below the dominant band, similar
to models m15psB12 and m39nrB12. In contrast to these models,
rotation and magnetic fields clearly affect the mode
structure. A detailed analysis of PNS eigenmodes for the ``millisecond magnetars'' formed in such models is called for in future, but is beyond the scope of the current, more descriptive study.

\subsubsection{Advanced Explosion Phase}
The magnetorotational explosion model m15afB12 is also remarkable for its  behaviour later on during the explosion phase. All the other exploding models exhibit familiar behaviour, with high-frequency emission tapering off a few 
hundred milliseconds after shock revival (though some explosion models have not been evolved sufficiently far into the explosion to show this yet), and some of them develop characteristic tails with modest amplitude offsets of a few $10\, \mathrm{cm}$ due to aspherical shock expansion \citep{murphy_gw_2009,muller_2013_gw}.
In m15afB12, the tail amplitude is unusually high, reaching
$350\, \mathrm{cm}$, and a similarly pronounced tail is developing
in model m39pfB12 as well. This tail is the result of the more highly prolate shock geometry in magnetorotational explosions compared to neutrino-driven explosions (cp.\ Figure~\ref{fig:entro}). The high-frequency emission subsides as expected due to the termination of accretion downflows that could excite strong PNS oscillations.

Model m39pfB12 still sustains strong high-frequency
emission after the onset of explosion. However, the
simulation time of this model is relatively short.
Model m15afB12, which runs beyond $0.6\, \mathrm{s}$
after bounce still shows a clear fingerprint of
a dominant f/g-mode in the spectrogram, but high-frequency
emission subsides to very low levels once accretion
onto the PNS ceases and ceases to excite PNS oscillations.

\subsection{Prospects of Detection}
To characterise the overall strength and frequency structure of the models as parameters that can potentially be constrained by future GW observations, we provide the total emitted energy $E_\mathrm{GW}$ in GWs, the peak amplitude 
$A_{20,\mathrm{max}}^\mathrm{E2}$ prior to the tail phase, and the
peak frequency $f_\mathrm{p}$ of the time-integrated energy
spectrum  in Table~\ref{tab:modelparams}. 
The energy emitted in GWs can be computed in axisymmetry  as \citep{mueller_gravitational_1997}, 
\begin{align}
E_{\mathrm{GW}}=\frac{c^{3}}{G} \frac{1}{32 \pi} \int_{-\infty}^{+\infty}\left(\frac{\mathrm{d} A_{20}^{\mathrm{E} 2}}{\mathrm{d} t}\right)^{2} \mathrm{d} t.
\end{align}
The time-integrated energy spectrum, which is required to determine $f_\mathrm{p}$
is calculated as \citep{muller_2013_gw},
\begin{align}\label{eq:spec}
\frac{\ud E}{\ud f}=\frac{c^{3}}{16 \pi G}(2 \pi f)^{2}\left|\int_{-\infty}^{\infty} e^{-2 \pi i f t} A_{20}^{\mathrm{E} 2}(t) \, \ud t\right| ^{2},
\end{align}
where $f$ is the frequency.

In most cases, $E_\mathrm{GW}$ is yet to plateau by the end of our simulations.  However, we already see trends emerging that are by and large compatible with previous
findings in the literature; i.e., exploding models
generally have higher GW energies than non-exploding ones \citep{muller_2013_gw,radice_characterizing_2019}, and there is a tendency for models which explode earlier to exhibit larger GW energies. These trends are not without exception and subject to significant scatter. For example, the energy
emitted in GWs is exceptionally high in model m39nrB0 even though it fails to explode. It is more than twice as high as for the exploding model m39nrB12, which cannot be explained by the slightly longer simulation time. 
We also see the $E_\mathrm{GW}$ is higher in all $39\,M_{\odot}$ models than in the $ 15\,M_{\odot}$ cases, again confirming the trend towards stronger GW emission for more massive progenitors
found previously in the literature
\citep{muller_2013_gw,radice_characterizing_2019}

There are no clear monotonic trends of the emitted GW energy with magnetisation or rotation. 
The indirect effects of magnetisation and rotation on GW emission through the explosion dynamics fall within stochastic model variations when we consider an integrated measure like $E_\mathrm{GW}$. 

\begin{figure*}
\centering
\begin{subfigure}{0.49\textwidth}
  \centering
  \includegraphics[width=1\linewidth]{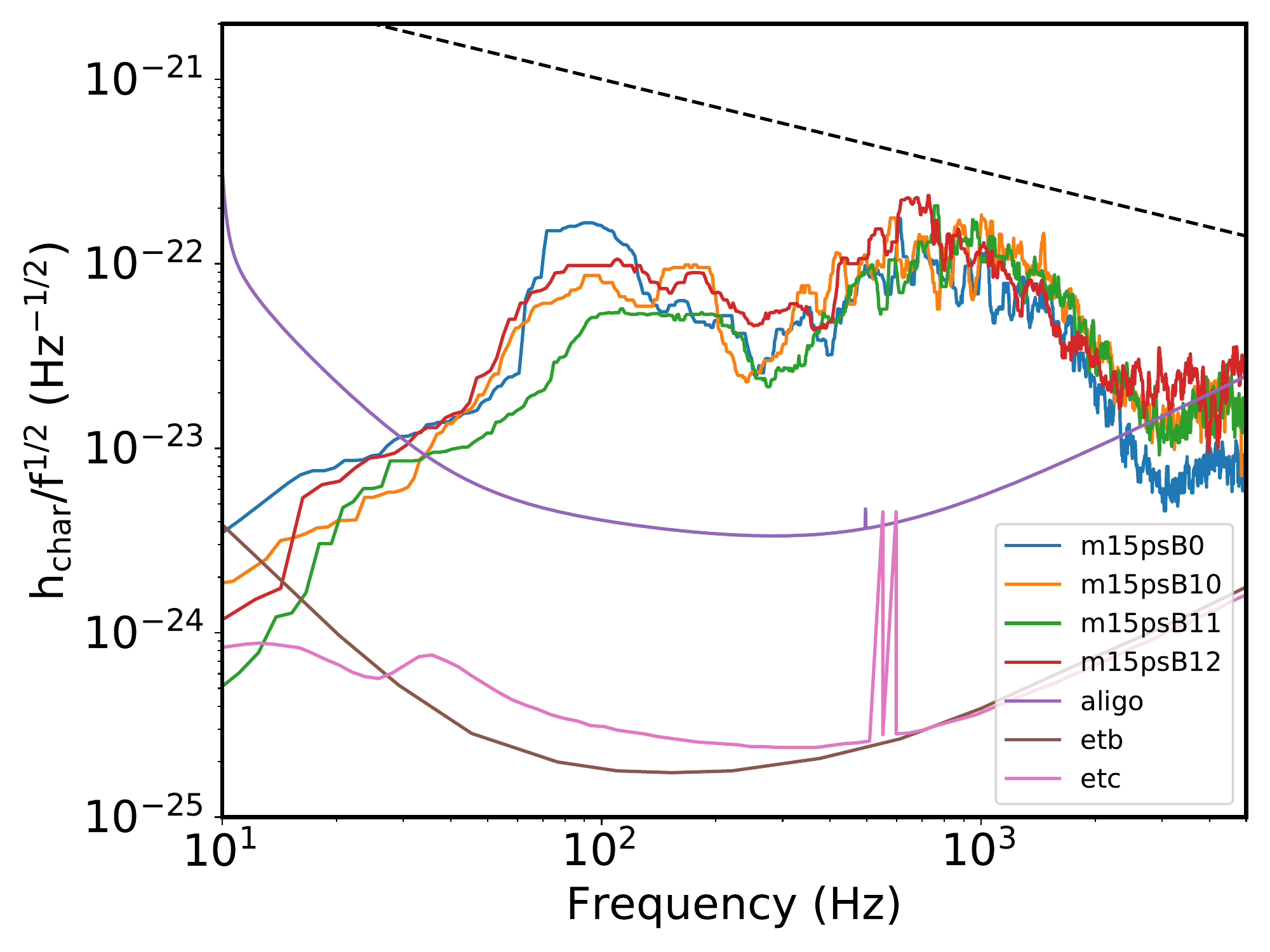}
  \caption{}
\end{subfigure}
\begin{subfigure}{0.49\textwidth}
  \centering
  \includegraphics[width=1\linewidth]{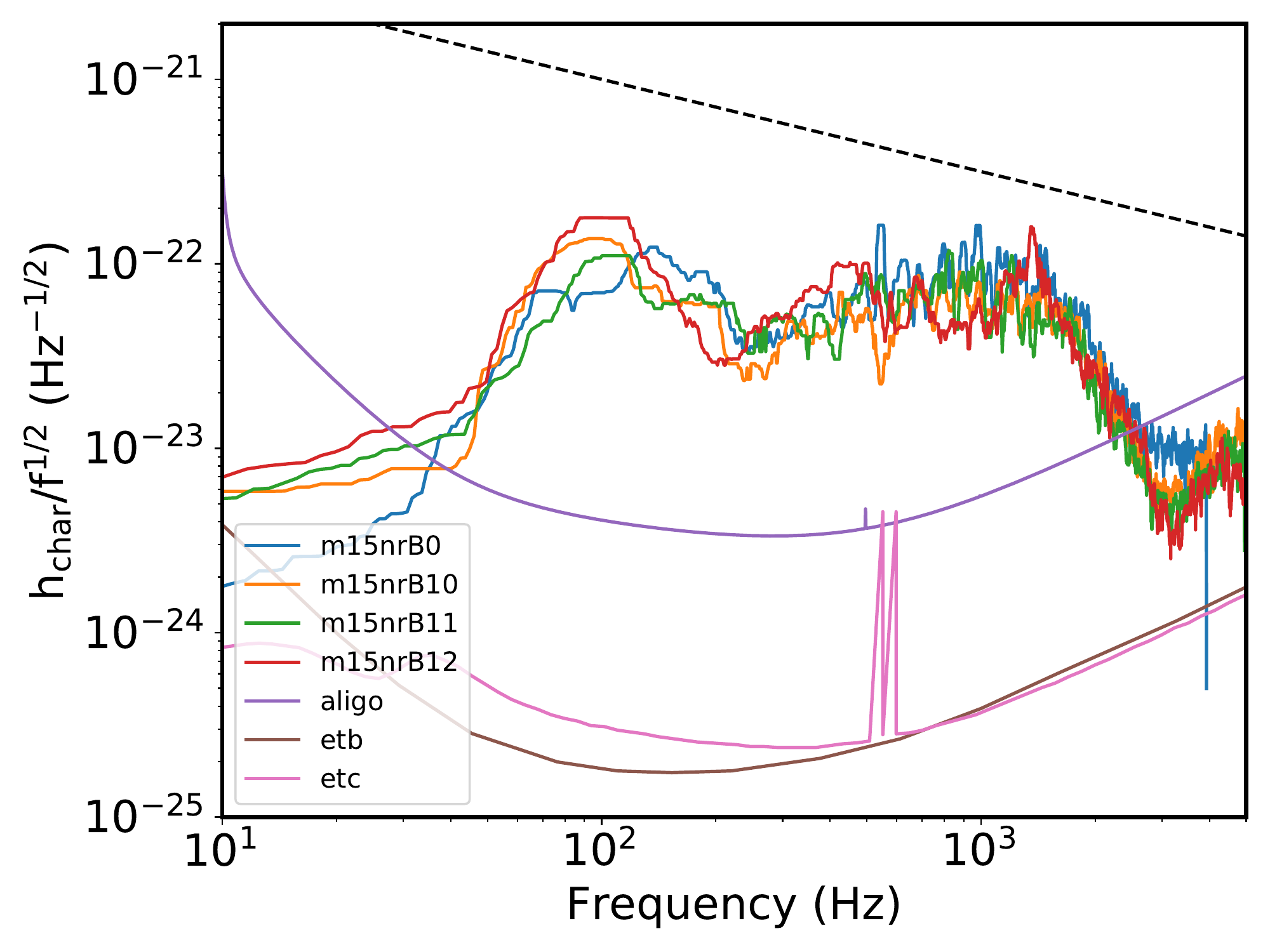}
  \caption{}
\end{subfigure}
\begin{subfigure}{0.49\textwidth}
  \centering
  \includegraphics[width=1\linewidth]{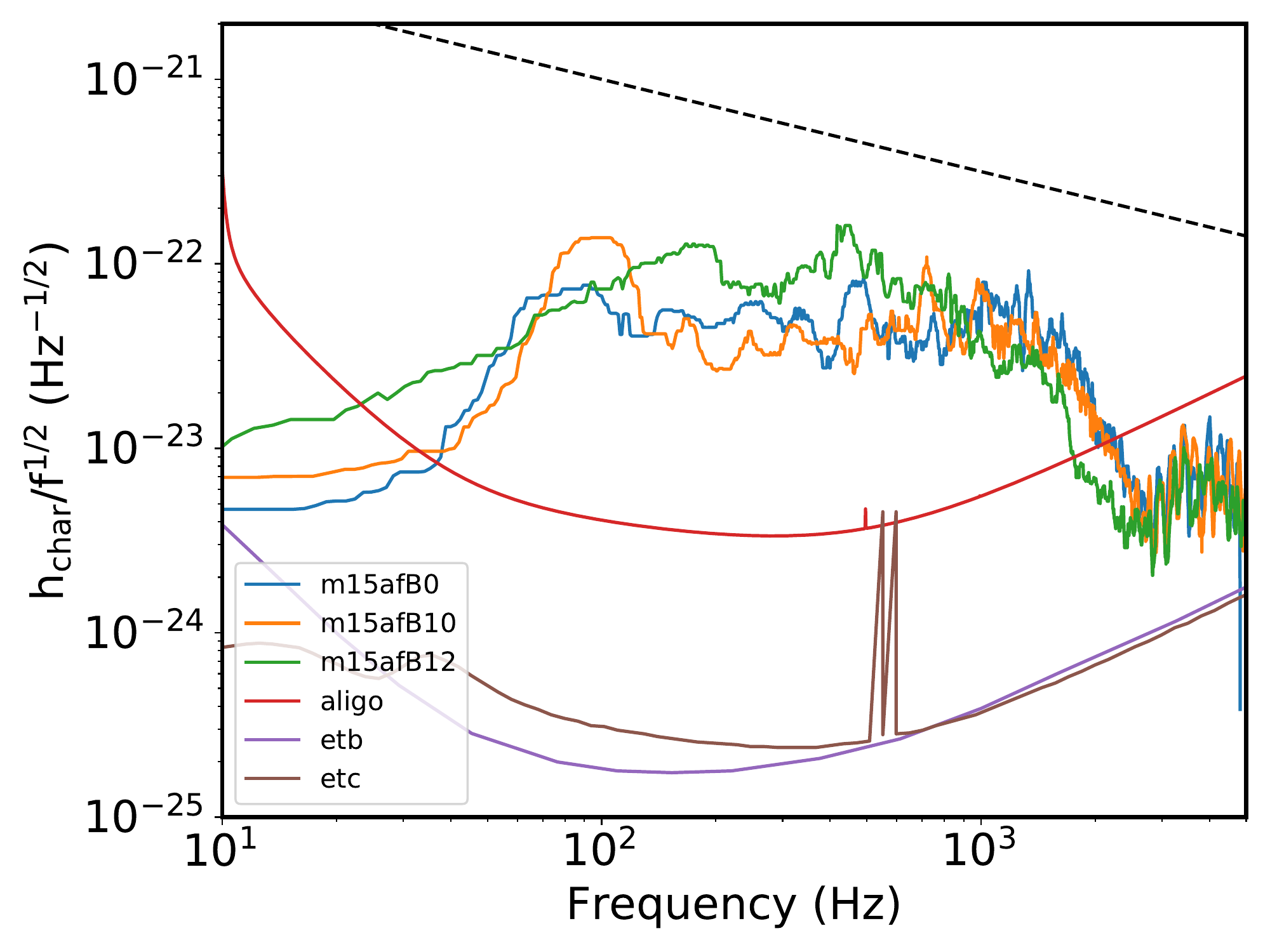}
  \caption{}
  \end{subfigure}
  \begin{subfigure}{0.49\textwidth}
  \centering
  \includegraphics[width=1\linewidth]{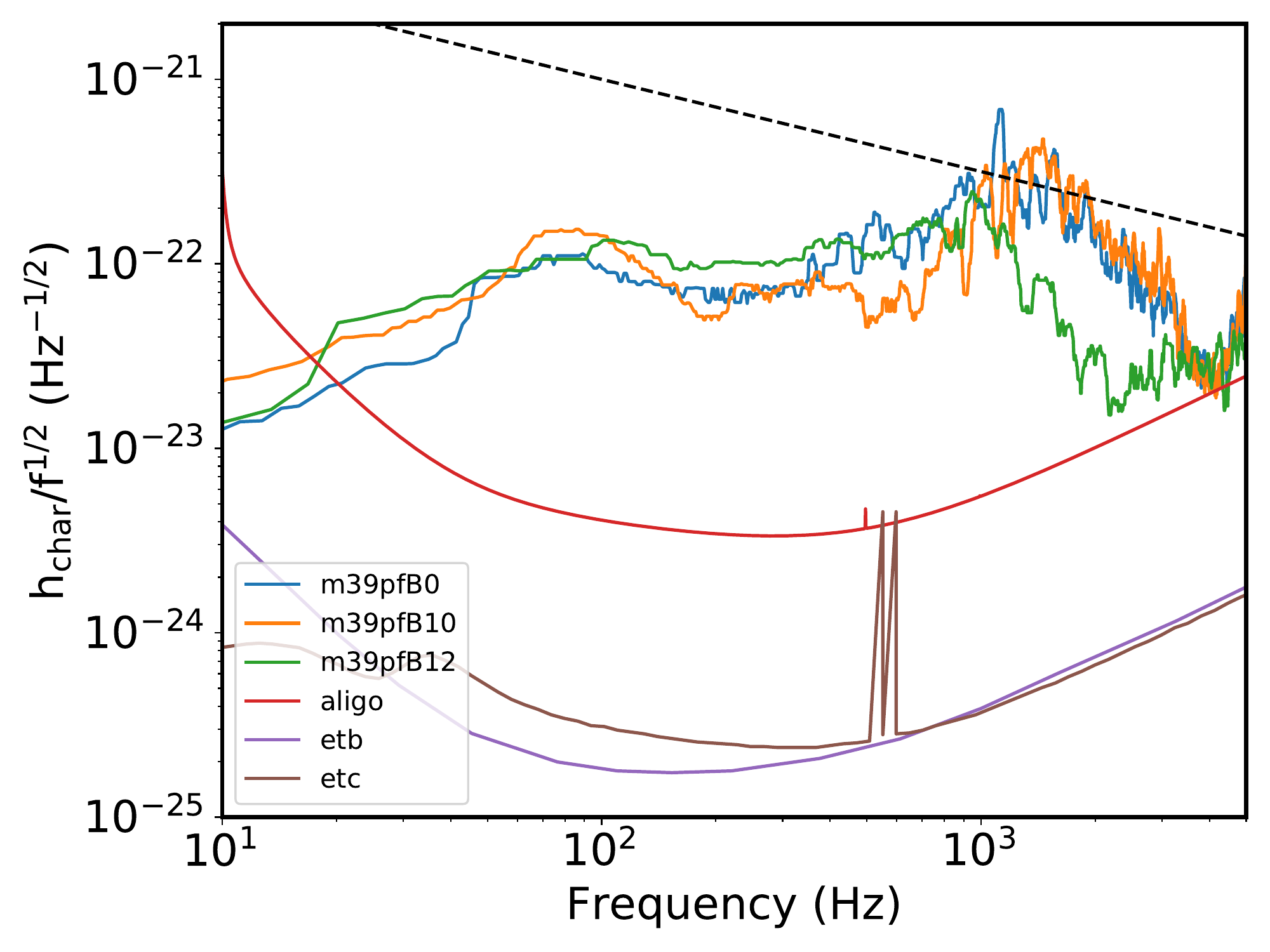}
  \caption{}
\end{subfigure}
\begin{subfigure}{0.49\textwidth}
  \centering
  \includegraphics[width=1\linewidth]{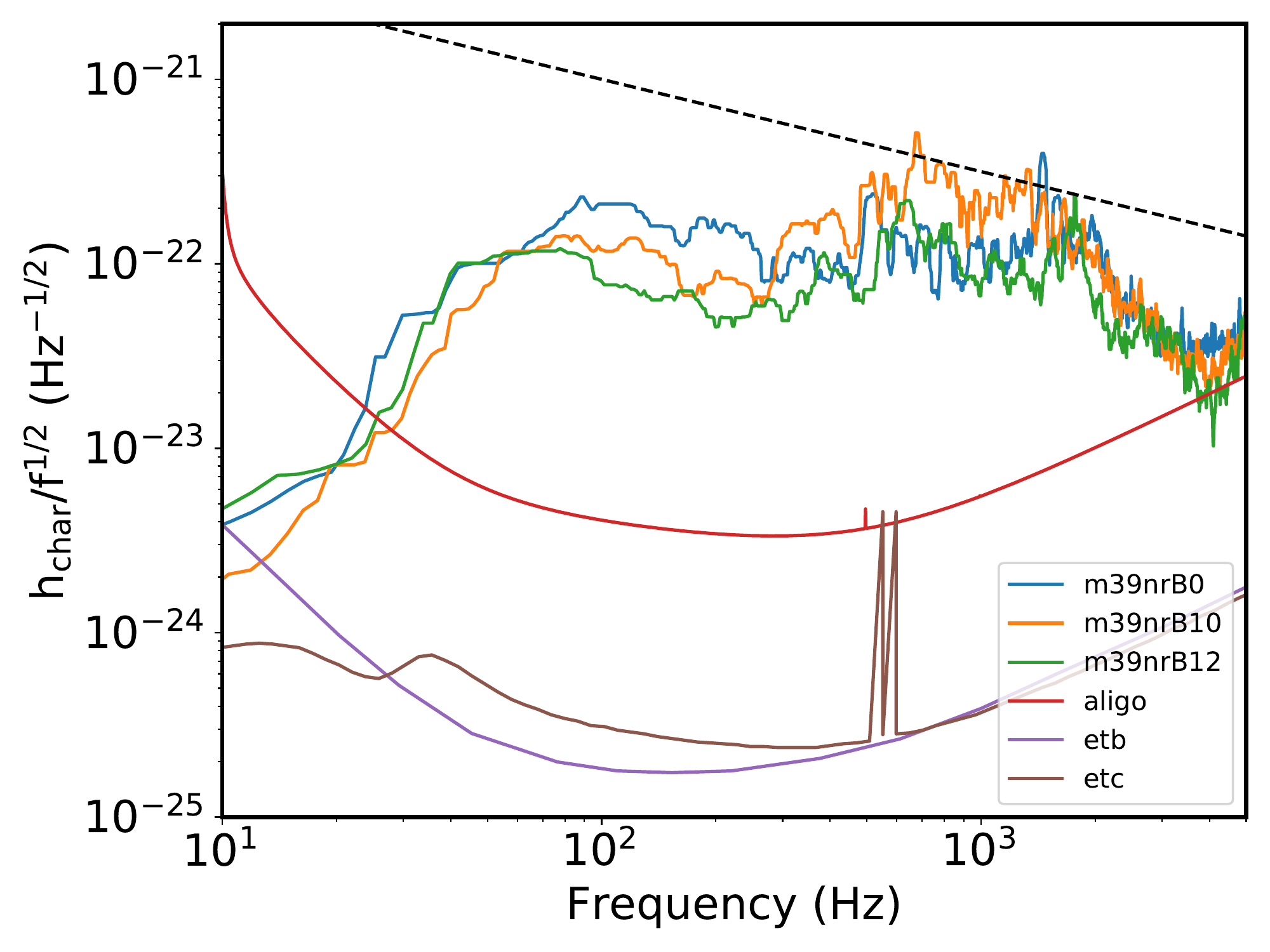}
  \caption{}
\end{subfigure}
 \caption{The smoothed characteristic GW strain 
 $h_\mathrm{char}$ divided by the square root of the frequency $f$ for all our models compared to the amplitude spectral density of the detector noise for
 Advanced LIGO at design sensitivity
 (aLIGO) and configuration ET-B
 and ET-C of the Einstein Telescope.
 The signal curves assume a distance of $10\, \mathrm{kpc}$ and an optimal orientation and location of the detector in the equatorial plane of the models. The slope of lines of constant spectral energy density
 $\ud E/\ud f$  is indicated by dashed black lines.
}
\label{fig:hchar}
\end{figure*}

There is also hardly any systematic effect on the peak
frequency $f_\mathrm{p}$ of the spectrum, and on the shape
of the spectrum, even for the magnetorotational models
m15afB12 and m39pfB12.
To illustrate the shape of the time-integrated spectra we plot the characteristic strain $h_\mathrm{char}$ as a quantity that is closely related to the energy spectrum $\ud E/\ud f$, and is also useful for determining the detectability of the waveforms \citep{flanagan_measuring_1998}:
\begin{align}\label{eq:hchar}
    h_{\mathrm{char}}(f)=\sqrt{\frac{2G}{\pi ^2 c^3 R^2}\dv{E}{f}}.
\end{align}
Figure~\ref{fig:hchar} shows $h_{\mathrm{char}}/f^{1/2}$
for all models at a distance of $10\, \mathrm{kpc}$ along with the square root of the power spectral density noise
$S(f)$ for Advanced LIGO
at design sensitivity (aLIGO; \citealp{aLIGO2015,aligosense}) and 
two possible configurations, ET-B and ET-C, for
the future Einstein Telescope \citep{hild_etb2008,hild_etc_2009}.

For the magnetorotational models m15afB12 and
m39pfB12, one would expect a strong reflection of the
tail signal in the spectrum;
previous studies often found a nearly
linear slope on the double-logarithmic scale as in
Figure~\ref{fig:hchar}, corresponding to a power law in frequency.
However, such a power-law spectrum is an artefact of using a
rectangular window with a finite duration for Fourier transforming
the signal and does not accurately capture the detector response
to a slowly varying tail signal. Using a rectangular window introduces
a spurious discontinuity in the Fourier-transformed periodic function. No
such discontinuity appears in reality; the real signal would vary smoothly
before and after the simulated time window. We experimented with various
methods such as continuous padding to verify that the low-frequency power
is significantly overestimated using the standard approach. For
pragmatic reasons, we follow our previous papers \citep{2019MNRAS.487.1178P,powell_20,powell_21}
and simply apply a high-pass Butterworth filter
that removes frequency components below $10\, \mathrm{Hz}$ and has a
smooth attenuation function above.
Model m15afB12 does still show higher power than the
other $15 M_\odot$ models at low frequency, and
it also has a rather low peak frequency of $f_\mathrm{p}=450\, \mathrm{Hz}$, but still does not
stand apart considerably from the other models.

All the  other models exhibit a more familiar shape  with a recognisable broad peak around $100\, \mathrm{Hz}$ in some models (due to the early quasi-periodic signal and later SASI activity) and a broad distribution of power at high frequencies that usually declines above $1\texttt{-}1.5\, \mathrm{kHz}$.

Based on the characteristic strain, we estimate the detectability of our GW models 
by computing the optimal signal-to-noise ratio (SNR) from $20\texttt{-}2000\, \mathrm{Hz}$ at the typical distance of $10$\,kpc for a galactic core-collapse supernova. 
Assuming optimal detector orientation and an observer in the equatorial plane, the SNR  can be computed as 
\citep{flanagan_measuring_1998}
\begin{align}\label{eq:SNR}
\langle \mathrm{SNR}^2 \rangle =
\frac{15}{8}
\int \frac{h_{\mathrm{char}}(f)^2}{h_{\mathrm{RMS}}(f)^2}
\,\ud \ln f
=\frac{15}{8}\int \frac{h_{\mathrm{char}}(f)^2}{f S(f)}
\,\ud \ln f,
\end{align}
where $h_{\mathrm{RMS}}(f)=\sqrt{f S(f)}$ in
terms of the the power spectral density the noise in the detector \citep{flanagan_measuring_1998,moore_15}. Note that an extra
factor $15/8$ appears for
a detector in the equatorial direction compared
to Equation~(5.2) of \citet{flanagan_measuring_1998}, which involves averaging over source directions (but not over sub-optimal detector orientations).
We note that this simple procedure ignores
complications  like non-stationarity,  non-Gaussianity and glitches 
\citep[for a recent discussion, see][]{szczepanczyk_21}, and there are also ambiguities due to different possible choices of windowing, padding, and high-pass/low-pass filtering to remove frequencies outside the detector band. Especially the contribution of low-frequency tails to signal detectability must be considered somewhat uncertain.
SNRs at a distance of $10\, \mathrm{kpc}$
and estimated detection distances based on a threshold SNR of 8
for Advanced LIGO at design sensitivity and configurations ET-B and ET-C of the Einstein Telescope are shown in Table~\ref{tab:snr}.
For a non-templatable signal, this condition  can predict somewhat too optimistic detection distances, but the criterion still yields indicative results \citep{flanagan_measuring_1998,abjgw2017} in the ballpark of more more sophisticated signal analysis methods
\citep{logue_12,hayama_15,gossan_16,powell_17, szczepanczyk_21}.

For Advanced LIGO, we obtain detection distances 
from a few $10\, \mathrm{kpc}$
up to about $140\, \mathrm{kpc}$
for most of our models.
This is within the typical range of detection distances for current CCSN simulations \citep{gossan_16,srivastava_19,szczepanczyk_21}, though maybe somewhat on the high side of the distribution. All models would thus be comfortably detectable within the Milky Way  with current instruments, and the $39 M_\odot$ models
would likely remain detectable in the Large Magellanic Cloud at $50\, \mathrm{kpc}$ even for non-optimal orientation. With the Einstein Telescope,
most waveforms would be detectable in M31
for optimal orientation, and in case of m39nrB0
and m39nrB10 possibly even in the nearest galaxies outside the Local Group (e.g., NGC~300 at $1.9\, \mathrm{Mpc}$). Interestingly, the magnetorotational models m15afB12 and m39pfB12 do \emph{not} have the highest detection distances. Because of better overlap with the sensitivity range of Advanced LIGO and the Einstein Telescope, the non-rotating models
m39nrB0 and m39nrB10 yield the highest SNRs. It must be borne in mind, however, that strong GW emission is still ongoing by the end of the simulation in case of
m39pfB12; and we also have not included the rotational bounce signal. Model m39pfB12 is arguably more representative of hypernova progenitors than m15afB12 
with its artificially imposed rotation profile.
It therefore stands to reason that this particular magnetorotational models would eventually reach similar detection distances as m39nrB0 and m39nrB10, and be detectable throughout the Local Group.
With a fraction of
$\mathord{\sim}1\%$ of CCSNe exploding
as broad-lined Ic ``hypernovae'' \citep{smith_11}, putatively powered by some form of magnetorotational explosion, 
and $\mathord{\sim}10$ CCSNe per century
\citep{mattila2001supernovae,srivastava_19} in the Local Group, this would suggest a chance of about $10\%$ for observing a hypernova in GWs in the next hundred years.

Variations in the maximum detection distance due to the impact of rotation and magnetic fields on the explosion dynamics are modest for the most part. 
For the $15\,M_\odot$ models, detection distances for Advanced LIGO fall between $45\, \mathrm{kpc}$ and $85\, \mathrm{kpc}$, i.e., they vary by less than a factor of two. 
The $39 M_\odot$ models, which generally have SNRs almost twice as high
as the $15\,M_\odot$ models, show even less
of an impact of rotation and magnetic fields on the detection distance, 
which varies
from $82\, \mathrm{kpc}$
to $146\, \mathrm{kpc}$.

It is noteworthy that the  detection distance for model m39pfB0 is, by and large, comparable to the corresponding 3D model of \citet{powell_20}.
A direct comparison between our model and theirs
is not straightforward because they employed
the CFC approximation, which leads to a
 systematic decrease of the f/g-mode frequency in general relativity \citep{muller_2013_gw}. Physically, there are different effects at play in the 2D and 3D models that partly compensate each other with respect to detectability, i.e., lower peak amplitudes in the 3D model, but also a lower f/g-mode frequency due the destabilisation of the PNS surface by a negative angular momentum gradient 
\citet{powell_20}, and there are two different polarisation modes in 3D. Overall the comparison between  \citet{powell_20} and our study illustrates that GW detection distances need not \emph{always} be significantly lower in 3D simulations than in 2D, provided that
3D models explode successfully and develop a pronounced bipolar geometry. However, in many of our models, especially the non-exploding ones, the assumption of 2D axisymmetry may overestimate
the wave amplitudes \citep{abjgw2017}: 3D simulations are always required for quantitative accuracy even if 2D models can already reveal trends in GW emission and important features
of the waveforms.

The small impact of magnetic fields and the rotation on waveform detectability, even in the extreme case of a magnetorotational explosion, is due to the nature of the prominent signal features that appear in these cases. The pronounced tail signal primarily adds some power at low frequencies. Thus power is added to the spectrum mostly outside the region where Advanced LIGO, Virgo \citep{2015CQGra..32b4001A}, and KAGRA \citep{2020arXiv200909305A} and planned third-generation detectors like the Einstein Telescope and Cosmic Explorer \citep{2019BAAS...51g..35R} will be most sensitive.

\begin{table*}
\centering
 \begin{tabular}{||c c c c c c c||} 
 \hline
  & \multicolumn{2}{c}{aLIGO} &
  \multicolumn{2}{c}{ET-B} &
  \multicolumn{2}{c}{ET-C} \\
 [0.5ex]
  Model &  SNR &  Range & SNR &  Range & SNR &  Range\\ [0.5ex]
 \hline\hline
  m15psB0 & 58 & 73\, kpc & 1140 & 1.43\, Mpc & 760 & 950\, kpc\\
 \hline
  m15psB10 & 57 & 72\, kpc & 1000 & 1.25\, Mpc & 760 & 950\, kpc \\ 
 \hline
 m15psB11 & 47 & 59\, kpc & 770 & 960\, kpc & 650 & 810\, kpc\\
 \hline
 m15psB12 & 70 & 87\, kpc & 1210 & 1.51\, Mpc & 890 & 1.12\, Mpc\\
 \hline
 m15afB0 & 37 & 46\, kpc & 690 & 860\, kpc & 500 & 620\, kpc\\
 \hline
 m15afB10 & 42 & 53\, kpc & 850 & 1.07\, Mpc & 560 & 710\, kpc\\
  \hline
 m15afB12 & 61 & 77\, kpc & 1130 & 1.41\, Mpc & 820 & 1.02\, Mpc\\
 \hline
  m15nrB0 & 52 & 65\, kpc & 950 & 1.18\, Mpc & 700 & 880\, kpc\\
 \hline
 m15nrB10 & 48 & 61\, kpc & 970 & 1.21\, Mpc & 650 & 810\, kpc\\
 \hline
  m15nrB11 & 45 & 56\, kpc & 840 & 1.05\, Mpc & 600 & 750\, kpc\\
 \hline
 m15nrB12 & 62 & 78\, kpc & 1250 & 1.57\, Mpc & 840 & 1.04\, Mpc\\
 \hline
 m39pfB0 & 94 & 118\, kpc & 1510 & 1.88\, Mpc & 1320 & 1.65\, Mpc\\
 \hline
 m39pfB10 & 81 & 101\, kpc & 1390 & 1.74\, Mpc & 1140 & 1.42\, Mpc\\
 \hline
 m39pfB12 & 87 & 109\, kpc & 1590 & 1.99\, Mpc & 1170 & 1.46\, Mpc\\
 \hline
m39nrB0 & 107 & 134\, kpc & 2080 & 2.6\, Mpc & 1440 & 1.8\, Mpc\\
 \hline
m39nrB10 & 117 & 146\, kpc & 1930 & 2.42\, Mpc & 1580 & 1.98\, Mpc\\
 \hline
m39nrB12 & 67 & 84\, kpc & 1200 & 1.50\, Mpc & 860 & 1.08\, Mpc\\
 \hline
\end{tabular}
\caption{Key waveform parameters of relevance for the detectability of the 17 models. SNR denotes the optimal signal-to-noise ratio of each model calculated according to Equation~\eqref{eq:SNR} over the frequency  from $20\,\mathrm{Hz}$ to $2\, \mathrm{kHZ}$ at a distance of $10\, \mathrm{kpc}$ assuming an optimal location and orientation of the detector for Advanced LIGO at design sensitivity
(aLIGO) and configurations ET-B and ET-C of the
 Einstein Telescope. For all three cases, ``range'' denotes the estimated maximum detection distance which is obtained by assuming a threshold signal-to-noise ratio of 8.
 }
\label{tab:snr}
\end{table*}

\section{Conclusions}
\label{sec:conclusions}
In this work, we conducted 17 different 2D CCSN simulations for $15\,M_{\odot}$ and $39\,M_{\odot}$ progenitor stars with the \textsc{CoCoNuT-FMT} code to study the impact
of rotation and magnetic fields on the post-bounce dynamics and GW emission. We considered fast, slow and non-rotating pre-collapse  models with
initial magnetic field strengths of
$0\, \mathrm{G}$, $10^{10}\, \mathrm{G}$, $10^{12} \, \mathrm{G}$ and in two cases $10^{11}\, \mathrm{G}$ in the progenitor core.

In most cases, we find that strong initial magnetic fields aid the development of explosions independent
of the initial rotation rate, in line with previous
studies of MHD effects in CCSN simulations
of non-rotating progenitors in 2D
\citep{obergaulinger_14} and 3D \citep{mueller_20b}, and of rapidly rotating progenitors \citep{winteler_12,moesta_14b,obergaulinger_17,obergaulinger_20,kuroda_magnetorotational_2020}. However, we also found one case where strong magnetic fields
delay the development of an explosion by inhibiting the growth of the turbulent kinetic energy in the gain region, similar to the recent study of \citet{matsumoto_20}. 
The impact of rotation on the post-bounce dynamics tends to be more non-monotonic, e.g., in the $15\,M_\odot$ models conditions are more favourable for moderate rotation and less favourable for rapid rotation and no rotation. Again, this is consistent with previous studies of rotational effects on the post-bounce dynamics \citep{summa_rotation-supported_2018}. For rapid rotation and strong initial magnetic fields, we find powerful magnetorotational explosions for both progenitors. The rapidly rotating, strongly magnetised
$15\,M_\odot$ model has reached an explosion energy of
$8.6 \times 10^{51}\, \mathrm{erg}$ by the end of the simulation, and is still growing at an unabated pace.

In all cases, the GW signals from our models exhibit similar patterns as in systematic studies of the
post-bounce GW emission from non-rotating, non-magnetised progenitors
\citep{murphy_gw_2009,muller_2013_gw,abjgw2017,Morogw2018,radice_characterizing_2019,2019MNRAS.487.1178P,mezzacappa_2020}
with a dominant high-frequency f/g-mode emission band whose power peaks around the onset of explosion, and in some case low-frequency emission from the SASI. Variations in peak amplitude, the total energy emitted in GWs, and the SNR in GW detectors are modest and can be traced to variations in explosion outcome (shock revival vs.\ failure) or the time and strength of explosion.

Deviations from the usual behaviour are found for rapidly rotating and strongly magnetised models, however. For the rapidly rotating $15\,M_\odot$ models, we find that the familiar analytic frequency relation
 $ f_{\text {peak }}
\approx (2 \pi)^{-1} G M_{\text {by}}/R_{\text {PNS}}^{2} (1.1 m_\mathrm{n}/\left\langle E_{\bar{v}_{e}}\right\rangle)^{1/2}$
for the dominant f/g-mode \citep{muller_2013_gw} no longer holds because the stabilising influence of a positive angular momentum gradient shifts the mode frequency upward by up to $20\%$. 

For  magnetorotational models, the trajectory of the dominant high-frequency emission band bears little resemblance to the standard frequency relation,
but a single dominant high-frequency emission mode,
likely a modified f/g-gmode remains visible in the spectrograms. In the magnetorotational explosion
of the $15 M_\odot$ progenitor, high-frequency emission
subsides to a low level some time after the onset
of the explosion because accretion on the proto-neutron
star essentially ceases.
Due to their strongly prolate explosion geometry,  the matter signal from magnetorotational explosions is characterised by a very pronounced
tail signal during the explosion phase. While such a strong tail in the matter signal appears unique to jet-driven explosions, similarly strong tails may arise from anisotropic neutrino emission and therefore cannot be easily used to identify magnetorotational explosions.

The estimated maximum detection distances for our models amount to several $10\, \mathrm{kpc}$
to $140\, \mathrm{kpc}$ for Advanced LIGO and $0.6\texttt{-}2.6 \, \mathrm{Mpc}$
for the Einstein Telescope, and fall within the range of predicted detection distances in the literature. The magnetorotational models do not stick out significantly in terms of detectability.
The highest detection distances are predicted
for neutrino-driven explosion models, although
this might yet change for longer simulations
because of continued GW emission in
the two magnetorotational models. Moreover, the
$15 M_\odot$ model may not be representative of
hypernova progenitors, as it is based on a ``normal''
supernova progenitor with an artificially imposed rotation profile. The magnetorotational explosion of the $39 M_\odot$
star would be detectable 
 throughout most of  the Local Group. Given the local supernova rate and hypernova fraction, this suggests a chance of $10\%$ per century for observing a hypernova in GWs.

Given the inherent limitations of axisymmetric models, our work is intended as an exploratory and descriptive study that seeks to identify effects of magnetic fields and rotation on the supernova GW signal that should then be followed up further in 3D simulations. 
Future 3D simulations should address the unusual  time-frequency structure of the GW signal and the underlying emission mechanisms in magnetorotational hypernova explosions in more detail. As in the case of rapidly rotating, non-magnetised models
\citep{2014PhRvD..89d4011K,andresen_19,shibagaki_20} genuinely new features may appear in 3D in the GW signals of hypernovae. In addition to the GW fingerprint of magnetoconvection and and $\alpha$-$\Omega$ dynamo in the PNS \citep{raynaud_21}, the magnetic fields generated by the magnetorotational instability \citep{balbus_91,akiyama_03} or dynamo amplification within the gain region or at the PNS surface may leave unexpected traces in the GW signals. Although the odds are stacked against a nearby hypernova explosion, the GW emission may have significant potential to elucidate the nature of an extreme supernova explosion in case of such a lucky event.

\section*{Acknowledgements}
BM acknowledges support by ARC Future Fellowship FT160100035.
JP is supported by the Australian Research Council (ARC) Centre of 
Excellence for Gravitational Wave Discovery (OzGrav), through project number 
CE170100004 and by the ARC Discovery Early Career Researcher Award (DECRA) project number DE210101050.
 This research
was undertaken with the assistance of resources and services from the
National Computational Infrastructure (NCI), which is supported by the
Australian Government.  It was supported by resources provided by the
Pawsey Supercomputing Centre with funding from the Australian
Government and the Government of Western Australia.
We acknowledge computer time allocations from NCMAS and ASTAC.
Some of this work was 
performed on the OzSTAR national facility at Swinburne University of Technology. 
OzSTAR is funded by Swinburne University of Technology and the National 
Collaborative Research Infrastructure Strategy (NCRIS).

\section*{Data Availability}

The data from our simulations will be made available upon reasonable requests made to the authors.

\bibliographystyle{mnras}
\bibliography{references.bib}

\end{document}